\documentclass[11pt, letterpaper]{article}
\usepackage[letterpaper,left=1in,right=1in,top=1in,bottom=1.0in,footskip=0.4in]{geometry}
\usepackage{verbatim}
\usepackage{titling}
\usepackage{cancel}
\usepackage{enumitem} 
\setlist{noitemsep,topsep=0pt,parsep=0pt} 
\usepackage{subfig}
\usepackage{mathtools} 
\usepackage{multirow}
\usepackage[all]{xy}
\usepackage{dashbox}

\usepackage{fancybox,framed}
\usepackage{tikz}
\usetikzlibrary{arrows,backgrounds,calc,fit,decorations.pathreplacing,decorations.markings,shapes.geometric}

\tikzset{every fit/.append style=text badly centered}

\usepackage{tyson}
\usepackage{bm}
\usepackage{lipsum}
\usepackage{blindtext}
\usepackage{chngcntr}

\numberwithin{figure}{section}
\numberwithin{table}{section}
\usepackage[bookmarks=true,hypertexnames=false]{hyperref}
\hypersetup{colorlinks=true, citecolor=blue, linkcolor=red, urlcolor=blue}
\makeatletter
\newcommand{\rmnum}[1]{\romannumeral #1}
\newcommand{\Rmnum}[1]{\expandafter\@slowromancap\romannumeral #1@}
\makeatother

\newcommand{\ii}{\mathfrak{i}}
\newcommand{\Holant}{\operatorname{Holant}}
\newcommand{\PlHolant}{\operatorname{Pl-Holant}}
\newcommand{\holant}[2]{\ensuremath{\Holant\left(#1\mid #2\right)}}
\newcommand{\plholant}[2]{\ensuremath{\PlHolant\left(#1\mid #2\right)}}

\newcommand{\CSP}{\operatorname{\#CSP}}
\newcommand{\PlCSP}{\operatorname{Pl-\#CSP}}

\newenvironment{remark}{\medskip{\bfseries \noindent Remark:}}{\par\medskip}{\par\medskip}

\def\borderColor{blue!60}
\def\scale{0.6}
\def\nodeDist{1.4cm}
\tikzstyle{internal} = [draw, fill, shape=circle]
\tikzstyle{external} = [shape=circle]
\tikzstyle{square}   = [draw, fill, rectangle]
\tikzstyle{triangle} = [draw, fill, regular polygon, regular polygon sides=3, inner sep=3pt]
\tikzstyle{pentagon} = [draw, fill, regular polygon, regular polygon sides=5, inner sep=2pt, minimum size=14pt]

\begin{document}

\title{\bf{New Planar P-time Computable Six-Vertex Models and a Complete Complexity Classification}}

\vspace{0.3in}

\author{Jin-Yi Cai\thanks{Department of Computer Sciences, University of Wisconsin-Madison. Supported by NSF CCF-1714275.
 } \\ {\tt jyc@cs.wisc.edu}
\and Zhiguo Fu\thanks{School of Information Science and Technology, Northeast Normal University. Supported by NSFC-61872076, 
 Natural Science Foundation of Jilin Province 20200201161JC and Fundamental Research Funds for Central Universities.}\\ \tt
fuzg432@nenu.edu.cn
\and
Shuai Shao$^\ast$ \\ \tt sh@cs.wisc.edu}

\date{}
\maketitle
\thispagestyle{empty}
\bibliographystyle{plain}

 \begin{abstract}
We discover new  P-time computable six-vertex models
on planar graphs beyond 
Kasteleyn's algorithm 
for counting planar perfect matchings.\footnote{This is also known as the FKT algorithm.
Fisher and  Temperley~\cite{TF61}, and Kasteleyn~\cite{Kasteleyn1961} independently discovered
this algorithm on the grid graph. Subsequently Kasteleyn~\cite{Kasteleyn1967} 
generalized this to all
planar graphs.}  We further prove that there are \emph{no more}: Together, they exhaust \emph{all} 
P-time computable six-vertex models on planar graphs, assuming 
\#P is not P. 
This leads to the following exact 
complexity
classification:
For \emph{every} parameter setting 
in ${\mathbb C}$ for the six-vertex model,
the partition function is
either (1) computable in 
P-time for every graph, or
(2) \#P-hard for general graphs but computable in 
P-time for
planar graphs, or
(3) \#P-hard  even for planar graphs.
The classification has an explicit criterion.
The new P-time cases
in (2) provably cannot be subsumed by Kasteleyn's algorithm.
They are obtained by a non-local connection to \#CSP,
defined in terms of
a ``loop space''. 

This is the first substantive advance toward a planar Holant classification
with not necessarily symmetric constraints.
%
We introduce M\"{o}bius transformation on  ${\mathbb C}$ as a powerful
new tool in hardness proofs for counting problems.
 \end{abstract}

\newpage
\setcounter{page}{1}

\section{Introduction}\label{sec:intro}

Partition functions are Sum-of-Product computations.
In physics, one considers a set of particles connected by some
bonds. Then physical laws impose various local constraints,
with suitable weights for valid local configurations.
If a global configuration $\sigma$ satisfies
all local constraints, the product of local weights
is the weight of $\sigma$, and the sum over all such
$\sigma$ is the value of the partition function.
The partition function encodes much information about a physical system.

By definition, a partition function is an exponential sized sum.
But in some cases, clever algorithms exist that can compute
it in P-time.
Well-known examples of partition functions from physics
include the Ising model, Potts model, hardcore gas and
the \emph{six-vertex model}~\cite{baxter2016exactly, bax}.
Most of these are spin systems~\cite{jerrum-sinclair,
goldberg-jerrum-patterson,goldberg-jerrum-potts,lu-ying-li}. If particles take
$+$/$-$ spins, each can be modeled by a  Boolean variable,
and local constraints are expressed by edge (binary) constraint functions.
These are nicely modeled by the \#CSP framework~\cite{bulatov,dyer-richerby, bulatov2012csp, cai-chen-lu-nonnegative-csp, caichen}.
Some physical systems are more naturally described as orientation problems,
and these can be modeled by Holant problems~\cite{Cai-Lu-Xia}, of which
\#CSP is a special case.
In this paper we study the six-vertex model,
which consists of orientation problems.

The six-vertex model has a  long history in 
physics.
Pauling in 1935 introduced the six-vertex model
to account for the residual entropy
of water ice~\cite{Pauling}.
Consider a large number of oxygen and hydrogen atoms
in a  1 to 2 ratio.
Each oxygen atom (O) is connected by a bond to four other neighboring oxygen
atoms (O), and each bond is occupied by one hydrogen atom (H).
Physical constraint requires that each (H) is closer to exactly one
of the two neighboring (O).
 Pauling argued~\cite{Pauling} that, furthermore, the allowed configurations
are such that at each oxygen (O) site,  exactly two hydrogen (H)
are closer to it, and the other two are farther away.
This can be naturally represented by a 4-regular graph.
The constraint on the placement of hydrogen atoms (H) can be represented by
an orientation of the edges of the graph, such that
at every vertex (O), the in-degree and out-degree are both 2.
In other words, this is an \emph{Eulerian orientation} \cite{Mihail-Winkler, cai2020beyond}.
Since there are ${4 \choose 2} =6$ local valid configurations,
this is called the six-vertex model.
In addition to  water ice,
potassium dihydrogen phosphate KH$_2$PO$_4$ (KDP) also
satisfies this model.
\vspace{-4ex}
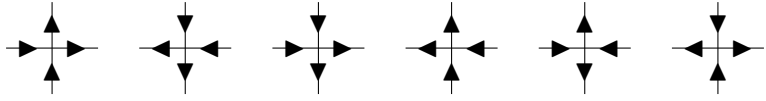
\begin{figure}[!hb]
\centering
\subfloat{
\begin{tikzpicture}[scale=0.4]
\node [external] (1) at (2, 0) {};
\node [external] (2) at (2, 4) {};
\node [external] (3) at (0, 2) {};
\node [external] (4) at (4, 2) {};
\draw (1) to (2);
\draw (3) to (4);
\node at (1.2, 2) {$\blacktriangleright$};
\node at (2.8, 2) {$\blacktriangleright$};
\node at (2, 1.2) {$\blacktriangle$};
\node at (2, 2.8) {$\blacktriangle$};
\end{tikzpicture}
}
\hspace*{-1.5em}
 \subfloat{
\begin{tikzpicture}[scale=0.4]
\node [external] (1) at (2, 0) {};
\node [external] (2) at (2, 4) {};
\node [external] (3) at (0, 2) {};
\node [external] (4) at (4, 2) {};
\draw (1) to (2);
\draw (3) to (4);
\node at (1.2, 2) {$\blacktriangleleft$};
\node at (2.8, 2) {$\blacktriangleleft$};
\node at (2, 1.2) {$\blacktriangledown$};
\node at (2, 2.8) {$\blacktriangledown$};
\end{tikzpicture}
}
\hspace*{-1.5em}
 \subfloat{
\begin{tikzpicture}[scale=0.4]
\node [external] (1) at (2, 0) {};
\node [external] (2) at (2, 4) {};
\node [external] (3) at (0, 2) {};
\node [external] (4) at (4, 2) {};
\draw (1) to (2);
\draw (3) to (4);
\node at (1.2, 2) {$\blacktriangleright$};
\node at (2.8, 2) {$\blacktriangleright$};
\node at (2, 1.2) {$\blacktriangledown$};
\node at (2, 2.8) {$\blacktriangledown$};
\end{tikzpicture}
}
\hspace*{-1.5em}
 \subfloat{
\begin{tikzpicture}[scale=0.4]
\node [external] (1) at (2, 0) {};
\node [external] (2) at (2, 4) {};
\node [external] (3) at (0, 2) {};
\node [external] (4) at (4, 2) {};
\draw (1) to (2);
\draw (3) to (4);
\node at (1.2, 2) {$\blacktriangleleft$};
\node at (2.8, 2) {$\blacktriangleleft$};
\node at (2, 1.2) {$\blacktriangle$};
\node at (2, 2.8) {$\blacktriangle$};
\end{tikzpicture}
}
\hspace*{-1.5em}
 \subfloat{
\begin{tikzpicture}[scale=0.4]
\node [external] (1) at (2, 0) {};
\node [external] (2) at (2, 4) {};
\node [external] (3) at (0, 2) {};
\node [external] (4) at (4, 2) {};
\draw (1) to (2);
\draw (3) to (4);
\node at (1.2, 2) {$\blacktriangleright$};
\node at (2.8, 2) {$\blacktriangleleft$};
\node at (2, 1.2) {$\blacktriangledown$};
\node at (2, 2.8) {$\blacktriangle$};
\end{tikzpicture}
}
\hspace*{-1.5em}
\subfloat{
\begin{tikzpicture}[scale=0.4]
\node [external] (1) at (2, 0) {};
\node [external] (2) at (2, 4) {};
\node [external] (3) at (0, 2) {};
\node [external] (4) at (4, 2) {};
\draw (1) to (2);
\draw (3) to (4);
\node at (1.2, 2) {$\blacktriangleleft$};
\node at (2.8, 2) {$\blacktriangleright$};
\node at (2, 1.2) {$\blacktriangle$};
\node at (2, 2.8) {$\blacktriangledown$};
\end{tikzpicture}
}
\caption{Valid configurations of the six-vertex model}
\label{fig:6vertex}
\end{figure}

The valid local configurations of the six-vertex model
are illustrated in Figure~\ref{fig:6vertex}.
%
%
The  energy $E$  of the system is determined by
six parameters $\epsilon_1, \epsilon_2, \ldots, \epsilon_6$
associated with each type of { local configuration}.
If there are  $n_i$ sites in  local configurations of type $i$,
 then
$E = n_1 \epsilon_1 + n_2 \epsilon_2 + \ldots +  n_6  \epsilon_6$.
Then the  partition function is $Z_{\rm Six} = \sum e^{-E/k_BT}$,
where the sum is  over all valid configurations, $k_B$
 is Boltzmann's constant, and $T$ is the system's temperature.
This is a \emph{sum-of-product}
computation where the sum is over all Eulerian orientations
of the graph, and the product is over all vertices where each
contributes a factor $c_i = c^{\epsilon_i}$
if it is in configuration $i$
($1 \le i \le 6$) for some constant $c$.

Some choices of the parameters
are well-studied.
For modeling ice ($\epsilon_1= \ldots =\epsilon_6=0$)
 on the square $N \times N$ lattice graph,
Lieb~\cite{Lieb} famously showed that, 
the value of the ``partition function per vertex''
$W = Z^{1/N^2}$ approaches $\left( \frac{4}{3} \right)^{3/2}
\approx 1.5396007\ldots$ (Lieb's square ice constant).
 This matched experimental
data $1.540 \pm 0.001$ so well that it is considered a triumph.
Other well-known six-vertex models include:
the KDP model of a ferroelectric
($\epsilon_1=\epsilon_2=0$,  and
$\epsilon_3 = \epsilon_4 = \epsilon_5 = \epsilon_6>0$),
 the Rys $F$ model of an antiferroelectric
($\epsilon_1=\epsilon_2= \epsilon_3 = \epsilon_4 >0$,
and $\epsilon_5 = \epsilon_6 = 0$).
Historically these are widely considered  among
the most significant applications ever made of statistical mechanics
to real substances.
In classical statistical mechanics the parameters are 
real numbers. However, it's meaningful to consider parameters over complex values. 
In quantum theory the parameters are generally complex valued.
Even in classical theory, for example,  Baxter  generalized the parameters to complex values to develop the ``commuting transfer matrix'' for
tackling the six-vertex model~\cite{bax}. 
Some other models can be transformed to a six-vertex model with
complex weights. There are books 
with sections (e.g., see section 2.5.2 of~\cite{GRS}) that are dedicated to this, for example, the Hamiltonian of a one dimensional
spin chain is simply an extension of the Hamiltonian of a six-vertex model with complex Boltzmann
weights.

The six-vertex model has broad connections to combinatorics. The resolution of the 
famous \emph{Alternating
Sign Matrix} conjecture is one example~\cite{Korepin,Mills-Robbins-Rumsey,Zeilberger,Kuperberg,Bressoud}.
Also, the Tutte polynomial on a planar graph
at 
the point 
$(3, 3)$ is precisely $1/2$ of  $Z_{\rm Six}$ on its medial  graph which is also a planar graph
with a specific weight assignment~\cite{tutte}.

Although Pauling most likely did not think of it in such terms,
the six-vertex model
can be expressed
perfectly as a family of Holant problems with 6 parameters, expressed by  signatures of arity 4.
Previously, without being able to account for the planar
restriction, it has been proved~\cite{cfx} that there is a complexity
dichotomy where 
the problem on general graphs
is either in P or \#P-hard. However, the more interesting
problem is what happens on planar structures where physicists
had discovered some remarkable algorithms, such as
the FKT algorithm~\cite{TF61, Kasteleyn1961, Kasteleyn1967}.
Due to the presence of nontrivial algorithms,  a
complete complexity classification in the planar case
is more difficult to achieve.
Not only are reductions to FKT  expected
to give  planar P-time computable cases that are
\#P-hard in general, but also a more substantial obstacle awaits us.
It turns out that there is \emph{another}  planar P-time computable case
that had not been discovered for the six-vertex model in all these decades,
till now. (Since our algorithm and its proof that it runs in P-time is valid for all planar graphs,
this certainly also applies to the grid case, which is traditionally the main concern for
physicists.)

The main theorem in this paper is
a complexity trichotomy  for the
six-vertex model: According to the
6 parameters from $\mathbb{C}$,
the   partition function $Z_{\rm Six}$ is
either (1) computable in P-time,
or (2) \#P-hard on general graphs but computable in P-time on
planar graphs, or (3) remains  \#P-hard on planar graphs.
The classification has an explicit criterion.
The planar tractable class (2) includes
those that depend on FKT,
and a previously unknown family. 
 Functions that are expressible as matchgates
 (denoted by $\mathscr{M}$) or those that are transformable
to matchgates
(denoted by $\widehat{\mathscr{M}}$) do constitute a family
of $Z_{\rm Six}$ in class (2).
This follows from the FKT and Valiant's holographic algorithms~\cite{Val08}.\footnote{It was known~\cite{fw1,fw2} that on the grid graph the
parameter settings that satisfy $cz = ax + by$
(using notations in Section~\ref{prelim}) is P-time computable;
in our theory this is in $\mathscr{M}$, and the proof
is: It follows by Matchgate Identities~\cite{jcbook}.}
However, beyond these,
we discover an additional family of
P-time computable $Z_{\rm Six}$ on planar graphs.
The  P-time tractability is via a
non-local reduction
to P-time computable \#CSP,
where the variables in  \#CSP correspond to carefully defined circuits in $G$.
The fact that this 
\#CSP problem is
 in P depends crucially on the global topological constraint 
imposed by the planarity of $G$ (but the \#CSP instances that this produces is not planar in general.)
The new tractable class  provably cannot be subsumed by FKT (even with a holographic transformation).

After carving out this last tractable family, we 
prove that everything else is \#P-hard, even for the planar case.
A powerful tool in hardness proofs  is interpolation~\cite{valiant:interpolation}.
Typically an interpolation proof can succeed
when certain quantities (such as ratios of eigenvalues)
are not roots of unity, lest the iteration repeat after
a bounded number of steps.
A sufficient condition is that these quantities
have complex norm not equal to $1$.
However, for some  constraint functions,
we can show that these constructions only produce
such quantities of norm equal to $ 1$.
To overcome this difficulty we introduce  a new technique in hardness proofs:
M\"{o}bius transformations.\footnote{M\"{o}bius transformations were previously used in the design of quantum algorithms for approximating the Potts model~\cite{potts}. Here we use M\"{o}bius transformations in a different way, which is for hardness proofs. These  M\"{o}bius transformations are maps
on $\mathbb{C}$; they are 
unrelated to  M\"{o}bius inversions
for partial orders, e.g., as used
in~\cite{dyer2006counting}. 
}
We explore properties of
M\"{o}bius transformations
that map unit circle to
 unit circle on $\mathbb{C}$,
and obtain a suitable M\"{o}bius transformation that generates
an infinite group. This allows our interpolation proof to succeed.

The classification of the six-vertex model is not only  interesting
in its own right, more importantly, it serves as a basic building block in  the
classification program for Holant problems on asymmetric signatures.
For Holant problems over general graphs, complexity dichotomies \cite{cai2011dichotomy, Cai-Lu-Xia-holant-c, Backens-Holant-plus, Backens-Holant-c, DBLP:conf/icalp/CaiF020} were proved when certain signatures of odd arity (e.g. unary signatures) are present.
When it comes to signature sets of only even arities, the situation is  more difficult. 
The six-vertex model is precisely an inductive base
case for Holant problems with asymmetric signatures of even arity.
Very recently a full dichotomy
for real-valued  Holant  problems on asymmetric signatures was achieved~\cite{real-holant-focs2021}. This theorem is for
general graphs without addressing planar tractability.
The very first step to achieve this dichotomy
~\cite{real-holant-focs2021} 
is 
 the 
 dichotomy of the six-vertex model without planarity~\cite{cfx}. After that,  complexity dichotomies were proved for eight-vertex models~\cite{cai-fu-eight}, and for counting weighed Eulerian orientation problems with a reversal symmetry condition~\cite{cai2020beyond}.
 Lin and Wang proved a dichotomy
 for nonnegative valued Holant~\cite{wang-lin}.
 The dichotomy~\cite{real-holant-focs2021} is built on top of all that;
 see Figure~\ref{fig:structure}.
 
  However, from the very beginning~\cite{Val08} the additional planar tractability afforded by the likes of the FKT algorithm is at the  very heart of 
 Holant problems. One can say this is the  raison d'\^{e}tre
 of holographic algorithms. This classification is already done for symmetric signatures~\cite{cai2015holant}.
We hope that the present work
will serve as the beginning step
toward achieving a classification of Holant problems including planar
tractability, without the symmetry assumption.
\begin{figure}[!h]
\centering
$$\xymatrixrowsep{1.5ex}
\xymatrix@C-=4ex{
 \framebox[28ex]{\txt{Holant with unaries~\cite{Backens-Holant-c}}}\ar[r]
&  \framebox[28ex]{\txt{Holant with odd arity~\cite{DBLP:conf/icalp/CaiF020}}}\ar[dr]  &  \framebox[28ex]{\txt{Non-negative Holant~\cite{wang-lin}}}\ar[d] \\
  \framebox[28ex]{\txt{
 Six-vertex model~\cite{cfx}}}\ar[dr]\ar[r] &  \framebox[28ex]{\txt{\#Eulerian orientations~\cite{cai2020beyond}}}\ar[r] & \framebox[28ex]{\txt{Real-valued Holant~\cite{real-holant-focs2021} }} \\
  &   \framebox[28ex]{\txt{
 Eight-vertex model~\cite{cai-fu-eight}}}\ar[ur]
&  \\}$$
\vspace{-1ex}
\caption{A partial map 
of the complexity classification program for Holant problems}
\label{fig:structure}
 \end{figure}
 \vspace{-3ex}

\section{Preliminaries and Notations}\label{prelim}
In this paper, ${\frak i}$ denotes $\sqrt{-1}$, a square root of $-1$. 
\subsection{Definitions and Notations}
A constraint function $f$, or a signature, of arity $k$
is a map $\{0,1\}^k  \rightarrow \mathbb{C}$.
Fix a set $\mathcal{F}$ of constraint functions. A signature grid
$\Omega=(G, \pi)$
 is a tuple, where $G = (V,E)$
is a graph, $\pi$ labels each $v\in V$ with a function
$f_v\in\mathcal{F}$ of arity ${\operatorname{deg}(v)}$,
and the incident edges
$E(v)$ at $v$ with input variables of $f_v$.
 We consider all 0-1 edge assignments $\sigma$,
each gives an evaluation
$\prod \limits_{v\in V}f_v(\sigma|_{E(v)})$, where $\sigma|_{E(v)}$
denotes the restriction of $\sigma$ to $E(v)$. The counting problem on the instance $\Omega$ is to compute
$$\Holant_{\Omega} =
\Holant(\Omega; \mathcal{F})=\sum\limits_{\sigma:E\rightarrow\{0, 1\}}\prod_{v\in V}f_v(\sigma|_{E_{(v)}}).$$
The Holant problem parameterized by the set $\mathcal{F}$ is denoted by Holant$(\mathcal{F})$. 
If $\mathcal{F}=\{f\}$ is a single set, for simplicity, we write $\{f\}$ as $f$ directly, and also we write $\{f, g\}$ as $f, g$.
When $G$ is a planar graph, the corresponding signature grid is called a planar signature grid. 
We use $\holant{\mathcal{F}}{\mathcal{G}}$ to denote the Holant problem over signature grids with a bipartite graph $H = (U,V,E)$,
where each vertex in $U$ or $V$ is assigned a signature in $\mathcal{F}$ or $\mathcal{G}$
respectively.
We list the values of a signature $f: \{0,1\}^k  \rightarrow \mathbb{C}$
as a vector of dimension $2^k$
in lexicographic
order.
Signatures in $\mathcal{F}$ are considered as row vectors (or covariant tensors);
signatures in $\mathcal{G}$ are considered as column vectors (or contravariant tensors).
Similarly,
$\plholant{\mathcal{F}}{\mathcal{G}}$ denotes the Holant problem over signature grids with a planar bipartite graph.


A signature $f$ of arity 4 has the signature matrix
$M(f)=M_{x_1x_2, x_4x_3}(f)=\left[\begin{smallmatrix}
f_{0000} & f_{0010} & f_{0001} & f_{0011}\\
f_{0100} & f_{0110} & f_{0101} & f_{0111}\\
f_{1000} & f_{1010} & f_{1001} & f_{1011}\\
f_{1100} & f_{1110} & f_{1101} & f_{1111}
\end{smallmatrix}\right]
$.
Notice the order reversal $x_4x_3$; this is for the convenience of
composing these signatures in a planar fashion.
If $(i,j,k,\ell)$ is a 
permutation of $(1,2,3,4)$,
then the $4 \times 4$ matrix $M_{x_ix_j, x_\ell x_{k}}(f)$ lists the 16 values
 with row  index $x_ix_j \in\{0, 1\}^2$
and column index
$x_{\ell}x_{k} \in\{0, 1\}^2$ in lexicographic order.

The planar six-vertex model is Pl-Holant$( \not =_2 \mid f)$,
where $M(f)=
\left[\begin{smallmatrix}
0 & 0 & 0 & a \\
0 & b & c & 0 \\
0 & z & y & 0 \\
x & 0 & 0 & 0
\end{smallmatrix}\right]$.
The \emph{outer matrix}  of $M(f)$ 
is the submatrix
$\left[\begin{smallmatrix}
M(f)_{1, 1} & M(f)_{1, 4}\\
M(f)_{4, 1} & M(f)_{4, 4}
\end{smallmatrix}\right]=
\left[\begin{smallmatrix}
0 & a\\
x & 0
\end{smallmatrix}\right]$, and is 
 denoted by $M_{\rm{Out}}(f)$.
The \emph{inner matrix} of $M(f)$ is
$\left[\begin{smallmatrix}
M(f)_{2, 2} & M(f)_{2, 3}\\
M(f)_{3, 2} & M(f)_{3, 3}
\end{smallmatrix}\right]=
\left[\begin{smallmatrix}
b & c\\
z & y
\end{smallmatrix}\right]$, and is
 denoted by $M_{\rm{In}}(f)$.
A binary signature $g$ has the signature matrix $M(g)=M_{x_1, x_2}(g)=\left[
\begin{smallmatrix}
g_{00} & g_{01}\\
g_{10} & g_{11}\\
\end{smallmatrix}
\right].$
Switching the order, $M_{x_2, x_1}(g)=\left[
\begin{smallmatrix}
g_{00} & g_{10}\\
g_{01} & g_{11}\\
\end{smallmatrix}
\right].$
We use $(\neq_2)$ to denote binary {\sc Disequality} signature $(0,1, 1, 0)^T$.
It has the signature matrix  $\left[\begin{smallmatrix}
0 & 1 \\
 1 & 0
\end{smallmatrix}\right]$. Let
$N=
\left[\begin{smallmatrix}
0 & 1 \\
 1 & 0
\end{smallmatrix}\right]
\otimes
\left[\begin{smallmatrix}
0 & 1 \\
 1 & 0
\end{smallmatrix}\right]
=
\left[\begin{smallmatrix}
 0 & 0 & 0 & 1 \\
 0 & 0 & 1 & 0 \\
 0 & 1 & 0 & 0 \\
 1 & 0 & 0 & 0 \\
\end{smallmatrix}\right]$.
Note that $N$ is the double {\sc Disequality} $(x_1 \not = x_4) \wedge 
(x_2 \not = x_3)$,
which is the function of connecting two pairs of edges by $(\not =_2)$.
A function is symmetric if its value depends only
on the Hamming weight of its input.
A symmetric function $f$ on $k$ Boolean variables can
be expressed as
$[f_0, f_1, \ldots, f_k]$,
where $f_w$ is the value of $f$ on inputs of Hamming weight $w$.
For example, $(=_k)$ is the {\sc Equality} signature $[1, 0, \ldots, 0, 1]$
(with $k-1$ many 0's) of arity $k$.
The support of a signature $f$ is the set of inputs on which $f$ is nonzero.


Counting constraint satisfaction problems (\#CSP)
can be defined as a special case of Holant problems.
An instance of $\CSP(\mathcal{F})$ is presented
as a bipartite graph.
There is one node for each variable and for each occurrence
of constraint functions respectively.
Connect a constraint node to  a variable node if the
variable appears in that occurrence
of constraint, with a labeling on the edges
for the order of these variables.
This bipartite graph is also known as the \emph{constraint graph}.
If we attach each variable node with an \textsc{Equality} function,
and consider every edge as a variable, then
the \#CSP is just the Holant problem on this bipartite graph.
Thus
$\CSP(\mathcal{F}) \equiv_T \holant{\mathcal{EQ}}{\mathcal{F}}$,
where $\mathcal{EQ} = \{{=}_1, {=}_2, {=}_3, \dotsc\}$ is the set of \textsc{Equality} signatures of all arities.
By restricting to planar constraint graphs,
we have the planar \#CSP framework,
which we denote by $\PlCSP$.
The construction above also shows that $\PlCSP(\mathcal{F}) \equiv_T \plholant{\mathcal{EQ}}{\mathcal{F}}$. 

\subsection{Gadget Construction}
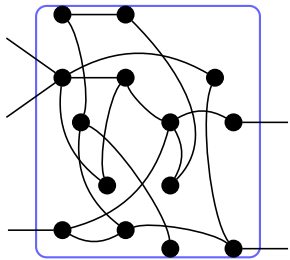
\begin{figure}[!htbp]
 \centering
 \begin{tikzpicture}[scale=\scale,transform shape,node distance=\nodeDist,semithick]
  \node[external]  (0)                     {};
  \node[internal]  (1) [below right of=0]  {};
  \node[external]  (2) [below left  of=1]  {};
  \node[internal]  (3) [above       of=1]  {};
  \node[internal]  (4) [right       of=3]  {};
  \node[internal]  (5) [below       of=4]  {};
  \node[internal]  (6) [below right of=5]  {};
  \node[internal]  (7) [right       of=6]  {};
  \node[internal]  (8) [below       of=6]  {};
  \node[internal]  (9) [below       of=8]  {};
  \node[internal] (10) [right       of=9]  {};
  \node[internal] (11) [above right of=6]  {};
  \node[internal] (12) [below left  of=8]  {};
  \node[internal] (13) [left        of=8]  {};
  \node[internal] (14) [below left  of=13] {};
  \node[external] (15) [left        of=14] {};
  \node[internal] (16) [below left  of=5]  {};
  \path let
         \p1 = (15),
         \p2 = (0)
        in
         node[external] (17) at (\x1, \y2) {};
  \path let
         \p1 = (15),
         \p2 = (2)
        in
         node[external] (18) at (\x1, \y2) {};
  \node[external] (19) [right of=7]  {};
  \node[external] (20) [right of=10] {};
  \path (1) edge                             (5)
            edge[bend left]                 (11)
            edge[bend right]                (13)
            edge node[near start] (e1) {}   (17)
            edge node[near start] (e2) {}   (18)
        (3) edge                             (4)
        (4) edge[out=-45,in=45]              (8)
        (5) edge[bend right, looseness=0.5] (13)
            edge[bend right, looseness=0.5]  (6)
        (6) edge[bend left]                  (8)
            edge[bend left]                  (7)
            edge[bend left]                 (14)
        (7) edge node[near start] (e3) {}   (19)
       (10) edge[bend right, looseness=0.5] (12)
            edge[bend left,  looseness=0.5] (11)
            edge node[near start] (e4) {}   (20)
       (12) edge[bend left]                 (16)
       (14) edge node[near start] (e5) {}   (15)
            edge[bend right]                (12)
       (16) edge[bend left,  looseness=0.5]  (9)
            edge[bend right, looseness=0.5]  (3);
  \begin{pgfonlayer}{background}
   \node[draw=\borderColor,thick,rounded corners,fit = (3) (4) (9) (e1) (e2) (e3) (e4) (e5),inner sep=0pt,transform shape=false] {};
  \end{pgfonlayer}
 \end{tikzpicture}
 \caption{(Figure 1 in \cite{guo-williams}) An $\mathcal{F}$-gate with 5 dangling edges.}
 \label{fig:Fgate}
\end{figure}

One basic tool used throughout the paper is gadget construction.
An $\mathcal{F}$-gate is similar to a signature grid $(G, \pi)$ for $\Holant(\mathcal{F})$ except that $G = (V,E,D)$ is a graph with internal edges $E$ and dangling edges $D$.
The dangling edges $D$ define input variables for the $\mathcal{F}$-gate.
We denote the regular edges in $E$ by $1, 2, \dotsc, m$ and the dangling edges in $D$ by $m+1, \dotsc, m+n$.
Then the  $\mathcal{F}$-gate  defines a function $f$
\[
f(y_1, \dotsc, y_n) = \sum_{\sigma: E \rightarrow\{0, 1\}} \prod_{v\in V}f_v(\hat{\sigma}\mid_{E(v)})
\]
where $(y_1, \dotsc, y_n) \in \{0, 1\}^n$ is an assignment on the dangling edges, $\hat{\sigma}$ is the extension of  $\sigma$ on $E$ by the assignment $(y_1, \ldots, y_m)$, and $f_v$ is the signature assigned at each vertex $v \in V$. (See Figure~\ref{fig:Fgate} for an example.)
This function $f$ is called the signature of the $\mathcal{F}$-gate. 
We say a signature $f$ is \emph{realizable} from a signature set $\mathcal{F}$ by gadget construction
if $f$ is the signature of an 
 $\mathcal{F}$-gate. 

An $\mathcal{F}$-gate is planar if the underlying graph $G$ is a planar graph,
and the dangling edges,
ordered counterclockwise corresponding to the order of the input variables,
are in the outer face in a planar embedding.
A planar $\mathcal{F}$-gate can be used in a planar signature grid as if it is just a single vertex with the particular signature.
If $f$ is realizable  by a planar $\mathcal{F}$-gate,
then we can freely add $f$ into $\mathcal{F}$ while preserving the complexity of the planar Holant problem, i.e., $\PlHolant(\mathcal{F}, f) \equiv_T \PlHolant(\mathcal{F})$.
The reduction from the right to the left is trivial. The reduction in the other direction is also simple. 
Given an instance of $\PlHolant(\mathcal{F}, f)$,
by replacing every occurrence of $f$ with the $\mathcal{F}$-gate,
we get an instance of $\PlHolant(\mathcal{F})$.

\begin{figure}[!htb]
 \centering
 \def\capWidth{6cm}
 \captionsetup[subfigure]{width=\capWidth}
 \tikzstyle{entry} = [internal, inner sep=2pt]
 \subfloat[A clockwise rotation $(x_1, x_2, x_3, x_4)
\rightarrow (x_2, x_3, x_4, x_1)$. The variables are
ordered counterclockwise, starting with $x_1$
indicated by the diamond.
]{
  \begin{tikzpicture}[scale=\scale,transform shape,node distance=\nodeDist,semithick]
   \node[external]  (0)                     {};
   \node[external]  (1) [right       of=0]  {};
   \node[internal]  (2) [below right of=1]  {};
   \node[external]  (3) [below left  of=2]  {};
   \node[external]  (4) [left        of=3]  {};
   \node[external]  (5) [above right of=2]  {};
   \node[external]  (6) [right       of=5]  {};
   \node[external]  (7) [below right of=2]  {};
   \node[external]  (8) [right       of=7]  {};
   \node[external]  (9) [below right of=6]  {};
   \node[external] (10) [right       of=9]  {};
   \node[external] (11) [above right of=10] {};
   \node[external] (12) [right       of=11] {};
   \node[internal] (13) [below right of=12] {};
   \node[external] (14) [below left  of=13] {};
   \node[external] (15) [left        of=14] {};
   \node[external] (16) [above right of=13] {};
   \node[external] (17) [right       of=16] {};
   \node[external] (18) [below right of=13] {};
   \node[external] (19) [right       of=18] {};
   \node[external] (191) [below       of=9] {};
   \node[external] (192) [below       of=191] {};
   \path  (0) edge[out=   0, in= 135, postaction={decorate, decoration={
                                                             markings,
                                                             mark=at position 0.4   with {\arrow[>=diamond,white] {>}; },
                                                             mark=at position 0.4   with {\arrow[>=open diamond]  {>}; },
                                                             mark=at position 0.999 with {\arrow[>=diamond,white] {>}; },
                                                             mark=at position 1     with {\arrow[>=open diamond]  {>}; } } }] (2)
          (2) edge[out=-135, in=   0]  (4)
              edge[out=  45, in= 180]  (6)
              edge[out= -45, in= 180]  (8)
         (11) edge[out=   0, in= 135, postaction={decorate, decoration={
                                                             markings,
                                                             mark=at position 0.4   with {\arrow[>=diamond,white] {>}; },
                                                             mark=at position 0.4   with {\arrow[>=open diamond]  {>}; } } }] (13)
         (15) edge[out=   0, in=-135] (13)
         (17) edge[out= 180, in=  45, postaction={decorate, decoration={
                                                             markings,
                                                             mark=at position 0.999 with {\arrow[>=diamond,white] {>}; },
                                                             mark=at position 1   with {\arrow[>=open diamond]  {>}; } } }] (13)
         (13) edge[out= -45, in= 180] (19);
   \path (9.west) edge[->, >=stealth] (10.east);
   \begin{pgfonlayer}{background}
    \node[draw=\borderColor,transform shape=false,thick,rounded corners,fit =  (1)  (3)  (5)  (7)] {};
    \node[draw=\borderColor,transform shape=false,thick,rounded corners,fit = (12) (14) (16) (18)] {};
   \end{pgfonlayer}
  \end{tikzpicture}}
 \qquad
 \subfloat[Movement of signature matrix entries under a clockwise rotation
of $\frac{\pi}{2}$.]{
  \makebox[\capWidth][c]{
   \begin{tikzpicture}[scale=\scale,transform shape,>=stealth,node distance=\nodeDist,semithick]
    \node[entry] (11)               {};
    \node[entry] (12) [right of=11] {};
    \node[entry] (13) [right of=12] {};
    \node[entry] (14) [right of=13] {};
    \node[entry] (21) [below of=11] {};
    \node[entry] (22) [right of=21] {};
    \node[entry] (23) [right of=22] {};
    \node[entry] (24) [right of=23] {};
    \node[entry] (31) [below of=21] {};
    \node[entry] (32) [right of=31] {};
    \node[entry] (33) [right of=32] {};
    \node[entry] (34) [right of=33] {};
    \node[entry] (41) [below of=31] {};
    \node[entry] (42) [right of=41] {};
    \node[entry] (43) [right of=42] {};
    \node[entry] (44) [right of=43] {};
    \node[external] (nw) [above left  of=11] {};
    \node[external] (ne) [above right of=14] {};
    \node[external] (sw) [below left  of=41] {};
    \node[external] (se) [below right of=44] {};
    \path (13) edge[->, dotted]                (12)
          (12) edge[->, dotted]                (21)
          (21) edge[->, dotted]                (31)
          (31) edge[->, dotted,out=65,in=-155] (13)
          (42) edge[->, dashed]                (43)
          (43) edge[->, dashed]                (34)
          (34) edge[->, dashed]                (24)
          (24) edge[->, dashed,out=-115,in=25] (42)
          (14) edge[->, very thick]            (22)
          (22) edge[->, very thick]            (41)
          (41) edge[->, very thick]            (33)
          (33) edge[->, very thick]            (14)
          (23) edge[<->]                      (32);
    \path (nw.west) edge (sw.west)
          (ne.east) edge (se.east)
          (nw.west) edge (nw.east)
          (sw.west) edge (sw.east)
          (ne.west) edge (ne.east)
          (se.west) edge (se.east);
   \end{tikzpicture}}}
 \caption{(Figure 2 in \cite{cai2015holant}) The movement of the entries in the signature matrix of an arity~4 signature under a clockwise rotation of the input edges.
  The Hamming weight two entries are in the two solid cycles (one has length 4 and the other one is a swap).}
 \label{fig:rotate_asymmetric_signature}
\end{figure}
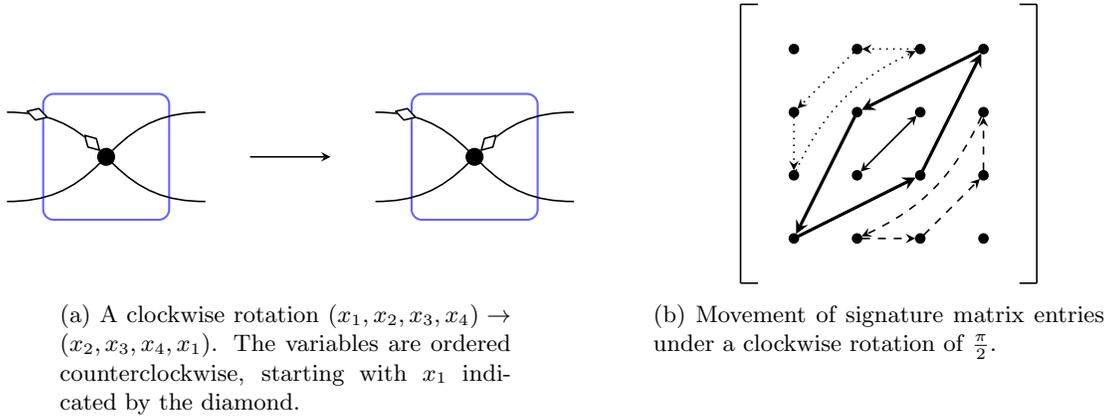

In this paper, we focus on planar graphs, and we assume the edges incident to a vertex are ordered counterclockwise. When connecting two signatures, we need to keep the counterclockwise order of the edges incident to each vertex. 
Given a signature $f$ with signature matrix $M_{x_1x_2, x_4x_3}(f)$,
we can rotate it to obtain, for any cyclic permutations
$(i, j, k, \ell)$ of $(1, 2, 3, 4)$,
the signature $f'$ with signature matrix $M_{x_1x_2, x_4x_3}(f')
= M_{x_ix_j, x_\ell x_{k}}(f)$.
There are four cyclic permutations of $(1, 2, 3, 4)$, 
so correspondingly, a signature $f$ has four rotated
forms, with $4\times4$ signature matrices
$M_{x_1x_2, x_4x_3}(f) =
\left[\begin{smallmatrix}
0 & 0 & 0 & a \\
0 & b & c & 0 \\
0 & z & y & 0 \\
x & 0 & 0 & 0
\end{smallmatrix}\right]$,
$M_{x_2x_3, x_1x_4}(f) =
\left[\begin{smallmatrix}
0 & 0 & 0 & y \\
0 & a & z & 0 \\
0 & c & x & 0 \\
b & 0 & 0 & 0
\end{smallmatrix}\right]$,
$M_{x_3x_4, x_2x_1}(f) =
\left[\begin{smallmatrix}
0 & 0 & 0 & x \\
0 & y & c & 0 \\
0 & z & b & 0 \\
a & 0 & 0 & 0
\end{smallmatrix}\right]$,
and $M_{x_4x_1, x_3x_2}(f) =
\left[\begin{smallmatrix}
0 & 0 & 0 & b \\
0 & x & z & 0 \\
0 & c & a & 0 \\
y & 0 & 0 & 0
\end{smallmatrix}\right]$. 
These are denoted as
$f$, $f^{\frac{\pi}{2}}$, $f^{{\pi}}$ and $f^{\frac{3\pi}{2}}$,
respectively. Thus $M_{x_1x_2, x_4x_3}(f^{\frac{\pi}{2}})
= M_{x_2x_3, x_1x_4}(f)$, etc.
Without other specification, $M(f)$ denotes $M_{x_1x_2, x_4x_3}(f)$.
Once we get one form, all four rotation forms can be freely used.
In the proof, after one construction,
 we may use this property to get a similar construction and conclude
 by quoting this rotational symmetry.
The movement of signature entries under a rotation is illustrated
in Figure~\ref{fig:rotate_asymmetric_signature} (Figure 2 in \cite{cai2015holant}).
Note that no matter in which signature matrix, the pair $(c, z)$ (and only
$(c, z)$) is always in the inner matrix. We call $(c, z)$ the inner pair, and $(a, x)$, $(b, y)$ 
the outer pairs. 

\begin{figure}[!htbp]
\centering
                \includegraphics[height=1.2in]{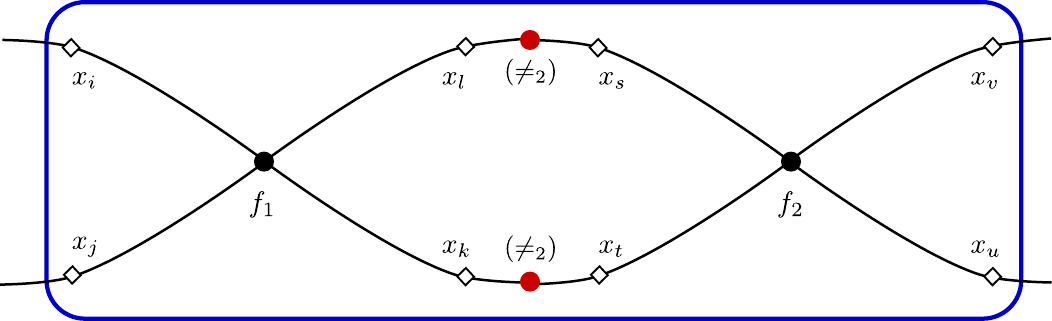}
        \caption{Connect variables $x_\ell$, $x_k$ of $f_1$ with variables $x_s$, $x_t$ of $f_2$ both using $(\neq_2)$.}
        \label{111}
        \end{figure}

There are three common gadgets we will use in this paper.
The first gadget construction is as follows.
Suppose $f_1$ and $f_2$ have signature matrices
 $ M_{x_ix_j, x_\ell x_{k}}(f_1)$ and $
M_{x_{s}x_{t}, x_{v}x_{u}}(f_2)$, where $(i, j, k, \ell)$ and $(s, t, u, v)$ are permutations of $(1, 2, 3, 4)$.
By connecting $x_\ell$ with $x_{s}$, $x_{k}$ with $x_t$,
both using  {\sc Disequality}  $(\not =_2)$, we get a signature
of arity 4 with the signature matrix
$M_{x_ix_j, x_\ell x_{k}}(f_1) N M_{x_{s}x_{t}, x_{v}x_{u}}(f_2)$
by matrix product
with row index $x_ix_j$ and column index $x_{v}x_{u}$ (See Figure \ref{111}).

\begin{figure}[!h]
 \centering
		\includegraphics[height=1.2in]{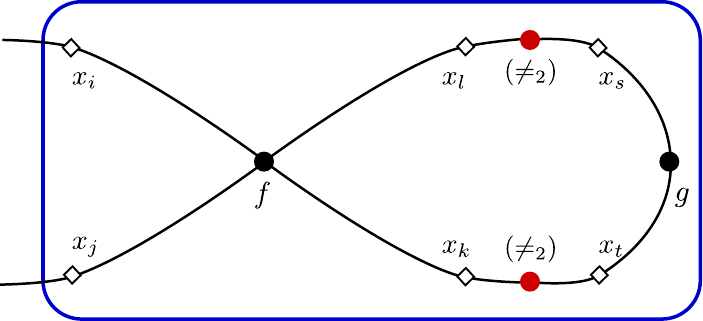}
	\caption{Connect variables $x_\ell$, $x_k$ of $f$ with variables $x_s$, $x_t$ of $g$ both using $(\neq_2)$.}\label{222}
	\end{figure}

 A binary signature $g$ has the signature vector $g(x_1, x_2)=(g_{00}, g_{01}, g_{10}, g_{11})^T$, 
and also $g(x_2, x_1)=(g_{00}, g_{10}, g_{01}, g_{11})^T$.  
Without other specification,  $g$  denotes $g(x_1, x_2)$. 
Let $f$  be a signature of arity $4$ with the signature matrix
 $M_{x_ix_j, x_\ell x_k}(f)$ and $(s, t)$ be a permutation of $(1, 2)$. 
The second gadget construction is as follows.
 By connecting $x_\ell$ with $x_s$ and $x_{k}$ with $x_t$, both using {\sc Disequality} $(\neq_{2})$, 
 we get a binary signature with the signature matrix $M_{x_i x_j, x_k x_{\ell}}Ng{(x_s, x_t)}$ as a matrix product with index $x_i x_j$ (See Figure \ref{222}). 
 If $g_{00}=g_{11}$, 
 then $N(g_{00}, g_{01}, g_{10}, g_{11})^T=(g_{11}, g_{10}, g_{01}, g_{00})^T=(g_{00}, g_{10}, g_{01}, g_{11})^T$, and similarly, $N(g_{00}, g_{10}, g_{01}, g_{11})^T=(g_{00}, g_{01}, g_{10}, g_{11})^T$. 
 Therefore, $M_{x_i x_j, x_\ell x_k}Ng{(x_s, x_t)}=M_{x_i x_j, x_\ell x_{k}}g{(x_t, x_s)}$, 
 which means that
 connecting variables $x_\ell$, $x_k$ of $f$ with, respectively,
 variables $x_s$, $x_t$ of $g$
 using $N$ is equivalent to connecting them directly without $N$. 
Hence, in the setting
Pl-Holant$( \not =_2 \mid f, g)$ we can form
$M_{x_i x_j, x_\ell x_{k}}(f)g{(x_t, x_s)}$,
which is technically $M_{x_i x_j, x_\ell x_{k}}Ng{(x_s, x_t)}$,
provided that $g_{00}=g_{11}$.
Note that for a  binary signature $g$,
we can rotate it by $180^\circ$ without violating
planarity, and so  both $g(x_s, x_t)$ and $g(x_t, x_s)$ can be freely used once we get one of them.

%

\begin{figure}[!htbp]
 \centering
		\includegraphics[height=1.2in]{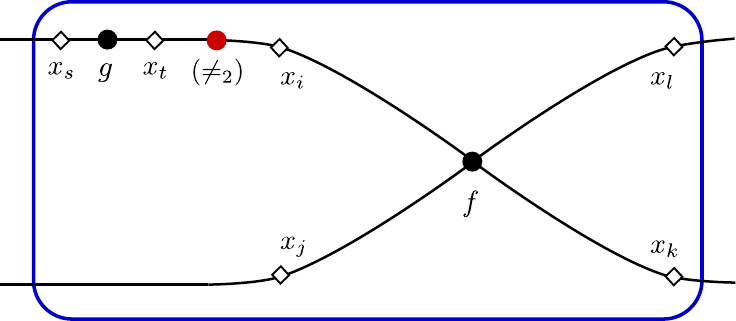}
	\caption{Connect variable $x_t$ of $g$ with variable $x_i$ of $f$ using $(\neq_2)$.}\label{333}
	\end{figure}

 A signature $f$ of arity $4$ also has the $2 \times 8$ signature matrix $$M_{x_1, x_2 x_4 x_3}(f)=
\left[\begin{matrix}
f_{0000} & f_{0010} & f_{0001} & f_{0011} &
f_{0100} & f_{0110} & f_{0101} & f_{0111}\\
f_{1000} & f_{1010} & f_{1001} & f_{1011} &
f_{1100} & f_{1110} & f_{1101} & f_{1111}
\end{matrix}\right].$$
Suppose the signature matrix of  $g$ is  $M_{x_s, x_t}(g)$ and the signature matrix of $f$ is $M_{x_i, x_j x_\ell x_k}(f)$. 
Our third  gadget construction is as follows.
By connecting $x_t$ with $x_i$ using {\sc Disequality} $(\neq_{2})$, 
we get a signature $h$ of arity $4$ with the signature matrix
$M_{x_s, x_t}(g)M(\neq_2)M_{x_i, x_j x_\ell x_k}(f)$
by matrix product with row index $x_s$ and column index $x_j x_\ell x_k$ (See Figure \ref{333}). 
We may change this form to a signature matrix with row index $x_s x_j$ and column index $x_\ell x_k$.
	
In particular, if $M_{y_1, y_2}(g)=
\left[\begin{smallmatrix}
0 & 1\\
t & 0\\
\end{smallmatrix}\right]$, then connecting $y_2$ with $x_1$ via $(\neq_2)$
gives
\begin{align*}
M_{y_1, x_2 x_4 x_3}(h)&=M_{y_1, y_2}(g)M(\neq_2)M_{x_1, x_2 x_4 x_3}(f)\\
&=\left[\begin{matrix}
0 & 1\\
t & 0\\
\end{matrix}\right]
\left[\begin{matrix}
0 & 1\\
1 & 0\\
\end{matrix}\right]
\left[\begin{matrix}
f_{0000} & f_{0010} & f_{0001} & f_{0011} &
f_{0100} & f_{0110} & f_{0101} & f_{0111}\\
f_{1000} & f_{1010} & f_{1001} & f_{1011} &
f_{1100} & f_{1110} & f_{1101} & f_{1111}
\end{matrix}\right]\\
&=\left[\begin{matrix}
f_{0000} & f_{0010} & f_{0001} & f_{0011} &
f_{0100} & f_{0110} & f_{0101} & f_{0111}\\
tf_{1000} & tf_{1010} & tf_{1001} & tf_{1011} &
tf_{1100} & tf_{1110} & tf_{1101} & tf_{1111}
\end{matrix}\right].
\end{align*}
If we rename the variable $y_1$ 
by $x_1$, then $M_{x_1 x_2, x_4 x_3}(h)=
\left[\begin{smallmatrix}
f_{0000} & f_{0010} & f_{0001} & f_{0011} \\
f_{0100} & f_{0110} & f_{0101} & f_{0111}\\
tf_{1000} & tf_{1010} & tf_{1001} & tf_{1011} \\
tf_{1100} & tf_{1110} & tf_{1101} & tf_{1111}
\end{smallmatrix}\right]$. 
That is, the new  signature has the matrix
obtained from multiplying $t$ to the last
 two rows of $M_{x_1 x_2, x_4 x_3}(f)$
 corresponding to  $x_1=1$.
Similarly we can modify the last two columns of $M_{x_1 x_2, x_4 x_3}(f)$.
Given $g=(0, 1, t, 0)^T$,
we call the modification from $M_{x_1 x_2, x_4 x_3}(f)$ to
\[
\left[\begin{matrix}
f_{0000} & f_{0010} & f_{0001} & f_{0011} \\
f_{0100} & f_{0110} & f_{0101} & f_{0111}\\
tf_{1000} & tf_{1010} & tf_{1001} & tf_{1011} \\
tf_{1100} & tf_{1110} & tf_{1101} & tf_{1111}
\end{matrix}\right]\]
the operation of $t$ scaling on  $x_1=1$.
Similarly we call the modification from $M_{x_1 x_2, x_4 x_3}(f)$ to
\[
\left[\begin{matrix}
f_{0000} & f_{0010} & tf_{0001} & tf_{0011} \\
f_{0100} & f_{0110} & tf_{0101} & tf_{0111}\\
f_{1000} & f_{1010} & tf_{1001} & tf_{1011} \\
f_{1100} & f_{1110} & tf_{1101} & tf_{1111}
\end{matrix}\right]\]
the operation of $t$ scaling on  $x_4=1$.

For any  scalar $c\neq 0$
and any set of signatures $\mathcal{F}$,
we have
 $\Holant(\mathcal{F}\cup \{f\})\equiv_{T}\Holant(\mathcal{F}\cup \{cf\})$,
and $\PlHolant(\mathcal{F}\cup \{f\})\equiv_{T}\PlHolant(\mathcal{F}\cup \{cf\})$.
Thus a scalar $c\neq 0$ does not change the complexity of a Holant problem.
Hence we can normalize any particular nonzero signature entry to be 1.

\subsection{Holographic Transformation}
To introduce the idea of holographic transformation,
it is convenient to consider bipartite graphs.
For a general graph,
we can always transform it into a bipartite graph while preserving the Holant value,
as follows.
For each edge in the graph,
we replace it by a path of length two.
(This operation is called the \emph{2-stretch} of the graph and yields the edge-vertex incidence graph.)
Each new vertex is assigned the binary \textsc{Equality} signature $=_2$. Thus, we have $\holant{=_2}{\mathcal{F}}\equiv_T \Holant(\mathcal{F})$.


For an invertible $2$-by-$2$ matrix $T \in {\rm GL}_2({\mathbb{C}})$
 and a signature $f$ of arity $n$, written as
a column vector (contravariant tensor) $f \in \mathbb{C}^{2^n}$, we denote by
$T^{-1}f = (T^{-1})^{\otimes n} f$ the transformed signature.
  For a signature set $\mathcal{F}$,
define $T^{-1} \mathcal{F} = \{T^{-1}f \mid  f \in \mathcal{F}\}$ the set of
transformed signatures.
For signatures written as
 row vectors (covariant tensors) we define
$f T$ and  $\mathcal{F} T$ similarly.
Whenever we write $T^{-1} f$ or $T^{-1} \mathcal{F}$,
we view the signatures as column vectors;
similarly for $f T$ or $\mathcal{F} T$ as row vectors.
In the special case of the Hadamard matrix
$H_2 = \frac{1}{\sqrt{2}} \left[\begin{smallmatrix} 1 & 1 \\ 1 & -1 \end{smallmatrix}\right]$,
we also define $\widehat{\mathcal{F}} = H_2  \mathcal{F}$.
Note that $H_2$ is orthogonal.
Since constant factors are immaterial, for convenience we sometime
drop the factor $\frac{1}{\sqrt{2}}$ when using $H_2$. 

Let $T \in {\rm GL}_2({\mathbb{C}})$.
The holographic transformation defined by $T$ is the following operation:
given a signature grid $\Omega = (H, \pi)$ of $\holant{\mathcal{F}}{\mathcal{G}}$,
for the same bipartite graph $H$,
we get a new signature grid $\Omega' = (H, \pi')$ of $\holant{\mathcal{F} T}{T^{-1} \mathcal{G}}$ by replacing each signature in
$\mathcal{F}$ or $\mathcal{G}$ with the corresponding signature in $\mathcal{F} T$ or $T^{-1} \mathcal{G}$. Valiant's Holant Theorem~\cite{Val08} states that the instances $\Omega$ and $\Omega'$ have the same Holant value. 
  This result also holds for planar instances.

\begin{theorem}[\cite{Val08}]
 For every $T \in {\rm GL}_2({\mathbb{C}})$,
  $\PlHolant(\mathcal{F} \mid \mathcal{G}) \equiv_T \PlHolant(\mathcal{F} T \mid T^{-1} \mathcal{G}).$
\end{theorem}

\begin{definition} \label{def:prelim:trans}
 We say a signature set $\mathcal{F}$ is $\mathscr{C}$-transformable
 if there exists a $T \in \rm{GL}_2(\mathbb{C})$ such that
 $(0, 1, 1, 0) T^{\otimes 2} \in \mathscr{C}$ and $T^{-1}\mathcal{F} \subseteq \mathscr{C}$.
\end{definition}
This definition is important because if $\PlHolant(\mathscr{C})$ is tractable,
then $\PlHolant(\neq_2 \mid\mathcal{F})$ is tractable for any $\mathscr{C}$-transformable set $\mathcal{F}$. 

\subsection{Polynomial Interpolation}
Polynomial interpolation is a powerful technique to prove \#P-hardness for counting problems. We use polynomial interpolation
to  prove the following lemmas.
\begin{lemma}\label{1111}
Let $f$ be a 4-ary signature with the signature matrix
$M(f)=\left[\begin{smallmatrix}
0 & 0& 0& 1\\
0 & b& 0& 0\\
0 & 0& b& 0\\
1 & 0& 0& 0\\
\end{smallmatrix}\right]$, where $b\neq 0$ is not a root of unity.
Let $\chi_1$ be a 4-ary signature with the signature matrix
$M(\chi_1)=\left[\begin{smallmatrix}
0 & 0& 0& 1\\
0 & 1& 0& 0\\
0 & 0& 1& 0\\
1 & 0& 0& 0\\
\end{smallmatrix}\right].$
Then for any signature set $\mathcal{F}$ containing $f$, we have $$\operatorname{Pl-Holant}(\neq_2\mid \mathcal{F}\cup \{\chi_1\})\leqslant_{T}\operatorname{Pl-Holant}(\neq_2\mid \mathcal{F}).$$
\end{lemma}
\begin{figure}[!htbp]
\centering
		\includegraphics[height=1.0in]{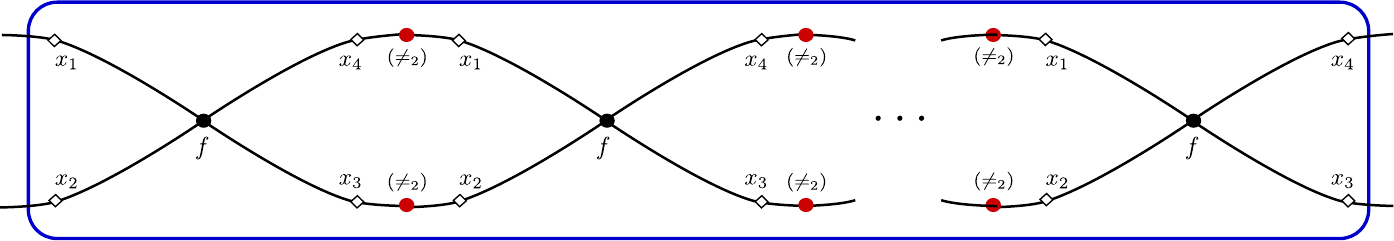}
	\caption{A chain of $2s+1$ many copies of $f$ linked by double {\sc Disequality} $N$}
	\label{2s+1}
	\end{figure}
{\bf Proof. }
We construct a series of gadgets $f_{2s+1}$ by a chain of $2s+1$ many copies of $f$ linked by the double {\sc Disequality} $N$ (See Figure~\ref{2s+1}).	
Clearly $f_{2s+1}$ has the following signature matrix 
$$M(f_{2s+1})=M(f)(NM(f))^{2s}=
\left[\begin{matrix}
0 & 0 & 0 & 1\\
0 & b^{2s+1} & 0 & 0\\
0 & 0 & b^{2s+1} & 0\\
1 & 0 & 0 & 0\\
\end{matrix}\right]
.$$ 
The matrix $M(f_{2s+1})$ has a good form for polynomial interpolation. 
Suppose $\chi_1$ appears $m$ times in an instance $\Omega$ of $\PlHolant(\neq_2 \mid \mathcal{F} \cup \{\chi_1\})$.
We replace each appearance of $\chi_1$ by a copy of
the gadget $f_{2s+1}$ to get an instance $\Omega_{2s+1}$ of $\operatorname{Pl-Holant}(\neq_2\mid \mathcal{F} \cup \{f_{2s+1}\})$, which is also an instance of $\operatorname{Pl-Holant}(\neq_2\mid \mathcal{F})$.
We divide $\Omega_{2s+1}$ into two parts.
One part consists of $m$ signatures $f_{2s+1}$ and its signature is represented by $(M(f_{2s+1}))^{\otimes{m}}$. 
Here we rewrite $(M(f_{2s+1}))^{\otimes{m}}$ as a column vector.
The other part is the rest of $\Omega_{2s+1}$ and its signature is represented by $A$ which is a tensor expressed as a row vector. 
Then, the Holant value of $\Omega_{2s+1}$ is the dot product $\langle A, (M(f_{2s+1}))^{\otimes{m}}\rangle$, which is a summation over $4m$ bits. 
That is, a sum over all $0, 1$ values for  the $4m$ edges connecting the two parts. 
We can stratify all $0, 1$ assignments of these $4m$ bits having a nonzero evaluation of a term in Pl-Holant$_{\Omega_{2s+1}}$ into the following categories:
\begin{itemize}
\item
There are $i$ many copies of $f_{2s+1}$ receiving inputs $0011$ or $1100$;
\item
There are $j$ many copies of $f_{2s+1}$ receiving inputs $0110$ or $1001$;
\end{itemize}
where $i+j=m$.

For any assignment in the category with parameter $(i,j)$, the evaluation of
$(M(f_{2s+1}))^{\otimes m}$ is clearly $b^{(2s+1)j}$. 
Let $a_{ij}$ be the summation of values of the part $A$ over all assignments in the category $(i, j)$. Note that $a_{ij}$ is independent from the value of $s$ since we view the gadget $f_{2s+1}$ as a block. Since $i+j=m$, we can denote $a_{ij}$ by $a_j$. Then, we rewrite the dot product summation and get 
$$\PlHolant_{\Omega_{2s+1}}=\langle A, (M(f_{2s+1}))^{\otimes{m}}\rangle=\sum_{0\leqslant j \leqslant m}a_jb^{(2s+1)j}.$$
Under this stratification, the Holant value of $\PlHolant(\Omega, \neq_2 \mid \mathcal{F} \cup \{\chi_1\})$ can be represented as 
$$\PlHolant_{\Omega} =\langle A, (M(\chi_1))^{\otimes{m}}\rangle=\sum_{0\leqslant j \leqslant m}a_j.$$
Since $b \neq 0$ is not a root of unity, 
the linear equation system has a nonsingular Vandermonde  matrix 
$$\left[\begin{matrix}
b^0 & b^1 & \cdots & b^m\\
(b^3)^0 & (b^3)^1 & \cdots & (b^3)^m\\
\vdots & \vdots & \vdots & \vdots \\
(b^{2m+1})^0 & (b^{2m+1})^1 & \cdots & (b^{2m+1})^m\\
\end{matrix}\right].$$ 
By oracle querying the values of $\PlHolant_{\Omega_{2s+1}}$,
we can solve the coefficients $a_j$ in polynomial time and obtain the value of $p(x)=\sum \limits_{0\leqslant j \leqslant m}a_jx^j$ for any $x$. Let $x=1$, we get $\PlHolant_{\Omega}$. Therefore, we have $\operatorname{Pl-Holant}(\neq_2\mid \mathcal{F}\cup \{\chi_1\})\leqslant_{T}\operatorname{Pl-Holant}(\neq_2\mid \mathcal{F}).$\qed

\begin{corollary}\label{111-1}
Let $f$ be a 4-ary signature with the signature matrix
$M(f)=\left[\begin{smallmatrix}
0 & 0& 0& 1\\
0 & b& 0& 0\\
0 & 0& b& 0\\
-1 & 0& 0& 0\\
\end{smallmatrix}\right]$, where $b \neq 0$ is not a root of unity.
Let $\chi_2$ be a 4-ary signature with the signature matrix
$M(\chi_2)=\left[\begin{smallmatrix}
0 & 0& 0& 1\\
0 & 1& 0& 0\\
0 & 0& 1& 0\\
-1 & 0& 0& 0\\
\end{smallmatrix}\right].$
Then for any signature set $\mathcal{F}$ containing $f$, we have $$\operatorname{Pl-Holant}(\neq_2\mid \mathcal{F}\cup \{\chi_2\})\leqslant_{T}\operatorname{Pl-Holant}(\neq_2\mid \mathcal{F}).$$
\end{corollary}

{\bf Proof.}
We still construct a series of gadgets $f_{2s+1}$ by a chain of odd
many copies of $f$ linked by the double {\sc Disequality} $N$. 
Clearly  $f_{2s+1}$ has the following signature matrix 
$$M(f_{2s+1})=M(f)(NM(f))^{2s}=
\left[\begin{matrix}
0 & 0 & 0 & 1\\
0 & b^{2s+1} & 0 & 0\\
0 & 0 & b^{2s+1} & 0\\
-1 & 0 & 0 & 0\\
\end{matrix}\right]
.$$ 
Suppose $\chi_2$ appears $m$ times in an instance $\Omega$ of $\PlHolant(\neq_2 \mid f\cup \chi_2)$.
We replace each appearance of $\chi_2$ by a copy of
the gadget $f_{2s+1}$ to get an instance $\Omega_{2s+1}$ of $\operatorname{Pl-Holant}(\neq_2\mid \mathcal{F} \cup \{f_{2s+1}\})$.
In the same way as in the proof of Lemma \ref{1111}, we divide $\Omega_{2s+1}$ into two parts. One part is represented by $(M(f_{2s+1}))^{\otimes{m}}$ and the other part is represented by $A$.
Then, the Holant value of $\Omega_{2s+1}$ is the dot product $\langle A, (M(f_{2s+1}))^{\otimes{m}}\rangle$.
We can stratify all $0, 1$ assignments of these $4m$ bits having a nonzero evaluation of a term in Pl-Holant$_{\Omega_{2s+1}}$ into the following categories:
\begin{itemize}
\item
There are $i$ many copies of $f_{2s+1}$ receiving inputs $0011$;
\item
There are $j$ many copies of $f_{2s+1}$ receiving inputs $0110$ or $1001$;
\item
There are $k$ many copies of $f_{2s+1}$ receiving inputs $1100$;
\end{itemize}
where $i+j+k=m$.

For any assignment in those categories with parameters $(i, j, k)$ where $k \equiv 0 \pmod 2$, the evaluation of
$(M(f_{2s+1}))^{\otimes m}$ is clearly $(-1)^kb^{(2s+1)j}=b^{(2s+1)j}$. And for any assignment in those categories with parameters $(i, j, k)$ where $k \equiv 1 \pmod 2$, the evaluation of
$(M(f_{2s+1}))^{\otimes m}$ is clearly $(-1)^kb^{(2s+1)j}=-b^{(2s+1)j}$. 
Since $i+j+k=m$, the index $i$ is determined by $j$ and $k$. Let $a_{j0}$ be the summation of values of the part $A$ over all assignments in those categories $(i, j, k)$ where $k \equiv 0 \pmod 2$, and $a_{j1}$ be the summation of values of the part $A$ over all assignments in those categories $(i, j, k)$ where $k \equiv 1 \pmod 2$. 
Note that $a_{j0}$ and $a_{j1}$ are independent from the value of $s$.
Let $a_j=a_{j0}-a_{j1}$.
Then, we rewrite the dot product summation and get 
$$\PlHolant_{\Omega_{2s+1}} =\langle A, (M(f_{2s+1}))^{\otimes{m}}\rangle=\sum_{0\leqslant j \leqslant m}(a_{j0}b^{(2s+1)j}-a_{j1}b^{(2s+1)j})=\sum_{0\leqslant j \leqslant m}a_jb^{(2s+1)j}.$$
Under this stratification, the Holant value of $\plholant{\Omega; \neq_2}{f \cup \chi_2}$ can be represented as 
$$\PlHolant_{\Omega}=\langle A, (M(\chi_2))^{\otimes{m}}\rangle=\sum_{0\leqslant j \leqslant m}(a_{j0}-a_{j1})=\sum_{0\leqslant j \leqslant m}a_j.$$
Since $b \neq 0$ is not a root of unity, 
the Vandermonde coefficient matrix 
has full rank. Hence we can solve for all the values
$a_j$ in polynomial time 
and obtain the value $\sum \limits_{0\leqslant j \leqslant m}a_j$,
and so we get $\PlHolant_{\Omega}$. Therefore, we have $\operatorname{Pl-Holant}(\neq_2\mid \mathcal{F}\cup \{\chi_2\})\leqslant_{T}\operatorname{Pl-Holant}(\neq_2\mid \mathcal{F}).$\qed

\begin{lemma}\label{binary-interpolation}
Let $g=(0, 1, t, 0)^T$ be a binary signature, where $t \neq 0$ is not a root of unity. Then for any binary signature $g'$ of the form $(0, 1, t', 0)^T$ and any signature set $\mathcal{F}$ containing $g$, we have 
$$\plholant{\neq_2}{\mathcal{F}\cup \{g'\}}\leqslant_T \plholant{\neq_2}{\mathcal{F}}.$$
Inductively, for any finite signature set $\mathcal{B}$ consisting of binary signatures of the form $(0, 1, t', 0)^T$ and any signature set $\mathcal{F}$ containing $g$, we have 
$$\plholant{\neq_2}{\mathcal{F}\cup \mathcal{B}}\leqslant_T \plholant{\neq_2}{\mathcal{F}}.$$
\end{lemma}
{\bf Proof. }
Note that $M(g)=\left[\begin{smallmatrix}
 0& 1\\
 t& 0\\
\end{smallmatrix}\right]$. 
 Connecting the variable $x_2$ of a copy of $g$ with the variable $x_1$ of another copy of $g$ using $(\neq_{2})$, we get a signature $g_2$ with the signature matrix
$$
M(g_2)=M_{x_1, x_2}(g)M(\neq_2)M_{x_1, x_2}(g)=
\left[\begin{matrix}
 0& 1\\
 t& 0\\
\end{matrix}\right]
\left[\begin{matrix}
 0& 1\\
 1& 0\\
\end{matrix}\right]
\left[\begin{matrix}
 0& 1\\
 t& 0\\
\end{matrix}\right]=
\left[\begin{matrix}
 0& 1\\
 t^2& 0\\
\end{matrix}\right]
.$$
That is, $g_2=(0, 1, t^2, 0)^T.$ 
Recursively, we can construct $g_s=(0, 1, t^s, 0)^T$ for $s \ge 1$.
Here, $g_1$ denotes $g$.
Given an instance $\Omega'$ of $\plholant{\neq_2}{\mathcal{F}\cup \{g'\}}$, 
in the same way  as in  the proof of Lemma \ref{1111}, we can replace each appearance of $g'$ by $g_s$ and get an instance $\Omega_s$ of $\plholant{\neq_2}{\mathcal{F}}$. 
Similarly, the Holant value of $\Omega_s$ can be represented as 
$$\PlHolant_{\Omega_s}=\sum_{0\leqslant j \leqslant m}a_j (t^s)^j,$$ while the Holant value of $\Omega'$ can be represented as $$\PlHolant_{\Omega'}=\sum_{0\leqslant j \leqslant m}a_j (t')^j.$$
Since $t\neq 0$ is not a root of unity, all $t^s$ are distinct, 
and so
the Vandermonde coefficient matrix 
has full rank. Hence, we can solve
for all $a_j$, and then
compute  $\sum \limits_{0\leqslant j \leqslant m}a_j(t')^j$.
So we get $\PlHolant_{\Omega'}$. Therefore, we have $\operatorname{Pl-Holant}(\neq_2\mid \mathcal{F}\cup \{g'\})\leqslant_{T}\operatorname{Pl-Holant}(\neq_2\mid \mathcal{F}).$ The second part of this lemma follows directly by the first part. \qed

\begin{remark}
Note that the reason why the interpolation can succeed is that we can construct polynomially many binary signatures $g_s$ of the form $(0, 1, t_s, 0)^T$, where all $t_s$ are distinct such that the Vandermonde coefficient matrix has full rank. According to this, we have the following corollary.
\end{remark}

\begin{corollary}\label{any-interpolation}
Given a signature set $\mathcal{F}$, 
if we can use $\mathcal{F}$ to construct polynomially many distinct binary signatures $g_s=(0, 1, t_s, 0)^T$, then for any finite signature set $\mathcal{B}$ consisting of binary signatures of the form $(0, 1, t', 0)^T$, we have 
$$\plholant{\neq_2}{\mathcal{F}\cup \mathcal{B}}\leqslant_T \plholant{\neq_2}{\mathcal{F}}.$$
\end{corollary}

In Lemma \ref{conformal5}, we will show how to construct polynomially many distinct binary signatures $g_s=(0, 1, t_s, 0)^T$ using M\"{o}bius transformations~\cite{ahlfors}.
A M\"{o}bius transformation of the extended complex plane $\widehat{\mathbb{C}}=\mathbb{C}\cup \{\infty\}$,
the complex plane plus a point at infinity, is a rational function of the form $\mathfrak{z} \mapsto \dfrac{a\mathfrak z+b}{c\mathfrak z+d}$ of 
a complex variable $\mathfrak z$,
where the coefficients $a, b, c, d$ are complex numbers satisfying $\det \left[\begin{smallmatrix}
a & b\\ c & d\\
\end{smallmatrix}\right]=ad-bc\neq 0$.
It is a bijective conformal map. 
In particular, a M\"{o}bius transformation mapping the unit circle $S^1=\{z \mid |z|=1\}$ to itself is of the form
$\varphi(\mathfrak z)=e^{\ii\theta}\dfrac{(\mathfrak{z}+\alpha)}{1+\bar{\alpha} \mathfrak{z}}$ denoted by $\mathcal{M}(\alpha, e^{\ii\theta})$, 
where $|\alpha|\neq 1$, or $\varphi(\mathfrak z)=e^{\ii\theta}/{\mathfrak{z}}$. 
When $|\alpha|< 1$, it maps the interior of $S^1$ to the interior, while when $|\alpha|> 1$, it maps the interior of $S^1$ to the exterior. A M\"{o}bius transformation is completely determined by its values on any $3$ distinct points
of $\widehat{\mathbb{C}}$.

An interpolation proof based on a lattice structure will be given in Lemma \ref{twins}, where the following lemma is used.

\begin{lemma} \label{interpolation} \cite{cfx}
Suppose $\alpha, \beta \in \mathbb{C} - \{0\}$, and
the lattice  $L = \{(j,k) \in \mathbb{Z}^{2} \mid  \alpha^j \beta^k=1\}$
has the form $L=\{(ns,nt) \mid n \in \mathbb{Z}\}$, where $s,t \in \mathbb{Z}$
 and $(s,t) \neq (0,0)$. Let  $\phi$ and $\psi$ be any numbers
satisfying $\phi^s \psi^t=1$.
If we are  given the values $N_\ell =
\sum_{j,k \ge 0,~j+k\leq m} (\alpha^j \beta^k)^{\ell} x_{j,k}$
for ${\ell}=1,2,  \ldots {m+2 \choose 2}$,
then we can compute $\sum_{j,k \ge 0,~j+k\leq m} \phi^j \psi^k x_{j,k}$
in polynomial time.
\end{lemma}

\subsection{Tractable Signature Sets}
We give some sets of signatures that are known
to define tractable counting problems.
These are called tractable signatures.
There are three families: affine signatures,
product-type signatures,
and
matchgate signatures ~\cite{caifu16}.

\paragraph{Affine Signatures}
\begin{definition}
For a signature $f$ of arity $n$,
the support of $f$ is
\[\operatorname{supp}(f)=\{(x_1, x_2, \ldots,  x_n)
\in \mathbb{Z}_2^n \mid f(x_1, x_2, \ldots, x_n)\neq 0\}.\]
\end{definition}

\begin{definition}\label{definition-affine}
 A signature $f(x_1, \ldots, x_n)$ of arity $n$
is \emph{affine} if it has the form
 \[
  \lambda \cdot \chi_{A X = 0} \cdot {\frak i} ^{Q(X)},
 \]
 where $\lambda \in \mathbb{C}$,
 $X = (x_1, x_2, \dotsc, x_n, 1)$,
 $A$ is a matrix over $\mathbb{Z}_2$,
 $Q(x_1, x_2, \ldots, x_n)\in \mathbb{Z}_4[x_1, x_2, \ldots, x_n]$
is a quadratic (total degree at most 2) multilinear polynomial
 with the additional requirement that the coefficients of all
 cross terms are even, i.e., $Q$ has the form
 \[Q(x_1, x_2, \ldots, x_n)=a_0+\displaystyle\sum_{k=1}^na_kx_k+\displaystyle\sum_{1\leq i<j\leq n}2b_{ij}x_ix_j,\]
 and $\chi$ is a 0-1 indicator function
 such that $\chi_{AX = 0}$ is~$1$ iff $A X = 0$.
 We use $\mathscr{A}$ to denote the set of all affine signatures.
\end{definition}

The following two lemmas follow directly from the definition.
\begin{lemma}\label{4affine}
Let $g$ be a binary signature with support of size $4$. Then, $g\in \mathscr{A}$ iff 
$g$ has the signature matrix
$M(g)= \lambda
\left[\begin{smallmatrix}
\ii^{a}& \ii^{b}\\
\ii^{c}& \ii^{d}
\end{smallmatrix}\right]$, for some nonzero $\lambda \in \mathbb{C}$,
$a,b,c,d \in \mathbb{N}$ and 
$a+b+c+d \equiv 0 \pmod 2$.
\end{lemma}

\begin{lemma}\label{2affine}
Let $h$ be a unary signature with support of size $2$. Then, $h\in \mathscr{A}$ iff 
$h$ has the form
$M(h)= \lambda \left[\begin{matrix}
\ii^{a}& \ii^{b}
\end{matrix}\right]$, for some nonzero $\lambda \in \mathbb{C}$,
and $a, b \in \mathbb{N}$.
\end{lemma}


\paragraph{Product-Type Signatures}
\begin{definition}
\label{definition-product-2}
 A signature on a set of variables $X$
 is of \emph{product type} if it can be expressed as a
product of unary functions,
 binary equality functions $([1,0,1])$,
and binary disequality functions $([0,1,0])$, each on one or two
variables of $X$.
 We use $\mathscr{P}$ to denote the set of product-type functions.
\end{definition}

Note that the variables of the functions in the product
need not be distinct. E.g., let 
$f(x,y,z)$
be listed as
$\left[\begin{smallmatrix}
f_{000} & f_{001} & f_{010} & f_{011}\\
f_{100} & f_{101} & f_{110} & f_{111}
\end{smallmatrix}\right]
=\left[\begin{smallmatrix}
0 & a & 0 & 0\\
0 & 0 & b & 0
\end{smallmatrix}\right]$.
$f$ is the product of $(=_2)(x,y)$, $(\neq_2)(x, z)$
and $[a,b](x)$.
Let $g$ be
$\left[\begin{smallmatrix}
 g_{00} & g_{01}\\
 g_{10} & g_{11}
\end{smallmatrix}\right]
=\left[\begin{smallmatrix}
0 & c \\
d & 0
\end{smallmatrix}\right]$, and
$h(x, y, z, w)=f(x, y, z)g(z, w)$, sharing a variable $z$.
Then $f, g, h\in\mathscr{P}$.

The following theorem is known~\cite{Cai-Lu-Xia-csp}, 
since $(\neq_2) \in \mathscr{A}
\cap \mathscr{P}$.

\begin{theorem}\label{aptractable}
  Let $\mathcal{F}$ be any set of complex-valued signatures in Boolean variables.  If $\mathcal{F} \subseteq \mathscr{A}$
 or   $\mathcal{F} \subseteq \mathscr{P}$,
 then $\Holant(\neq_2 \mid \mathcal{F})$ is tractable.
\end{theorem}

\paragraph{Matchgate Signatures \vspace{.15in}
\\}

\hspace{.1in}
Matchgates were introduced by Valiant~\cite{val02a, val02b} to give polynomial-time algorithms for a collection of counting problems over planar graphs.
As the name suggests,
problems expressible by matchgates can be reduced to computing a weighted sum of perfect matchings.
The latter problem is tractable over planar graphs by Kasteleyn's algorithm~\cite{Kasteleyn1967},
a.k.a.~the FKT algorithm~\cite{TF61,Kasteleyn1961}.
These counting problems are naturally expressed in the Holant framework using \emph{matchgate signatures}.
We use $\mathscr{M}$ to denote the set of all matchgate signatures;
thus $\PlHolant(\mathscr{M})$ is tractable, as well as
$\PlHolant(\neq_2 \mid \mathscr{M})$.

The parity of a signature is even (resp.~odd) if its support is on entries of even (resp.~odd) Hamming weight.
We say a signature satisfies the even (resp. odd) Parity Condition
if all entries of odd  (resp. even) weight are zero. For signatures of arity 
at most $4$, the matchgate signatures are characterized by the following lemma.

\begin{lemma}[\cite{val02b,jcbook}]\label{matchgate4}
If $f$ has arity $\leqslant 3$, then $f\in\mathscr{M}$ iff $f$ satisfies
the Parity Condition.

If $f$ has arity 4 and $f$ satisfies the even Parity Condition, i.e.,
\[M_{x_1x_2, x_4x_3}(f)=\begin{bmatrix}
f_{0000} & 0 & 0 & f_{0011}\\
0 & f_{0110} & f_{0101} & 0\\
0 & f_{1010} & f_{1001} & 0\\
f_{1100} & 0 & 0 & f_{1111}\\
\end{bmatrix},\]
then
$f\in\mathscr{M}$ iff $$\det M_{\rm{Out}}(f)= \det M_{\rm{In}}(f).$$
\end{lemma}

By this matchgate identity, we have the following corollary.
\begin{corollary}\label{matchgateinvariant}
Given a signature $f$ of arity $4$, two $2$-by-$2$ matrices $D_{\lambda}=\left[\begin{smallmatrix}
1 & 0\\
0 & \lambda\\
\end{smallmatrix}\right] (\lambda\neq 0)$ and $M(\neq_2)=\left[\begin{smallmatrix}
0 & 1\\
1 & 0\\
\end{smallmatrix}\right]$, if $f\in\mathscr{M}$, then $D_{\lambda}f$ and  $M(\neq_2)f\in\mathscr{M}$.
\end{corollary}

{\bf Proof.} Since $f\in \mathscr{M}$, by Lemma \ref{matchgate4} we know  $f$ satisfies the Parity Condition. We only consider that $f$ satisfies even parity. The proof for $f$ satisfying odd parity is similar and we omitted it here. Suppose $f$ has the signature matrix
$M(f)=\left[\begin{smallmatrix}
d & 0 & 0 & a \\
0 & b & c & 0 \\
0 & z & y & 0 \\
x & 0 & 0 & w \\
\end{smallmatrix}\right]$. Then, we have  $M(D_{\lambda}f)=\left[\begin{smallmatrix}
d & 0 & 0 & \lambda^2a \\
0 & \lambda^2b & \lambda^2c & 0 \\
0 & \lambda^2z & \lambda^2y & 0 \\
\lambda^2x & 0 & 0 & \lambda^4w \\
\end{smallmatrix}\right]$ and $M(M(\neq_2)f)=\left[\begin{smallmatrix}
w & 0 & 0 & x \\
0 & y & z & 0 \\
0 & c & b & 0 \\
a & 0 & 0 & d \\
\end{smallmatrix}\right]$. 
Clearly, $D_{\lambda}f$ and $M(\neq_2)f$ also satisfy even parity.
Moreover, we have $$\det M_{\rm{Out}}(D_{\lambda}f)=\lambda^4\det M_{\rm{Out}}(f),  \det M_{\rm{In}}(D_{\lambda}f)=\lambda^4\det M_{\rm{In}}(f),$$ and $$\det M_{\rm{Out}}(M(\neq_2)f)=\det M_{\rm{Out}}(f), \det M_{\rm{In}}(M(\neq_2)f)=\det M_{\rm{In}}(f).$$
Since
$\det M_{\rm{Out}}(f)= \det M_{\rm{In}}(f),$ we have $$\det M_{\rm{Out}}(D_{\lambda}f)= \det M_{\rm{In}}(D_{\lambda}f), \text{ and }\det M_{\rm{Out}}(M(\neq_2)f)= \det M_{\rm{In}}(M(\neq_2)f).$$ That is, $D_{\lambda}f$ and  $M(\neq_2)f\in\mathscr{M}.$ \qed

Holographic transformations extend the reach of the FKT algorithm further,
as stated below.
By Definition \ref{def:prelim:trans}, a signature set $\mathcal{F}$ is $\mathscr{M}$-transformable
 if there exists a $T \in \rm{GL}_2(\mathbb{C})$ such that
 $(0, 1, 1, 0) T^{\otimes 2} \in \mathscr{M}$ and $T^{-1}\mathcal{F} \subseteq \mathscr{M}$.
 
\begin{theorem} \label{mtractable}
 Let $\mathcal{F}$ be any set of complex-valued signatures in Boolean variables. 
 If $\mathcal{F}$ is $\mathscr{M}$-transformable,
 then $\PlHolant(\neq_2 \mid \mathcal{F})$ is tractable. 
\end{theorem}

 We will show that for the six-vertex model, $\mathscr{M}$-transformable signatures are exactly characterized by $\mathscr{M}$ and $\widehat{\mathscr{M}}$.
 Recall the signature class $\widehat{\mathscr{M}}=H_2\mathscr{M}$, where  $H_2=\frac{1}{\sqrt{2}} \left[\begin{smallmatrix} 1 & 1 \\ 1 & -1 \end{smallmatrix}\right]$.
 We first give the following simple lemma.
 
  \begin{lemma}\label{matchhat}
 For any signature $f$ with the signature matrix $M(f)=\left[\begin{smallmatrix}
0 & 0 & 0 & a \\
0 & b & c & 0 \\
0 & z & y & 0 \\
x & 0 & 0 & 0
\end{smallmatrix}\right]$, and a $2$-by-$2$ matrix $D_{\lambda}=\left[\begin{smallmatrix}
1 & 0\\
0 & \lambda\\
\end{smallmatrix}\right]$, where $\lambda\neq 0$, we have $f\in\mathscr{\widehat M}$ iff $(D_{\lambda})^{\otimes 4}f\in\mathscr{\widehat M}$.
 \end{lemma}
 
 {\bf Proof.} Note that $M((D_{\lambda})^{\otimes 4}f)=\left[\begin{smallmatrix}
0 & 0 & 0 & \lambda^2a \\
0 & \lambda^2b & \lambda^2c & 0 \\
0 & \lambda^2z & \lambda^2y & 0 \\
\lambda^2x & 0 & 0 & 0
\end{smallmatrix}\right]=\lambda^2M(f).$ That is, $(D_{\lambda})^{\otimes 4}f=\lambda^2 f$. Thus, $f\in\mathscr{\widehat M}$ is equivalent  to $(D_{\lambda})^{\otimes 4}f=\lambda^2 f\in\mathscr{\widehat M}$.  \qed
 
\begin{lemma}\label{mtransform}
A signature $f$ with the signature matrix $M(f)=
\left[\begin{smallmatrix}
0 & 0 & 0 & a \\
0 & b & c & 0 \\
0 & z & y & 0 \\
x & 0 & 0 & 0
\end{smallmatrix}\right]$ is $\mathscr{M}$-transformable iff $f \in \mathscr{\widehat{M}}$. 

\end{lemma}
{\bf Proof.}
The reverse direction is obvious, since $(\not =_2) I^{\otimes 2} \in \mathscr{{M}}$, and
$(\not =_2) H_2^{\otimes 2} \in \mathscr{{M}}$.

Suppose $f$ is $\mathscr{M}$-transformable. By definition, there is $T \in \rm{GL}_2(\mathbb{C})$ such that
 \[(0, 1, 1, 0) T^{\otimes 2} \in \mathscr{M}~~~\mbox{and}~~~(T^{-1})^{\otimes 4}f \in \mathscr{M}.\]
 Let $T=\left[\begin{smallmatrix}
 \lambda & \mu \\
 \nu & \xi\\
 \end{smallmatrix}\right]$. We have
 $$(0, 1, 1, 0) T^{\otimes 2}=(2\lambda\nu, \lambda\xi+\mu\nu, \lambda\xi+\mu\nu, 2\mu\xi)\in \mathscr{M}.$$
 By Lemma \ref{matchgate4}, we have $\lambda\nu=\mu\xi=0$ or $\lambda\xi+\mu\nu=0$. 
 
 If $\lambda\nu=\mu\xi=0$, since $T \in \rm{GL}_2(\mathbb{C})$, we have  $\mu=\nu=0$ while $\lambda, \xi \neq 0$, or $\lambda=\xi=0$ while $\mu, \nu\neq 0$. That is,
 $T=\left[\begin{smallmatrix}
 \lambda & 0 \\
 0 & \xi\\
 \end{smallmatrix}\right]$ $(\lambda, \xi \neq 0)$, or $T=\left[\begin{smallmatrix}
 0 & \mu \\
 \nu & 0\\
 \end{smallmatrix}\right]$ $(\mu, \nu\neq 0)$. By normalization, we may assume $\lambda=1$ or $\mu=1$. That is, $$T=\left[\begin{matrix}
 1 & 0 \\
 0 & \xi\\
 \end{matrix}\right] (\xi \neq 0), \text{ or } T=\left[\begin{matrix}
 0 & 1 \\
 \nu & 0\\
 \end{matrix}\right]=\left[\begin{matrix}
 1 & 0 \\
 0 & \nu\\
 \end{matrix}\right]\left[\begin{matrix}
 0 & 1 \\
 1 & 0\\
 \end{matrix}\right] (\nu\neq 0).$$ 
 For any $\alpha\neq 0$, we use $D_{\alpha}$ to denote the matrix $\left[\begin{smallmatrix}
 1 & 0 \\
 0 & \alpha\\
 \end{smallmatrix}\right]$ and we know $D_{\alpha}^{-1}=D_{1/\alpha}.$
 Then, $T=D_{\xi}$ or $T=D_{\nu}M(\neq_2)$. 
 By Corollary \ref{matchgateinvariant}, we know $f\in \mathscr{M}$ given $T^{-1}f \in \mathscr{M}$.
 
 Otherwise, $\lambda\xi+\mu\nu=0$. Since $T \in \rm{GL}_2(\mathbb{C})$, we know $\det T= \lambda\xi-\mu\nu\neq0$. Thus,  $\lambda\xi\mu\nu\neq0$. By normalization, we may assume $\lambda=1$ and hence, $\xi=-\mu\nu$. That is 
 $$T=\left[\begin{matrix}
 1 & \mu \\
 \nu & -\mu\nu\\
 \end{matrix}\right]=\left[\begin{matrix}
 1 & 0 \\
 0 & \nu\\
 \end{matrix}\right]\left[\begin{matrix}
 1 & 1 \\
 1 & -1\\
 \end{matrix}\right]\left[\begin{matrix}
 1 & 0 \\
 0 & \mu\\
 \end{matrix}\right]=D_{\nu}H_2D_{\mu}.$$
 Hence, we have $T^{-1}=\frac{1}{2}D_{1/\mu}H_2D_{1/\nu}$ and we know $D_{1/\mu}H_2D_{1/\nu}f \in \mathscr{M}$. 
 We have $D_{\mu}\mathscr{M} = \mathscr{M}$.
 Hence $H_2D_{1/\nu}f \in \mathscr{M}$. Thus, we have $$D_{1/\nu}f \in
 \widehat{\mathscr{M}}.$$
 By Lemma \ref{matchhat}, we have $f\in \widehat{\mathscr{M}}$ given $D_{1/\nu}^{\otimes 4}f\in\widehat{\mathscr{M}}$. \qed
 
  For signatures of special forms, we give the following three characterizations of $\widehat{\mathscr{M}}$. They follow directly from the definition.
\begin{lemma}\label{binary-m}
A  binary
 signature $g$ with the signature matrix $M(g)=\left[\begin{smallmatrix}
g_{00} & g_{01}\\
g_{10} & g_{11}\\
\end{smallmatrix}\right]$ is in $\mathscr{\widehat{M}}$ iff $g_{00}=\epsilon g_{11}$ and  $g_{01}=\epsilon g_{10}$, where $\epsilon=\pm 1$.
\end{lemma}
\begin{lemma}\label{ary4-m}
A signature $f$ with the signature matrix $M(f)=
\left[\begin{smallmatrix}
0 & 0 & 0 & 0 \\
0 & b & c & 0 \\
0 & z & y & 0 \\
0 & 0 & 0 & 0
\end{smallmatrix}\right]$ is in $\mathscr{\widehat{M}}$ iff $b=\epsilon y$ and  $c=\epsilon z$, where $\epsilon=\pm 1$.
\end{lemma}
\begin{lemma}\label{notmatchhat}
If $f$ has the signature matrix $M(f)=
\left[\begin{smallmatrix}
0 & 0 & 0 & a \\
0 & b & 0 & 0 \\
0 & 0 & y & 0 \\
x & 0 & 0 & 0
\end{smallmatrix}\right]$, where $abxy\neq 0$, then $f \notin \mathscr{\widehat{M}}$.
\end{lemma}

\subsection{Known Dichotomies and Hardness Results}

 \begin{definition}
 A 4-ary signature is non-singular redundant iff in one of its  four $4\times4$ signature matrices,
  the middle two rows are identical and the
 middle two columns are identical, and the determinant
\[ \det \left[\begin{matrix}
f_{0000} & f_{0010} & f_{0011}\\
f_{0100} & f_{0110} & f_{0111}\\
f_{1100} & f_{1110} & f_{1111}
\end{matrix}\right] \not = 0. \]
 \end{definition}
\begin{theorem}\cite{caiguowilliams13}
\label{redundant}
If $f$ is a non-singular redundant signature, 
 then $\operatorname{Pl-Holant}(\neq_2|f)$ is $\SHARPP$-hard. 
\end{theorem}
 \begin{theorem}\cite{tutte}\label{tutte}
 Let $G$ be a connected plane graph and $\mathcal{EO}(H)$ be the set of all Eulerian orientations
 of the medial graph $H=H(G)$ which is a 4-regular planar graph. Then 
 \[
 \sum_{O\in\mathcal{EO}(H)}2^{\beta(O)}=2T(G; 3, 3),
 \]
 where $T$ is the Tutte polynomial,
and $\beta(O)$ is the number of saddle vertices in the orientation $O$, i.e., vertices in 
 which the edges are oriented ``in, out, in, out" in cyclic order.
 \end{theorem}
 \begin{remark}
 Note that $\sum_{O\in\mathcal{EO}(H)}2^{\beta(O)}$
 can be expressed as Pl-Holant$(\neq_2\mid f)$ on $H$, where $f$
has the signature matrix
$M(f)=\left[\begin{smallmatrix}
0 & 0 & 0 & 1\\
 0 & 1 & 2 & 0\\
 0 & 2 & 1 & 0\\
 1 & 0 & 0 & 0
 \end{smallmatrix}\right]$.
 Therefore,  Pl-Holant$(\neq_2\mid f)$ is $\SHARPP$-hard.
\end{remark}
\begin{theorem}\label{cspdic}\cite{caifu16}
 Let $\mathcal{F}$ be any set of complex-valued signatures in Boolean variables.
 Then $\PlCSP(\mathcal{F})$ is $\SHARPP$-hard unless
 $\mathcal{F} \subseteq \mathscr{A}$,
 $\mathcal{F} \subseteq \mathscr{P}$, or
 $\mathcal{F} \subseteq \widehat{\mathscr{M}}$,
 in which case the problem is computable in polynomial time. If $\mathcal{F} \subseteq \mathscr{A}$
 or $\mathcal{F} \subseteq \mathscr{P}$, then {\rm \#CSP}$(\mathcal{F})$ is computable in polynomial time without planarity; otherwise {\rm \#CSP}$(\mathcal{F})$ is  $\SHARPP$-hard.
 \end{theorem}
 \begin{theorem}\label{nonplanardic}\cite{cfx}
  Let $f$ be a 4-ary signature with the signature matrix
$M(f)=\left[\begin{smallmatrix}
0 & 0 & 0 & a\\
 0 & b & c & 0\\
 0 & z & y & 0\\
 x & 0 & 0 & 0
 \end{smallmatrix}\right]$, then
\holant{\neq_2}{f} is \#P-hard except for the following cases:
\begin{itemize}
\item $f\in\mathscr{P}$;
\item $f\in\mathscr{A}$;
\item there is a zero in each pair $(a,x), (b,y), (c,z)$;
\end{itemize}
in which cases \holant{\neq_2}{f} is computable in polynomial time.
 \end{theorem}

\section{Main Theorem, Proof Outline and Sample Problems}\label{sec:main}

\begin{theorem}\label{main}
Let $f$ be a signature with the signature matrix
$M(f) =
\left[\begin{smallmatrix}
0 & 0& 0& a\\
0 & b& c& 0\\
0 & z& y& 0\\
x & 0& 0& 0
\end{smallmatrix}\right]$, where $a,b,c,x,y,z \in \mathbb{C}$.
Then $\operatorname{Pl-Holant}(\neq_2\mid f)$ is
polynomial time 
computable in the following cases, and \#P-hard otherwise:
\begin{enumerate}
    \item\label{con1} $f \in \mathscr{P}$ or $\mathscr{A}$;
    \item\label{con2} There is a zero in each pair $(a, x)$, $(b, y)$, $(c, z)$;
    \item\label{con3} $f \in \mathscr{M}$ or $\widehat{\mathscr{M}}$;
    \item\label{con4} $c=z=0$ and 
    \begin{itemize}
\item[$\operatorname{(\rmnum{1})}$.] 
$(ax)^2=(by)^2$, or
\item[$\operatorname{(\rmnum{2})}$.] 
$x=a\ii^\alpha, b=a\sqrt{\ii}^\beta$, and $y=a\sqrt{\ii}^\gamma$, 
where $\alpha, \beta, \gamma \in \mathbb{N}$, and $\beta\equiv \gamma \pmod 2;$
\end{itemize}
\end{enumerate}
If $f$ satisfies condition \ref{con1} or \ref{con2}, then $\operatorname{Holant}(\neq_2\mid f)$ is computable in polynomial time without the planarity
restriction; otherwise (the non-planar) $\operatorname{Holant}(\neq_2\mid f)$ is \#P-hard.
\end{theorem}

Let ${\sf N}$ be  the number of zeros
among $a,b,c,x,y,z$. The following division of all cases
into Cases I, II, III and IV may not appear to be the most obvious, but it
is done to simplify the organization of the proof.
We define:
\begin{description}
\item{Case I:}
There is exactly one zero in each pair.
\item{Case II:}
There is a zero pair.
\item{Case III:}
${\sf N}=2$ and having no zero pair,
or ${\sf N}=1$ and the zero is in an outer pair.
\item{Case IV:}
${\sf N}=1$ and the
zero is in an inner pair, or ${\sf N}=0$.
\end{description}

Cases I, II, III and IV  are clearly disjoint.
To see that they cover all cases,
note that
 if ${\sf N} \geqslant 3$, then 
either there is a zero pair (in Case II), or
${\sf N}=3$ and each pair has exactly one zero (in Case I).
If  ${\sf N}=2$, then either it has a zero pair (in Case II), or
 it has no zero pair (in Case III).
If ${\sf N}=1$,  then either the single zero is in  an outer pair
(in Case III), or the single zero is in an inner pair (Case IV).
If ${\sf N}=0$ it is in Case IV.

Also note that if  ${\sf N}=2$ and it has no zero pair,
then the two zeros are in different pairs,
which implies that there is a zero in an outer pair. 
So in Case III, there is a zero in an outer pair regardless
${\sf N}=1$ or ${\sf N}=2$. In Case III an outer pair
has exactly one zero, and the other two pairs together have at most one
zero.



In Case II, depending on whether the zero pair is
inner or outer we have two different connections to \#CSP.
A previously established connection to
\#CSP (see~\cite{cfx}) can be adapted in the planar setting to handle
the case with a zero outer pair.
This connection is a local transformation, and we
observe that it preserves planarity.
A significantly more involved non-local connection to \#CSP is
discovered in this paper when the inner pair is zero (and no outer pair 
is zero).
We show that by the support structure of the signature 
we can define a set of
circuits, which
forms a partition of the edge set.
There are exactly two
valid configurations along each such circuit,
corresponding
to its two cyclic orientations.
These circuits may intersect in complicated ways, including
self-intersections. But we can define a \#CSP problem,
where
the variables are these circuits,  and their edge functions
exactly account for  the intersections. 
 We show that
 $\operatorname{Pl-Holant}(\neq_2\mid f)$ is equivalent to these
 \#CSP problems, which are
\emph{non-planar} in general.
 However, crucially, because
$\operatorname{Pl-Holant}(\neq_2\mid f)$ is planar, every two
such circuits must intersect an even number of times.
 Due to the planarity of
 $\operatorname{Pl-Holant}(\neq_2\mid f)$ we can exactly carve out
a new class of tractable problems via this non-local \#CSP connection,
by the kind of constraint functions they produce in the \#CSP problems.

For the proof of \#P-hardness in this paper, one particularly difficult case is
in Lemma \ref{conformal5}. 
This is where we introduce M\"{o}bius transformations
to prove dichotomy theorems for counting problems.
In this case, all constructible  binary
signatures correspond to points
on the unit circle $S^1$, and any iteration of the construction
amounts to mapping this point by a M\"{o}bius transformation
which preserves $S^1$.

The following is an outline on how Case I to Case IV are handled.
\begin{enumerate}
    \item[\Rmnum{1}.]\label{case1} 
    There is exactly one zero in each pair. 
    In this case, $\holant{\neq_2}{f}$ is tractable, proved in 
 \cite{cfx}.

    \item[\Rmnum{2}.]\label{case1} There is a zero pair:
    \begin{enumerate}
        \item[1.]\label{case1.1}  An outer pair $(a, x)$ or $(b, y)$ is a zero pair. We prove that
$\operatorname{Pl-Holant}(\neq_2\mid f)$ is tractable if
 $f \in \mathscr{P}, \mathscr{A}, \mathscr{M}$ or $\widehat{\mathscr{M}}$, 
and is \#P-hard  otherwise.
        
        In this Case \Rmnum{2}.1, we can rotate the signature $f$ such that the matrix $M_{\text{Out}}(f)$ is the zero matrix. Let $M(\widetilde {f}_{\rm {In}})=M_{\rm{In}}(f)\left[\begin{smallmatrix}
0 & 1\\
1 & 0\\
\end{smallmatrix}\right]$.
        We reduce Pl-\#CSP$(\widetilde f_{\rm In})$
 to \plholant{\neq_2}{f} via a local replacement (Lemma \ref{outerlemma}). 
We apply the dichotomy of Pl-\#CSP to get \#P-hardness (Theorem \ref{outer}).
Tractability of  \plholant{\neq_2}{f} follows from known tractable signatures.  
        
\item[2.]\label{case1.2}  The inner pair $(c, z)$ is a zero pair and no outer pair is a zero pair.
We prove that
 $\operatorname{Pl-Holant}(\neq_2\mid f)$ is \#P-hard unless $f$ 
satisfies condition~\ref{con4}, 
in which case it is tractable. 

This is the non-local reduction described above.
The tractable condition~\ref{con4} is previously unknown.
(Curiously, in Case \Rmnum{2}.2, condition~\ref{con4} subsumes
 $f \in \mathscr{M}$.)

    \end{enumerate}
    \item[\Rmnum{3}.]\label{case3} 
    \begin{enumerate}   
   \item[1.] There are exactly two zeros and they are in different pairs; 
   \item[2.] There is exactly one zero and it is in an outer pair.
   \end{enumerate}
 We prove that
   $\operatorname{Pl-Holant}(\neq_2\mid f)$ is \#P-hard unless $f \in \mathscr{M}$, in which case it is tractable. 
   
In Case III, there exists an outer pair which contains  a single zero.
By connecting two copies of the
signature $f$, we can construct a $4$-ary signature $f_1$ such that one outer pair is a zero pair. When $f \notin \mathscr{M}$, we can realize a signature $M(g)=\left[\begin{smallmatrix}
   0 & 0& 0& 0\\
   0 & 0& 1& 0\\
   0 & 1& 0& 0\\
   0 & 0& 0& 0\\
   \end{smallmatrix}\right]$ by interpolation using $f_1$ (Lemma \ref{fullrank}). 
This $g$ can help us ``extract'' the inner matrix of $M(f)$.
By connecting $f$ and $g$, we can construct a signature that belongs to Case \Rmnum{2}.
We then prove \#P-hardness using the result of Case \Rmnum{2} (Theorem \ref{onezero}).

    
    \item[\Rmnum{4}.]\label{case4} 
    \begin{enumerate}    
    \item[1.] There is exactly one zero and it is in the inner pair;
    \item[2.] All values in $\{a, x, b, y, c, z\}$ are nonzero.
    \end{enumerate}
We prove that
     $\operatorname{Pl-Holant}(\neq_2\mid f)$ is  \#P-hard unless $f \in \mathscr{M}$, in which case it is tractable. 
   
Assume $f  \not \in \mathscr{M}$.
The main idea is to use M\"{o}bius transformations.
However, there are some settings where we cannot do so, either because we don't
have the initial signature to start the process, or the matrix that
would define the M\"{o}bius transformation is singular.
   So we first treat the following two special cases.
      \begin{itemize}
      \item If $a=\epsilon x$, $b=\epsilon y$ and $c=\epsilon z$, where $\epsilon = \pm 1$, by interpolation based on a 
 lattice structure, either we can realize a non-singular redundant signature or
reduce from the evaluation of  the Tutte polynomial at  $(3,3)$, 
both of
 which are \#P-hard (Lemma \ref{twins}).
         \item If
$\det\left[\begin{smallmatrix}
b & c \\
z & y
\end{smallmatrix}\right] =0$
or
$\det\left[\begin{smallmatrix}
a & z \\
c & x
\end{smallmatrix}\right] =0$,
then either we can realize a non-singular redundant signature or a signature that is \#P-hard by Lemma \ref{twins} (Lemma \ref{degenerate}).
    \end{itemize}
      If $f$ does not belong to the above two cases, we 
want to realize binary signatures of the form $(0, 1, t, 0)^T$,
      for arbitrary values of $t$. If this can be done,
      by carefully choosing the values of $t$,
      we can construct a signature that belongs to Case \Rmnum{3}
and it is \#P-hard when $f \notin \mathscr{M}$  (Lemma \ref{t<>1}).
      We realize binary signatures by connecting $f$ with $(\neq_2)$. 
This corresponds naturally to a M\"{o}bius transformation.
      By discussing the following different forms of binary signatures we get, 
      we can either realize arbitrary $(0, 1, t, 0)^T$ or a signature belonging
 to Case \Rmnum{2}.2 that does not satisfy condition~\ref{con4}, therefore
is \#P-hard (Theorem \ref{nonzero}).

    \begin{itemize}
         \item If we can get a signature of the form $g = (0, 1, t, 0)^T$ where $t\neq0$ is not a root of unity, then by connecting a chain of $g$, we can 
get polynomially many distinct binary signatures $g_i=(0, 1, t^i, 0)^T$.
         Then, by interpolation, we can realize 
 arbitrary binary signatures of the form $(0, 1, t', 0)^T$.
         
         \item Suppose we can get a signature of the form $(0, 1, t, 0)^T$, where $t\neq0$ is an $n$-th primitive root of unity $(n\geqslant 5)$. Now, we only have $n$ many distinct signatures $g_i=(0, 1, t^i, 0)^T$.
         But we can relate $f$ to two M\"{o}bius transformations 
due to 
$\det\left[\begin{smallmatrix}
b & c \\
z & y
\end{smallmatrix}\right] \not =0$
and
$\det\left[\begin{smallmatrix}
a & z \\
c & x
\end{smallmatrix}\right] \not =0$.
         For each M\"{o}bius transformation $\varphi$, we can realize the signatures $g= (0, 1, \varphi(t^i), 0)^T$. 
If $|\varphi(t^i)|\neq 0, 1$ or $\infty$ for some $i$, 
then this is treated above, as this $\varphi(t^i)$ is
nonzero and not a root of unity.
Otherwise, since $\varphi$ is a bijection on the extended complex plane
$\widehat{\mathbb{C}}$, it can map at most two points of  $S^1$ to
$0$ or $\infty$.
         Hence, $|\varphi(t^i)|= 1$ for at least three distinct $t^i$.
But a M\"{o}bius transformation is determined by any three distinct points.
This implies that $\varphi$  maps  $S^1$ to itself.
 Such mappings $\varphi$  have a known special form 
$e^{\ii\theta}\dfrac{\mathfrak{z}+\alpha}{1+\bar{\alpha}\mathfrak{z}}$
(or $e^{\ii\theta}/\mathfrak{z}$, but the latter form actually
cannot occur in our context.)
By exploiting its property we can construct a signature $f'$ such that 
its corresponding M\"{o}bius transformation $\varphi'$ 
defines an infinite group. This implies
that $\varphi'^k(t)$ are all distinct. Then, we can get polynomially
 many distinct binary signatures $(0, 1, \varphi'^k(t), 0)$, and 
realize arbitrary binary signatures of the form $(0, 1, t', 0)^T$ (Lemma \ref{conformal5}).
         
         \item Suppose we can get a signature of the form $(0, 1, t, 0)^T$ where $t\neq0$ is an $n$-th primitive root of unity $(n=3, 4)$. Then we can either relate it to two M\"{o}bius transformations mapping the unit circle to itself, or 
realize a double  pinning $(0, 1, 0, 0)^T = (1, 0)^T \otimes (0, 1)^T$ (Corollary \ref{conformal4}).
         \item Suppose we can get a signature of the form $(0, 1, 0, 0)^T$.
 By connecting $f$ with it, we can get new signatures of the form $(0, 1, t, 0)^T$. Similarly, by analyzing the value of $t$, 
         we can either realize arbitrary binary signatures of the form $(0, 1, s, 0)^T$, or realize a signature that belongs to Case \Rmnum{2}.2, which is \#P-hard (Lemma \ref{pin}).
         
         \item Suppose we can only get signatures of the form $(0, 1, \pm 1, 0)$. That implies $a=\epsilon x$, $b=\epsilon y$ and $c=\epsilon z$, where $\epsilon=\pm 1$.  This has been treated before.
    \end{itemize}
    
\end{enumerate}

As Case I has already been proved tractable in~\cite{cfx}, we only deal
with Cases II, III and IV, and they
are each dealt with in the next three sections.
Before we start the proof, we first illustrate the scope of Theorem \ref{main} by several concrete problems.

\vspace{1ex}

$\mathbf{Problem ~1:}$ \#{\sc EO} on 4-Regular Planar Graphs.

$\mathbf{Input:}$ A 4-regular planar graph $G$.

$\mathbf{Output:}$ The number of Eulerian orientations of $G$,
i.e., the number of orientations of $G$ such that at every vertex
the in-degree and out-degree are equal.

This problem can be expressed as Pl-Holant$(\neq_2|f)$, where $f$
has the signature matrix 
$M(f)=\left[\begin{smallmatrix} 
0 & 0 & 0 & 1\\
 0 & 1 & 1 & 0\\
 0 & 1 & 1 & 0\\
 1 & 0 & 0 & 0
 \end{smallmatrix}\right]$. 
Huang and Lu proved this problem is \#P-complete~\cite{Huang-Lu}.
Theorem \ref{main} confirms this.

\vspace{2ex}

$\mathbf{Problem ~2:}$ Pl-$T(G; 3, 3)$.

$\mathbf{Input:}$ A planar graph $G$.

$\mathbf{Output:}$ The value of the Tutte polynomial $T(G; x, y)$  at $(3, 3)$.

Let $G_m$ be the medial graph of $G$, then $G_m$ is a 4-regular planar graph. By Theorem \ref{tutte}, we have  \[
 \sum_{O\in\mathcal{EO}(G_m)}2^{\beta(O)}=2T(G; 3, 3),
 \]
 where $\beta(O)$ is the number of saddle vertices in the orientation $O$.
 Note that $\sum_{O\in\mathcal{EO}(G_m)}2^{\beta(O)}$
 can be expressed as Pl-Holant$(\neq_2|f)$, where $f$
has the signature matrix
$M(f)=\left[\begin{smallmatrix}
0 & 0 & 0 & 1\\
 0 & 1 & 2 & 0\\
 0 & 2 & 1 & 0\\
 1 & 0 & 0 & 0
 \end{smallmatrix}\right]$.
Theorem \ref{main} confirms that this problem is \#P-hard.

\vspace{2ex}

Compared to the six-vertex model over general graphs, 
the planar version has new tractable problems 
due to the FKT algorithm under holographic transformations.
This tractable class can give highly nontrivial problems. 
For example, we consider the following problem.

\vspace{2ex}

$\mathbf{Problem ~3:}$ {\sc SmallPell}

$\mathbf{Input:}$ A planar 4-regular graph $G$ and a 4-ary signature $f$, where $f$ has the signature matrix 
\[M(f)=\left[\begin{smallmatrix}
317830805723707970 & -283823304736008960{\mathfrak i} & 283823304736008960{\mathfrak i}  & 317830805723707968\\
 -283823304736008960{\mathfrak i} & -253454564065438270 & 253454564065438272 & -283823304736008960{\mathfrak i}\\
 283823304736008960{\mathfrak i} & 253454564065438272 & -253454564065438270 & 283823304736008960{\mathfrak i}\\
 317830805723707968 & -283823304736008960{\mathfrak i} & 283823304736008960{\mathfrak i} & 317830805723707970
 \end{smallmatrix}\right].\]

$\mathbf{Output:}$ The evaluation of $\operatorname{Pl-Holant}(f)$ on $G$.

 By the holographic transformation $Z=\frac{1}{\sqrt{2}}
\left[\begin{smallmatrix}
 1 & 1\\
 \mathfrak i & -\mathfrak i
 \end{smallmatrix}\right]$, we have
 \[
 \operatorname{Pl-Holant}(f)\equiv_{T} \operatorname{Pl-Holant}(\neq_2|\widehat{f}),
 \]
where \[M(\widehat{f})=\left[\begin{smallmatrix}
0 & 0 & 0 & 1\\
 0 & 569465989630582080 & 32188120829134849 & 0\\
 0 & 32188120829134849 &  1819380158564160 & 0\\
 1 & 0 & 0 & 0
 \end{smallmatrix}\right].\]
 Since $(32188120829134849, 1819380158564160)$ is a solution of Pell's equation 
 $x^2-313y^2=1$, we have $\widehat{f}\in\mathscr{M}$
by Matchgate Identities~\cite{val02b}.
 By Theorem \ref{main}, $\operatorname{Pl-Holant}(f)$ can be computed in polynomial time.


\vspace{2ex}

In addition to matchgates and matchgates-transformable
signatures,
Theorem \ref{main} gives a new class of
 tractable problems on planar graphs. 
They are provably not contained in any
previously known tractable classes.
For example, we consider the following problem.

\vspace{1ex}

\noindent$\mathbf{Problem ~4:}$\label{problem4} Pl-Holant$(\neq_2| f)$, where $f$ has the signature matrix $M({f})=
\left[\begin{smallmatrix}
 0 & 0 & 0 & 1\\
 0 & \sqrt{\mathfrak i} & 0 & 0\\
 0 & 0 & \sqrt{\mathfrak i} & 0\\
 1 & 0 & 0 & 0
 \end{smallmatrix}\right]$.

\noindent$\mathbf{Input:}$ An instance of Pl-Holant$(\neq_2| f)$.

\noindent$\mathbf{Output:}$ The evaluation of this instance.

 By Theorem \ref{main} (condition \ref{con4} (\rmnum{2})),
 Pl-Holant$(\neq_2| f)$ can be computed in polynomial time.
 Note that Holant$(\neq_2|{f})$ is \#P-hard without the planar restriction.
It can be shown that ${f}$ is neither in
$\mathscr{M}$ nor $\mathscr{M}$-transformable.
By Lemma \ref{matchgate4} we know ${f} \not \in \mathscr{M}$, and by Lemma \ref{notmatchhat} we know ${f} \not \in \widehat{\mathscr{M}}$. By Lemma \ref{mtransform}, this implies ${f}$ is neither in
$\mathscr{M}$ nor $\mathscr{M}$-transformable.
 
Therefore,
the tractability is not derivable from the Kasteleyn's algorithm
or a holographic transformation to it. Hence, condition~\ref{con4} of
Theorem~\ref{main}  defines a new 
component
of planar tractability complementing the Kasteleyn's algorithm.
\emph{Furthermore}, it is an essential component because
with it the picture is complete.
\section{Case \Rmnum{2}\label{one-zero-pair}: One Zero Pair}
If an outer pair is a zero pair, by rotational symmetry, we may assume $(a, x)$ is a zero pair. 
\begin{definition}
Given a 4-ary signature $f$ with the signature matrix 
\begin{equation}\label{outermatrix}
M(f)=\left[\begin{matrix}
0 & 0& 0& 0\\
0 & b& c& 0\\
0 & z& y& 0\\
0 & 0& 0& 0\\
\end{matrix}\right], 
\end{equation} 
we denote by $\widetilde{f}_{\rm {In}}$  the binary signature with
 $M(\widetilde {f}_{\rm {In}})=M_{\rm{In}}(f)\left[\begin{smallmatrix}
0 & 1\\
1 & 0\\
\end{smallmatrix}\right]=\left[\begin{smallmatrix}
c & b\\
y & z\\
\end{smallmatrix}\right].$ 
Given a set $\mathcal{ F}$ consisting of signatures of the form {\rm (\ref{outermatrix})}, we define $\mathcal{\widetilde F}_{\rm {In}} = \{\widetilde f_{\rm {In}} \mid f\in \mathcal{F}\}$.
\end{definition}


\begin{lemma}\label{outerlemma}
For any set  $\mathcal{F}$  of signatures of the
 form {\rm (\ref{outermatrix})},
 $$\PlCSP(\mathcal{\widetilde {F}}_{\rm In})\leqslant_{T}\plholant{\neq_2}{\mathcal{F}}.$$
\end{lemma}

{\bf Proof.}
We adapt a  proof
from~\cite{cfx}, making sure that 
the reduction preserves planarity.  
This need to preserve planarity necessitates the twist introduced
in the definition of $\widetilde f_{\rm {In}}$ and $\mathcal{\widetilde F}_{\rm {In}}$.
We prove this reduction
in two steps.
In each step, we begin with a signature grid and end with a new signature grid such that the Holant values of both signature grids are the same.

\begin{figure}[!htpb]
\centering
\subfloat[]{
\begin{tikzpicture}[scale=0.4]
\node [internal, scale=0.6] at (1, 1.73) {};
\node at (1.8, 1.73) {$u'$};
\node [internal, scale=0.6] at (1, 5.73) {};
\node at (1.8, 5.73) {$u$};
\node [external] at (-1.3, 0) {};
\node [external] at (2, 0) {};
\node [external] at (-0.4, 4.33) {};
\node [external] at (2.4, 4.33) {};
\node [external] at (-0.4, 7.13) {};
\node [external] at (2.4, 7.13) {};
\node [external] at (0, 0) {};
\draw (0, 0)--(1, 1.73) [postaction={decorate, decoration={
                                        markings,
                                        mark=at position 0.63 with {\arrow[>=diamond,white] {>}; },
                                        mark=at position 0.63 with {\arrow[>=open diamond]  {>}; } } }];
\draw (2, 0)--(1, 1.73) [postaction={decorate, decoration={
                                        markings,
                                        mark=at position 0.63 with {\arrow[>=diamond,white] {>}; },
                                        mark=at position 0.63 with {\arrow[>=open diamond]  {>}; } } }];
\draw (1, 5.73)--(-0.4, 4.33) [postaction={decorate, decoration={
                                        markings,
                                        mark=at position 0.63 with {\arrow[>=diamond,white] {>}; },
                                        mark=at position 0.63 with {\arrow[>=open diamond]  {>}; } } }];
\draw (1, 5.73)--(-0.4, 7.13) [postaction={decorate, decoration={
                                        markings,
                                        mark=at position 0.63 with {\arrow[>=diamond,white] {>}; },
                                        mark=at position 0.63 with {\arrow[>=open diamond]  {>}; } } }];
\draw (1, 5.73)--(2.4, 4.33) [postaction={decorate, decoration={
                                        markings,
                                        mark=at position 0.63 with {\arrow[>=diamond,white] {>}; },
                                        mark=at position 0.63 with {\arrow[>=open diamond]  {>}; } } }];
\draw (1, 5.73)--(2.4, 7.13) [postaction={decorate, decoration={
                                        markings,
                                        mark=at position 0.63 with {\arrow[>=diamond,white] {>}; },
                                        mark=at position 0.63 with {\arrow[>=open diamond]  {>}; } } }];
\draw (1, 5.73)--(1, 1.73) [postaction={decorate, decoration={
                                        markings,
                                        mark=at position 0.63 with {\arrow[>=diamond,white] {>}; },
                                        mark=at position 0.63 with {\arrow[>=open diamond]  {>}; } } }];
\end{tikzpicture}
}
\qquad
\subfloat[]{
\begin{tikzpicture}[scale=0.33]
\draw [thick, ->](2, 3.73)--(4, 3.73);
\draw [thick, ->](15.5, 3.73)--(17.5, 3.73);
\node [internal, scale=0.6] (51) at (8.2, 7.56) {};
\node [internal, scale=0.6] (52) at (10.8, 7.56) {};
\node [internal, scale=0.6]  (53) at (9.5, 4.23) {};
\node [internal, scale=0.6] (54) at (7.5, 5.56) {};
\node [internal, scale=0.6] (55) at (11.5, 5.56) {};
\node at (9.5, 6) {$u$};
\node at (9.4, 1.2) {$u'$};
\node [external] (51a) at (6.8, 8.96) {};
\node [external] (52a) at (12.4, 8.96) {};
\node [external] (54a) at (6.1, 4.16) {};
\node [external] (55a) at (12.9, 4.16) {};
\draw (51)--(52);
\draw (51)--(54);
\draw (55)--(52);
\draw (53)--(54);
\draw (53)--(55);
\draw (51)--(51a) [postaction={decorate, decoration={
                                        markings,
                                        mark=at position 0.83 with {\arrow[>=diamond,white] {>}; },
                                        mark=at position 0.83 with {\arrow[>=open diamond]  {>}; } } }];
\draw (52)--(52a) [postaction={decorate, decoration={
                                        markings,
                                        mark=at position 0.83 with {\arrow[>=diamond,white] {>}; },
                                        mark=at position 0.83 with {\arrow[>=open diamond]  {>}; } } }];
\draw (54)--(54a) [postaction={decorate, decoration={
                                        markings,
                                        mark=at position 0.83 with {\arrow[>=diamond,white] {>}; },
                                        mark=at position 0.83 with {\arrow[>=open diamond]  {>}; } } }];
\draw (55)--(55a) [postaction={decorate, decoration={
                                        markings,
                                        mark=at position 0.83 with {\arrow[>=diamond,white] {>}; },
                                        mark=at position 0.83 with {\arrow[>=open diamond]  {>}; } } }];
\node [internal, scale=0.6]  (31) at (9.5, 2.23) {};
\node [internal, scale=0.6]  (32) at (8.5, 0.53) {};
\node [internal, scale=0.6]  (33) at (10.5, 0.53) {};
\node [external]  (32a) at (7.1, -0.87) {};
\node [external]  (33a) at (11.9, -0.87) {};
\draw (31)--(32);
\draw (31)--(33);
\draw (32)--(33);
\draw (32)--(32a) [postaction={decorate, decoration={
                                        markings,
                                        mark=at position 0.83 with {\arrow[>=diamond,white] {>}; },
                                        mark=at position 0.83 with {\arrow[>=open diamond]  {>}; } } }];
\draw (33)--(33a) [postaction={decorate, decoration={
                                        markings,
                                        mark=at position 0.83 with {\arrow[>=diamond,white] {>}; },
                                        mark=at position 0.83 with {\arrow[>=open diamond]  {>}; } } }];
\draw (53)--(31) [postaction={decorate, decoration={
                                        markings,
                                        mark=at position 0.83 with {\arrow[>=diamond,white] {>}; },
                                        mark=at position 0.83 with {\arrow[>=open diamond]  {>}; } } }];
%
\end{tikzpicture}
\label{holant-csp-b}
}
\qquad
\subfloat[]{
\begin{tikzpicture}[scale=0.35]
\node [triangle, scale=0.4] (c51) at (18.2, 7.56) {};
\node [external] (c51a) at (18.4, 9.56) {};
\node [external] (c51b) at (17, 9.2) {};
\draw [densely dashed] (c51) to [bend left=20] (c51a);
\draw [densely dashed] (c51) to [bend right=20] (c51b);
\node [triangle, scale=0.4] (c52) at (20.8, 7.56) {};
\node [external] (c52a) at (20.6, 9.56) {};
\node [external] (c52b) at (22, 9.2) {};
\draw [densely dashed] (c52) to [bend right=20] (c52a);
\draw [densely dashed] (c52) to [bend left=20] (c52b);
\node [triangle, scale=0.4]  (c53) at (19.5, 4.23) {};
\node [triangle, scale=0.4] (c54) at (17.5, 5.56) {};
\node [external] (c54a) at (15.5, 5.56) {};
\node [external] (c54b) at (16, 4.16) {};
\draw [densely dashed] (c54) to [bend left=20] (c54a);
\draw [densely dashed] (c54) to [bend right=20] (c54b);
\node [triangle, scale=0.4] (c55) at (21.5, 5.56) {};
\node [external] (c55a) at (23.5, 5.56) {};
\node [external] (c55b) at (23, 4.16) {};
\draw [densely dashed] (c55) to [bend right=20] (c55a);
\draw [densely dashed] (c55) to [bend left=20] (c55b);
\node [external] (c51a) at (16.8, 8.96) {};
\node [external] (c52a) at (22.4, 8.96) {};
\node [external] (c54a) at (16.1, 4.16) {};
\node [external] (c55a) at (22.9, 4.16) {};
\draw (c51) to [bend right=40](c52)[postaction={decorate, decoration={
                                        markings,
                                        mark=at position 0.5 with {\arrow[>=square,white, scale=0.7] {>}; },
                                        mark=at position 0.5 with {\arrow[>=open square, scale=0.7]  {>}; } } }];
\draw (c51) to [bend left=40](c54) [postaction={decorate, decoration={
                                        markings,
                                        mark=at position 0.5 with {\arrow[>=square,white, scale=0.7] {>}; },
                                        mark=at position 0.5 with {\arrow[>=open square, scale=0.7]  {>}; } } }];
\draw (c55) to [bend left=40](c52) [postaction={decorate, decoration={
                                        markings,
                                        mark=at position 0.5 with {\arrow[>=square,white, scale=0.7] {>}; },
                                        mark=at position 0.5 with {\arrow[>=open square, scale=0.7]  {>}; } } }];
\draw (c53) to [bend right=40](c54)[postaction={decorate, decoration={
                                        markings,
                                        mark=at position 0.5 with {\arrow[>=square,white, scale=0.7] {>}; },
                                        mark=at position 0.5 with {\arrow[>=open square, scale=0.7]  {>}; } } }];
\draw (c53) to [bend left=40](c55)[postaction={decorate, decoration={
                                        markings,
                                        mark=at position 0.5 with {\arrow[>=square,white, scale=0.7] {>}; },
                                        mark=at position 0.5 with {\arrow[>=open square, scale=0.7]  {>}; } } }];
\node [triangle, scale=0.4]  (c32) at (18, 2.53) {};
\node [external]  (c32a) at (16, 2.53) {};
\node [external]  (c32b) at (16.5, 1.13) {};
\draw [densely dashed] (c32) to [bend left=20](c32a);
\draw [densely dashed] (c32) to [bend right=20](c32b);
\node [triangle, scale=0.4]  (c33) at (21, 2.53) {};
\node [external]  (c33a) at (23, 2.53) {};
\node [external]  (c33b) at (22.5, 1.13) {};
\draw [densely dashed] (c33) to [bend right=20](c33a);
\draw [densely dashed] (c33) to [bend left=20](c33b);
\draw (c53) to [bend left=20](c32) [postaction={decorate, decoration={
                                        markings,
                                        mark=at position 0.5 with {\arrow[>=square,white, scale=0.7] {>}; },
                                        mark=at position 0.5 with {\arrow[>=open square, scale=0.7]  {>}; } } }];
\draw (c32) to [bend left=20](c33) [postaction={decorate, decoration={
                                        markings,
                                        mark=at position 0.5 with {\arrow[>=square,white, scale=0.7] {>}; },
                                        mark=at position 0.5 with {\arrow[>=open square, scale=0.7]  {>}; } } }];
                                        \draw (c53) to [bend right=20](c33) [postaction={decorate, decoration={
                                        markings,
                                        mark=at position 0.5 with {\arrow[>=square,white, scale=0.7] {>}; },
                                        mark=at position 0.5 with {\arrow[>=open square, scale=0.7]  {>}; } } }];
\end{tikzpicture}
\label{holant-csp-c}
}
\caption{\small{The reduction from \#Pl-CSP$(\mathcal{\widetilde F}_{\rm In})$ 
to Pl-Holant$(\neq_2\mid \mathcal{F})$.
 The circle vertices are labeled by $(=_d)$, where $d$ is the degree of the corresponding vertex,
 the diamond vertices are labeled by some $\widetilde f_{\rm {In}} \in \mathcal{\widetilde F}_{\rm In}$,
 the triangle vertices are labeled by the corresponding $f \in \mathcal{F}$,
 and the square vertices are labeled by $(\neq_2)$.
 }}
 \label{holant-csp}
\end{figure}
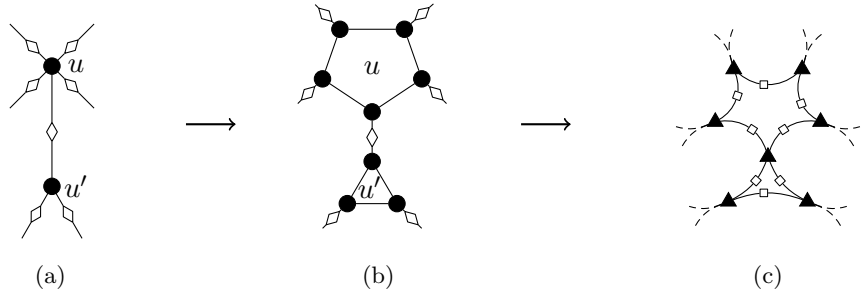

For step one,  let $G=(U, V, E)$ be a planar bipartite  graph
representing  an instance of 
\[\PlCSP(\mathcal{\widetilde F}_{\rm In})
= \plholant{\mathcal{EQ}}{\mathcal{\widetilde F}_{\rm In}},\]
 where each $u \in U$ is a variable,
and each $v \in V$ has degree two and is labeled by some $\widetilde f_{\rm {In}} \in \mathcal{\widetilde F}_{\rm In}$.
  We define a cyclic order of the edges incident to each vertex
$u \in U$, and split $u$ into $k = \deg(u)$ vertices.
 Then we connect the $k$
edges originally incident to $u$ to these $k$ new vertices so that each vertex is incident to exactly one edge. We also connect these $k$ new
 vertices in a cycle according to the cyclic order
(see Figure~\ref{holant-csp-b}).

Thus, in effect we have replaced $u$ by a cycle of length $k = \deg(u)$.
(If $k=1$ then there is a self-loop. If  $k=2$ then the cycle consists of
two parallel edges.)
    Each of  $k$ vertices has degree 3, and we label them by $(=_3)$.
This defines a signature grid for a planar holant problem,
since the construction preserves planarity.
Also clearly this does not change the value of the partition function. The resulting graph has the following properties: (1) every vertex has either degree 2 or degree 3; (2) each degree 2 vertex is connected
to degree 3 vertices;
(3) each degree 3 vertex is connected to exactly one degree 2 vertex.
\input{4cases}

Now step two. For every $v\in V$, $v$ has degree 2 and
is labeled by some  $\widetilde f_{\rm {In}} \in \mathcal{\widetilde F}_{\rm In}$.
 We contract the two edges incident to $v$
to produce a new vertex $v'$. 
The resulting graph $G'=(V', E')$ is 4-regular and planar.
We put a node on every edge of $G'$ (these are all edges of the
cycles created in step one) and label it by $(\neq_2)$
(see Figure~\ref{holant-csp-c}).
Next, we assign a copy of the corresponding $f$ to every $v'\in V'$.
The input variables $x_1, x_2, x_3, x_4$ are carefully
  assigned at each  copy of $f$ (as illustrated in Figure~\ref{four-cases}) such that there are exactly two configurations
 to each original cycle, which correspond to cyclic orientations,
  due to the $(\neq_2)$ on it and the
support set of $f$. These cyclic orientations correspond to  the $\{0, 1\}$ assignments at
the original variable $u \in U$.
Under this one-to-one correspondence, the value of $\widetilde f_{\rm In}$ is
perfectly mirrored by the value of $f$.
Therefore, we have
$\PlCSP(\mathcal{\widetilde F}_{\rm In})\leqslant_{T}\plholant{\neq_2}{\mathcal{F}}.$

\begin{figure}[!htbp]
\centering
		\includegraphics[height=2.3in]{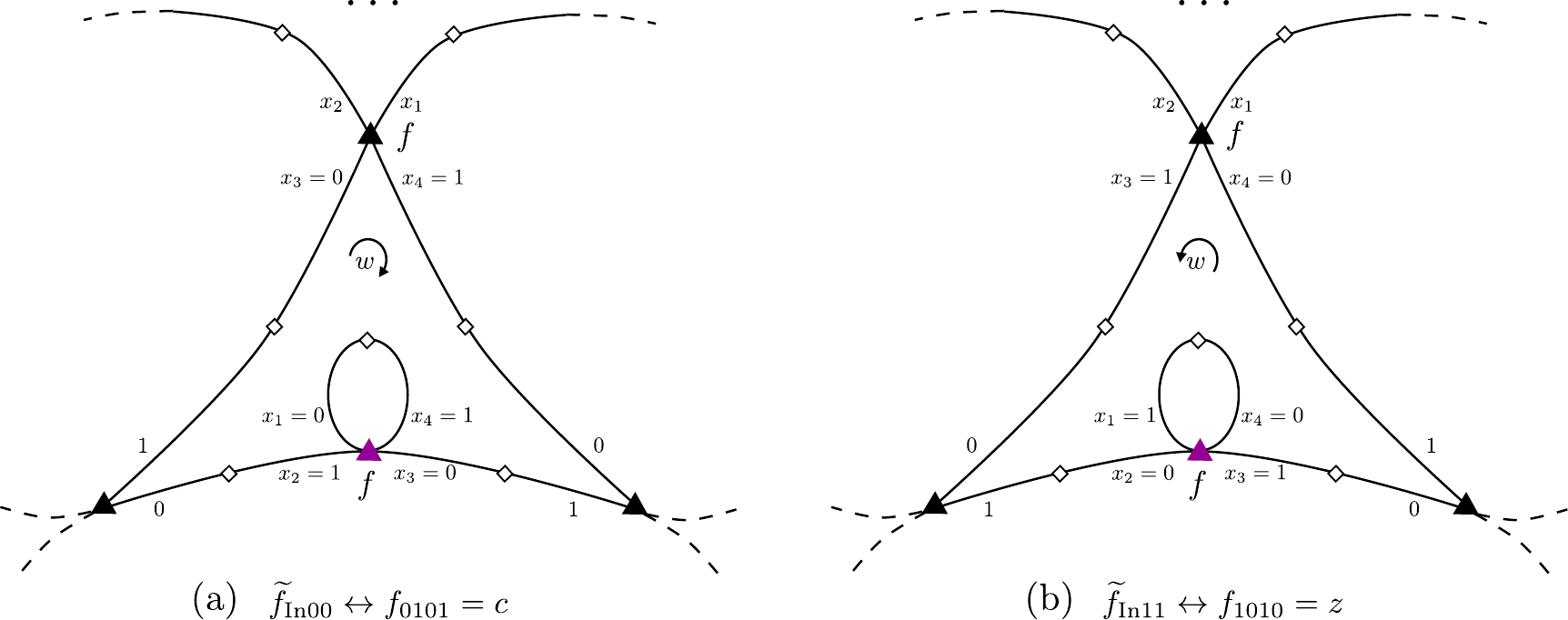}
	\caption{A self-loop on the cycle representing variable $w$ is
created for each constraint ${\widetilde f}_{\rm In}(w,w)$.
This creates a degree 4 vertex labeled by $f$,
with four input variables $(x_1,x_2,x_3,x_4)$ as described.
Note that the self-loop is created locally on the cycle such that it 
does not affect anything having to do with other cycles.
 Base on the support of $f$, the values $x_1 \not = x_2$ and $x_3 \not =
 x_4$. By the  $(\neq)$ on the loop, we also have $x_1 \not = x_4$.
Hence $(x_1, x_2)=(x_3, x_4)=(0, 1)$ or $(1, 0)$.
It is clear that the former corrresponds to $w=0$ (clockwise orientation), and
the latter corrresponds to $w=1$ (counterclockwise orientation). 
This is consistent with the association in Figure \ref{four-cases}.}
	\label{figure:self-loop-local-unary}
	\end{figure}

There is also the possibility that the binary constraint
${\widetilde f}_{\rm In}$ is applied to a single variable, say $w$,
resulting in a unary constraint that takes value ${\widetilde f}_{\rm In}(0,0)
=c$ if $w=0$ and ${\widetilde f}_{\rm In}(1,1)
=z$ if $w=1$. To reflect that, we simply introduce a self-loop
on the cycle representing the variable $w$ for every such occurrence,
as illustrated in Figure~\ref{figure:self-loop-local-unary}.
It is clear that the values $c$ and $z$ are perfectly mirrored by 
the values  that the local copy $f$ takes under the two orientations
for the cycle corresponding to  $w=0$ and $1$.
\qed


\begin{theorem} \label{outer}
Let $f$ be a 4-ary signature of the form {\rm (\ref{outermatrix})}.
Then $\operatorname{Pl-Holant}(\neq_2\mid f)$
is {\rm\#}P-hard unless
$f \in \mathscr{P}$, $f \in \mathscr{A}$, or $f \in \mathscr{\widehat{M}}$, 
  in which cases the problem is tractable. 
\end{theorem}

{\bf Proof.} 
Tractability follows from Theorems \ref{aptractable} and \ref{mtractable}.
For any $f$ of the form {\rm (\ref{outermatrix})}, 
note that the support of $f$ is contained in $(x_1\neq x_2) \wedge 
(x_3\neq x_4)$.
We have 
 $$f(x_1, x_2, x_3, x_4)=\widetilde f_{\rm In}(x_1, x_3)\cdot \chi_{x_1\neq x_2}\cdot \chi_{x_3\neq x_4},$$ 
where $\chi$ is the 0-1 indicator function.
Thus, ${\widetilde f_{\rm In}} \in \mathscr{P}$ or $\mathscr{A}$ is equivalent to  ${f} \in  \mathscr{P}$ or $\mathscr{A}$.
In addition, by Lemmas \ref{binary-m} and \ref{ary4-m}, ${\widetilde f_{\rm In}} \in \mathscr{\widehat M}$ is equivalent to  $f \in \mathscr{\widehat M}$.
Therefore, if $f \notin$  $\mathscr{P}, \mathscr{A}$ or $\mathscr{\widehat{M}}$, then ${\widetilde f_{\rm In}}$ $\notin$ $\mathscr{P}, \mathscr{A}$ or $\mathscr{\widehat{M}}$. 
By Theorem \ref{cspdic}, $\PlCSP({\widetilde f_{\rm In}})$ is \#P-hard, and then by Lemma \ref{outerlemma}, $\plholant{\neq_2}{f}$ is \#P-hard. \qed 

\begin{remark}
  One may observe that if $f \in \mathscr{M}$, then \plholant{\neq_2}{f} is also tractable as $f$ and $(=_2)$ are both realized by matchgates. However, Theorem \ref{outer} already accounted for this case because for  signature $f$ of the form (\ref{outermatrix}), $f \in \mathscr{M}$ implies $f \in \mathscr{P}$.
\end{remark}

Now, we consider the case that the inner pair is a zero pair and no outer pair is a zero pair.
Note that a signature in the form (\ref{innermatrix}) has support
contained in $(x_1 \neq x_3) \wedge (x_2 \neq x_4)$.
\begin{definition}\label{defn:Gf-Hf}

Given a 4-ary signature $f$ with the signature matrix 
\begin{equation}\label{innermatrix}
M(f)=\left[\begin{matrix}
0 & 0& 0& a\\
0 & b& 0& 0\\
0 & 0& y& 0\\
x & 0& 0& 0\\
\end{matrix}\right], 
\end{equation} 
where $(a, x) \neq (0, 0)$ and $(b, y) \neq (0, 0)$,
let $\mathcal{G}_f$ denote the set of all binary signatures $g_{{}_f}$ of the form $$M(g_{{}_f})=\left[ \begin{matrix}
a^{k_1+\ell_1}y^{k_2+\ell_2}x^{k_3+\ell_3}b^{k_4+\ell_4} & a^{k_2+\ell_4}y^{k_3+\ell_1}x^{k_4+\ell_2}b^{k_1+\ell_3}\\
a^{k_4+\ell_2}y^{k_1+\ell_3}x^{k_2+\ell_4}b^{k_3+\ell_1} & a^{k_3+\ell_3}y^{k_4+\ell_4}x^{k_1+\ell_1}b^{k_2+\ell_2}\\ \end{matrix} \right],$$
satisfying $k = \ell$, 
where $k =\sum_{i=1}^4 k_i, \ell = \sum_{i=1}^4 \ell_i$
and $k_1, k_2, k_3, k_4, \ell_1, \ell_2, \ell_3, \ell_4 \in \mathbb{N}$.
Let $\mathcal{H}_f$ denote the set of all unary signatures $h_{{}_f}$ of the form
$$M(h_{{}_f})=\left[ \begin{matrix}
a^{m_1}y^{m_2}x^{m_3}b^{m_4} & a^{m_3}y^{m_4}x^{m_1}b^{m_2} \end{matrix} \right],$$ where $m_1, m_2, m_3, m_4 \in \mathbb{N}$. 

Let $k=k_1=\ell_1=\ell=1$, we get a specific signature $g_{1_f} \in
\mathcal{G}_f$, with $M(g_{1_f})=\left[\begin{smallmatrix}
a^2 & by\\
by & x^2\\
\end{smallmatrix}\right]$. Let $k=k_1=\ell_3=\ell=1$, we get 
another specific signature $g_{2_f} \in \mathcal{G}_f$, with
 $M(g_{2_f})=\left[\begin{smallmatrix}
ax & b^2\\
y^2 & ax\\
\end{smallmatrix}\right]$.
\end{definition}
\begin{remark}
  For any $i,j \in \{1, 2, 3, 4\}$, let $k=k_i=\ell_j=\ell=1$, we can  get $16$  signatures in $\mathcal{G}_f$ that have similar signature matrices to $M(g_{1_f})$ and $M(g_{2_f})$. For example,
Choosing $k=k_3=\ell_1=\ell=1$,
we get $g'_{2_f}(x_1, x_2)$
with the signature matrix 
$M(g'_{2_f})=\left[\begin{smallmatrix}
ax & y^2\\
b^2 & ax\\
\end{smallmatrix}\right]$.
Indeed $g'_{2_f}(x_1, x_2) = g_{2_f}(x_2, x_1)$.
 In fact, $\mathcal{G}_f$ is the closure by the 
Hadamard product (entry-wise product)  of these 16 basic signature matrices.
\end{remark}
\begin{lemma} \label{csp}
Let $f$ be a signature of the form {\rm (\ref{innermatrix})}. Then,
\begin{equation}\label{<}
\operatorname{Pl-Holant}(\neq_2\mid f) \leqslant_{T} \operatorname{\#CSP}(\mathcal{G}_f\cup \mathcal{H}_f),
\end{equation} 
If $a^2= x^2 \neq 0$, $b^2= y^2 \neq 0$ and $\left(\frac{b}{a}\right)^8\neq 1$, then 
\begin{equation}\label{>}
\operatorname{\#CSP}(g_{1_f}, g_{2_f}) \leqslant_{T} \operatorname{Pl-Holant}(\neq_2\mid f).
\end{equation}
\end{lemma} 

{\bf Proof. } We divide the proof into two parts: We show the reduction~(\ref{<}) in Part \Rmnum{1}, and the reduction~(\ref{>}) in Part \Rmnum{2}.

{\bf Part \Rmnum{1}:} 
Suppose $\Omega=(G, \pi)$ is a given instance
of $\operatorname{Pl-Holant}(\neq_2\mid f)$,
where $G = (U, V, E)$ is a plane bipartite graph. Every vertex $v \in V$
has degree 4, 
and we list its incident four edges in counterclockwise order.
Two edges both incident to a vertex $v \in V$
 are called adjacent if they are adjacent in this cyclic order,
and non-adjacent otherwise.
Two edges in $G$ are called \emph{$2$-ary edge twins}
 if they are both incident to a vertex $u \in U$ (of degree $2$), 
and \emph{$4$-ary edge twins} if they are non-adjacent
but both incident to a vertex $v \in V$ (of degree $4$).
Both $2$-ary edge twins and $4$-ary edge twins are called \emph{edge twins}. 

Each edge has a unique $2$-ary edge twin at its endpoint in $U$ of degree $2$
 and a unique $4$-ary edge twin at its endpoint in $V$ of degree $4$.  
The reflexive and transitive closure of the symmetric binary relation 
\emph{edge twin} forms a partition of $E$ as an edge disjoint union
of \emph{circuits}:
$C_1, C_2, \ldots, C_k$.
Note that $C_i$ may include repeated vertices called self-intersection vertices, but no repeated edges. 
We arbitrarily pick an edge $e_i$ of $C_i$ to be the \emph{leader edge}
 of $C_i$. 
Given the leader edge $e_i = (u, v)$ of $C_i$, with $u \in U$ and $v \in V$,
the direction from $u$ to $v$
defines an orientation of the circuit $C_i$.
\footnote{This default orientation should not be confused with the
orientation in the proof of Lemma~\ref{outerlemma}.}
For any edge twins $\{e, e'\}$, this orientation defines one edge, say $e'$, 
as the successor of the other if $e'$ comes right after $e$ in the
orientation.
 When we list the assignments of edges in a circuit,
we list successive values of successors, starting with  the leader edge.

For any nonzero term in the sum
$$\operatorname{Pl-Holant}_\Omega=\sum_{\sigma:E\rightarrow \{0, 1\}}\prod_{w\in U \cup V}f_w(\sigma \mid_{E_{(w)}}),$$
the assignment of all edges $\sigma:E\rightarrow \{0, 1\}$ can be 
uniquely extended from its restriction on
 leader edges $\sigma':\{e_1, e_2, \cdots e_k\}\rightarrow \{0, 1\}$.
This is because the support of $f$ is 
contained in $(x_1 \neq x_3) \wedge (x_2 \neq x_4)$.
Thus, at each vertex $v \in V$, 
$f_v(\sigma \mid_{E_{(v)}})\neq 0$ only if each pair of edge twins in $E_{(v)}$ is assigned value $(0, 1)$ or $(1, 0)$. 
The same is true for any vertex $u \in U$ of degree 2,
which is labeled $(\neq_2)$.
Thus, if the leader edge  $e_i$  in  $C_i$ takes value  $0$
or 1 respectively,  
then all edges on $C_i$ must
take values $(0, 1, 0, 1, \cdots, 0, 1)$ or
 $(1, 0, 1, 0, \cdots, 1, 0)$ respectively
on successive successor edges,
  starting with $e_i$.
In particular,  all pairs of $4$-ary edge twins in $C_i$ take assignment $(0, 1)$ when $e_{i}=0$ and  $(1, 0)$ when $e_{i}=1$
(listing the value of the successor second).
Then, we have
$$\operatorname{Pl-Holant}_\Omega=\sum_{\sigma': \{e_1, \cdots, e_k\}\rightarrow \{0, 1\}}\prod_{v\in V}f_v(\widehat{\sigma'} \mid_{E_{(v)}}),$$
where $\widehat{\sigma'}$ denotes the unique extension of $\sigma'$.

For all $1 \le i < j \le k$,
let $V_{i, j}=C_i\cap C_j$ denote
 the set of all intersection vertices between $C_i$ and $C_j$. 
Denote by $\sigma'_{(e_i, e_j)}$ an assignment  
 $\{e_i, e_j\} \rightarrow \{0, 1\}$.
Define  a binary function $g_{i, j}$ on $e_i$ and $e_j$ as follows:
For any $b, b' \in \{0, 1\}$, let
$$g_{i,j}(b, b') 
= \prod_{v\in V_{i, j}}f_v(\widehat{\sigma'_{(e_i, e_j)}} \mid_{E_{(v)}}),$$
where $\widehat{\sigma'_{(e_i, e_j)}}$ is the unique extension
of $\sigma'_{(e_i, e_j)}$
on the union of edge sets of $C_i$ and $C_j$ as described above,
and  $\sigma'_{(e_i, e_j)}$ is the unique assignment on
$\{e_i, e_j\}$ such that $e_i \mapsto b$ and $e_j \mapsto b'$.
%
Since all edges incident to vertices in $V_{i, j}$ are either  in $C_i$ or  $C_j$, the assignment values of these edges are determined by
 $\sigma'_{(e_i, e_j)}$.
Hence, $g_{i, j}$ is well-defined.  

We show that $g_{i, j}\in \mathcal{G}_f$ by induction on the number $n$ of self-intersection vertices in $C_i$. 
Note that in this proof, $i$ and $j$ (with $i<j$)
 are not treated symmetrically.

For each vertex  $v \in V_{i,j}$, consider the two pairs of edge twins
 incident to it.
We label the edge twins in $C_i$ by the variables $(x_1, x_3)$
such that $x_3$ is the successor of $x_1$ in the orientation of $C_i$.
 Hence, for all $v  \in V_{i,j}$,
 these variables $(x_1, x_3)$ take the same assignment $(0, 1)$ when $e_i=0$ and $(1, 0)$ when $e_i=1$.
Then,  label  the edge twins in  $C_j$ at $v$
by $(x_2, x_4)$ so that the 4 edges at $v$ are
ordered  $(x_1, x_2, x_3, x_4)$ in counterclockwise order.
This choice of $(x_2, x_4)$ is unique given the labeling $(x_1, x_3)$.

\begin{figure}[!htbp]
\centering
                \includegraphics[height=2in]{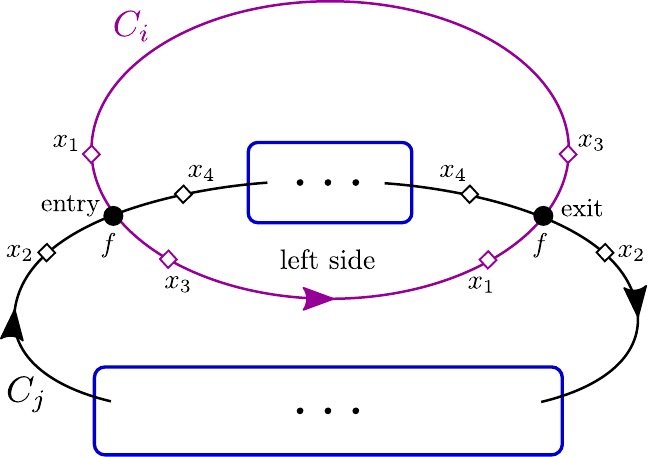}
        \caption{Intersection vertices between $C_i$ and $C_j$}
        \label{in-out}
        \end{figure}

As we traverse  $C_i$ according to the orientation of $C_i$,
locally there is a  notion of the \emph{left side} of $C_i$.
At any vertex $v \in C_i \cap C_j$,
if we take the traversal of $C_j$
 according to the orientation of $C_j$,
 it either 
\emph{comes into} or \emph{goes out} of the left side of $C_i$.
We call $v \in C_i \cap C_j$
of the former kind ``entry-vertices'', and the  latter
kind ``exit-vertices'' (see Figure \ref{in-out}).

 At any entry-vertex  $v \in V_{i,j}$,
 the variable  $x_4$ is the successor of $x_2$, 
while at any exit-vertex   $x_2$ is  the successor of $x_4$.
Therefore, at entry-vertices, variables $(x_2, x_4)$ take assignment $(0, 1)$ 
when $e_j=0$ and $(1, 0)$ when $e_j=1$, while at exit-vertices they take assignment $(1, 0)$ and $(0, 1)$ respectively instead. 


\begin{table}[!hbtp]
\renewcommand{\arraystretch}{1.2}
\centering
\begin{tabular}{|c|c|c|c|c|c|c|c|c|c|c|}
\hline
\multirow{2}{*}{$(e_i, e_j)$} & \multicolumn{5}{|c|}{entry-vertices } &\multicolumn{5}{|c|}{exit-vertices } \\
\cline{2-11}
& $(x_1, x_2, x_3, x_4)$ & $f$ & $f^{\frac{\pi}{2}}$ & $f^{\pi}$ & $f^{\frac{3\pi}{2}}$ & $(x_1, x_2, x_3, x_4)$ &  $f$ & $f^{\frac{\pi}{2}}$ & $f^{\pi}$ & $f^{\frac{3\pi}{2}}$ \\
\hline
$(0, 0)$ & $(0, 0, 1, 1)$ & $a$ & $y$ & $x$ & $b$ & $(0, 1, 1, 0)$ & $b$ & $a$ & $y$ & $x$ \\
\hline

$(0, 1)$ & $(0, 1, 1, 0)$ & $b$ & $a$ & $y$ & $x$ & $(0, 0, 1, 1)$ & $a$ & $y$ & $x$ & $b$ \\
\hline
$(1, 1)$ & $(1, 1, 0, 0)$ & $x$ & $b$ & $a$ & $y$ & $(1, 0, 0, 1)$ & $y$ & $x$ & $b$ & $a$ \\
\hline
$(1, 0)$ & $(1, 0, 0, 1)$ & $y$ & $x$ & $b$ & $a$ & $(1, 1, 0, 0)$ & $x$ & $b$ & $a$ & $y$\\

\hline
\end{tabular}
\caption{The values of $f$ and its rotated copies
 at intersection vertices}\label{gvalue1}
\end{table}

Table~\ref{gvalue1} summarizes the values of $f$ and its rotated copies
at intersection vertices $V_{i,j}$.
According to the 4 different assignments of $(e_i, e_j)$ 
as listed in column 1 of the table, column 2  and column 7 (indexed by $(x_1, x_2, x_3, x_4)$) list the assignments of $(x_1, x_2, x_3, x_4)$ at entry-vertices and exit-vertices separately. 
With respect to this local labeling of
$(x_1, x_2, x_3, x_4)$, the signature $f$ has four rotated forms:  
$$M
(f)=\left[\begin{smallmatrix}
0 & 0& 0& a\\
0 & b& 0& 0\\
0 & 0& y& 0\\
x & 0& 0& 0\\
\end{smallmatrix}\right], 
M
(f^{\frac{\pi}{2}})=\left[\begin{smallmatrix}
0 & 0& 0& y\\
0 & a& 0& 0\\
0 & 0& x& 0\\
b & 0& 0& 0\\
\end{smallmatrix}\right],
M
(f^{\pi})=\left[\begin{smallmatrix}
0 & 0& 0& x\\
0 & y& 0& 0\\
0 & 0& b& 0\\
a & 0& 0& 0\\
\end{smallmatrix}\right]
\text{ and }M
(f^{\frac{3\pi}{2}})=\left[\begin{smallmatrix}
0 & 0& 0& b\\
0 & x& 0& 0\\
0 & 0& a& 0\\
y & 0& 0& 0\\
\end{smallmatrix}\right].$$
columns 3, 4, 5, 6 and columns 8, 9, 10, 11 list the corresponding values of
the  signature $f$ in four forms $f$, $f^{\frac{\pi}{2}}$, $f^{\pi}$ and $f^{\frac{3\pi}{2}}$ respectively.

Suppose there are $k_1, k_2, k_3$ and $k_4$ many entry-vertices
 assigned $f$, $f^{\frac{\pi}{2}}$, $f^{\pi}$, and $f^{\frac{3\pi}{2}}$, 
respectively,
and there are $\ell_1, \ell_2, \ell_3$ and  $\ell_4$ many exit-vertices
 assigned  $f^{\frac{\pi}{2}}$, $f^{\pi}$, $f^{\frac{3\pi}{2}}$ and
$f$, respectively.
 Then, according to the assignments of $(e_i, e_j)$, the values of $g_{i, j}$ are listed in Table \ref{gtable}, and its signature matrix is given below:

$$M(g_{i, j})=\left[ \begin{matrix}
a^{k_1+\ell_1}y^{k_2+\ell_2}x^{k_3+\ell_3}b^{k_4+\ell_4} & a^{k_2+\ell_4}y^{k_3+\ell_1}x^{k_4+\ell_2}b^{k_1+\ell_3}\\
a^{k_4+\ell_2}y^{k_1+\ell_3}x^{k_2+\ell_4}b^{k_3+\ell_1} & a^{k_3+\ell_3}y^{k_4+\ell_4}x^{k_1+\ell_1}b^{k_2+\ell_2}\\ \end{matrix} \right].$$

\begin{table}[!hbtp]
\renewcommand{\arraystretch}{1.2}
\addtolength{\tabcolsep}{8pt}
\centering
\begin{tabular}{|c|c|}
\hline
$(e_i, e_j)$ & $g_{i, j}(e_i, e_j)=f^{k_1}(f^{\frac{\pi}{2}} )^{k_2}(f^{\pi} )^{k_3}(f^{\frac{3\pi}{2}} )^{k_4} (f^{\frac{\pi}{2}} )^{\ell_1}(f^{\pi})^{\ell_2}(f^{\frac{3\pi}{2}})^{\ell_3}f^{\ell_4}$ \\
\hline
$(0, 0)$ & $a^{k_1}y^{k_2}x^{k_3}b^{k_4}a^{\ell_1}y^{\ell_2}x^{\ell_3}b^{\ell_4}$\\
\hline
$(0, 1)$ & $b^{k_1}a^{k_2}y^{k_3}x^{k_4}y^{\ell_1}x^{\ell_2}b^{\ell_3}a^{\ell_4}$\\
\hline
$(1, 1)$ & $x^{k_1}b^{k_2}a^{k_3}y^{k_4}x^{\ell_1}b^{\ell_2}a^{\ell_3}y^{\ell_4}$\\
\hline
$(1, 0)$ & $y^{k_1}x^{k_2}b^{k_3}a^{k_4}b^{\ell_1}a^{\ell_2}y^{\ell_3}x^{\ell_4}$\\
\hline
\end{tabular}
\caption{The values of $g_{i, j}$}\label{gtable}
\end{table}

Our proof that $g_{i, j}\in \mathcal{G}_f$ is based on the
assertion that the number of ``entry-vertices'' and  ``exit-vertices''
are equal, namely $\sum_{i=1}^{4}k_i= \sum_{i=1}^{4}\ell_i$.

\begin{itemize}

\item First, consider the base case $n=0$. That is, $C_i$ is a simple cycle without self-intersection. By the Jordan Curve Theorem, $C_i$ divides the plane 
into two regions, an interior region and an exterior  region.
In this case, as we traverse  $C_i$ according to the orientation of $C_i$, 
 the left side of the traversal is always the same region; we call
it  ${\sf L}_i$
(which could be either the interior or the  exterior  region,
depending on the choice of the leader edge $e_i$).
If we traverse  $C_j$  according to the orientation of $C_j$,
we enter and exit the region ${\sf L}_i$ an equal number of times.
 Therefore there is an equal 
 number of ``entry-vertices'' and  ``exit-vertices''.
%
%
Hence 
 $\sum_{i=1}^{4}k_i= \sum_{i=1}^{4}\ell_i$.
It follows that $g_{i, j} \in \mathcal{G}_f$ by the definition 
of $\mathcal{G}_f$.

\item Inductively, suppose $g_{i, j}\in \mathcal{G}_f$ holds for any circuit $C_i$ with at most $n$ self-intersections. 
Let $C_i$  have $n+1$ self-intersections.
We decompose
$C_i$ into two edge-disjoint circuits,
 each of which has at most $n$ self-intersections (See Figure \ref{decompose}). 
 
 \begin{figure}[!htbp]
\centering
		\includegraphics[height=1.0in]{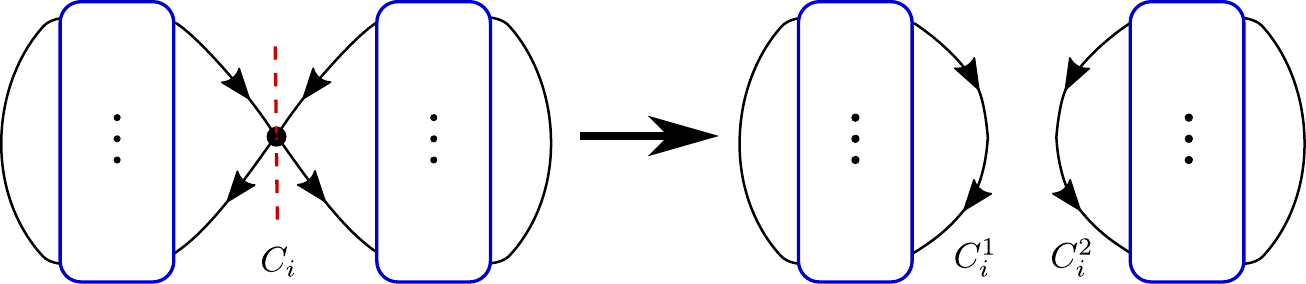}
	\caption{Decompose $C_i$ into $C^1_i$ and $C^2_i$.}
	\label{decompose}
	\end{figure}
	
Take any self-intersection vertex $v^*$ of $C_i$.
There are two pairs of 4-ary edge twins
$\{e, e'\}$ and $\{\overline{e}, \overline{e}'\}$,
 where  $e'$ is the successor of $e$
and $\overline{e}'$ is the successor of $\overline{e}$.
Note that $e$ and $\overline{e}$ are oriented  toward $v^*$,
and $e'$ and $\overline{e}'$ are oriented away from $v^*$.
By the definition of edge twins,
$\{e, \overline{e}\}$ are adjacent,
and $\{e', \overline{e}'\}$ are adjacent.
 We can break $C_i$ into two 
oriented circuits $C^1_i$ and $C^2_i$,
by splitting $v^*$ into two vertices, and let $e'$ follow $\overline{e}$
and let $\overline{e}'$ follow $e$.
Let the mapping $\gamma: [0,1] \rightarrow \mathbb{R}^2$,
such that $\gamma(0)=\gamma(1/2) = \gamma(1) = v^*$,
represent the traversal of $C_i$.
Then we can define two mappings $\gamma^1, \gamma^2: 
[0,1] \rightarrow \mathbb{R}^2$,
such that $\gamma^1(t) = \gamma(t/2)$ and
$\gamma^2(t) = \gamma((t+1)/2)$.
Then $\{\gamma^1, \gamma^2\}$ represent $\{C^1_i, C^2_i\}$
respectively.
It follows that $C_i$ is the edge disjoint union of
 $C^1_i$ and $C^2_i$ and they both inherit
the same orientation from that of $C_i$.
Any vertex in $V_{i, j}$ is distinct from
a self intersection point of $C_i$
and thus $V_{i, j}$ is  a disjoint union
$V^1_{i,j}\cup V^2_{i,j}$, where 
$V^1_{i,j} =C^1_i \cap C_j$ and $V^2_{i,j}= C^2_i \cap C_j$.

Since  $C^1_i$  inherits the orientation from $C_i$,
the orientation on   $C^1_i$ is consistent
with the orientation starting by choosing a leader edge on  $C^1_i$.
The same is true for the orientation on   $C^2_i$.
Thus, by induction, on each $C^1_i \cap C_j$ and $C^2_i \cap C_j$
there are an equal number of
  ``entry-vertices'' and  ``exit-vertices''.
Hence
 $\sum_{i=1}^{4}k_i= \sum_{i=1}^{4}\ell_i$,
and so  $g_{i, j} \in \mathcal{G}_f$, completing the
induction.

\end{itemize}

Let $V_i$ 
be the set of all self-intersections of $C_i$. 
Let $\sigma'_{(e_i)}$ denote the restriction of $\sigma'$ on $\{e_i\}$.
Define a unary function $h_{i}$ on $e_i$ as follows:
For any $b \in \{0, 1\}$, let
$$h_{i}(b) = \prod_{v\in V_i}f_v(\widehat{\sigma'_{(e_i)}})\mid_{E_{(v)}}),$$
where $\widehat{\sigma'_{(e_i)}}$ is the unique extension
of $\sigma'_{(e_i)}$ on the  edge set of $C_i$,
and  $\sigma'_{(e_i)}$ is the unique assignment on
$\{e_i\}$ such that $e_i \mapsto b$. 
The assignment of those edges incident to vertices in $V_{i}$ can be 
uniquely extended from the assignment $\sigma'_{(e_i)}$.
Hence, $h_i$ is well-defined. We show that $h_i \in \mathcal{H}_f$.

For each vertex in $V_i$, since it is a self-intersection vertex, the two pairs of edge twins incident to it are both in $C_i$. 
We still first
 label each pair of edge twins by a pair of variables  $(x_1, x_3)$ obeying the orientation of $C_i$. That is, $x_3$ is always the successor of $x_1$.
Now by the definition of 4-ary edge twins,
the two edges labeled $x_1$ are adjacent. Hence
at each vertex in $V_i$, starting from one $x_1$,
the four incident edges  are labeled by $(x_1, x_1, x_3, x_3)$ 
in counterclockwise order.
We pick the pair of variables $(x_1, x_3)$ that appear in
the second and fourth positions in this listing
and change them to $(x_2, x_4)$,
 so that the four edges are now labeled by $(x_1, x_2, x_3, x_4)$
 in counterclockwise order. Clearly, $(x_2, x_4)$ and $(x_1, x_3)$ take the same assignment. That is, at each vertex in $V_i$, the assignment of $(x_1, x_2, x_3, x_4)$  is $(0, 0, 1, 1)$ when $e_i=0$, and $(1, 1, 0, 0)$ when $e_i=1$. Under this labeling, the
signature $f$ still has four rotated forms. The values of 
these four forms are listed in Table \ref{self}. 


\begin{table}[!htp]
\renewcommand{\arraystretch}{1.2}
\addtolength{\tabcolsep}{8pt}
\centering
\begin{tabular}{|c|c|c|c|c|c|}
\hline
$e_i$ & $(x_1, x_2, x_3, x_4)$ &  $f$ & $f^{\frac{\pi}{2}}$ & $f^{\pi}$ & $f^{\frac{3\pi}{2}}$ \\
\hline
$0$ & $(0, 0, 1, 1)$ & $a$ &$y$ & $x$ &$b$\\
\hline
$1$ & $(1, 1, 0, 0)$ & $x$ &$b$ & $a$ &$y$\\
\hline
\end{tabular}
\caption{The values of $f$ and its rotated forms
at self-intersection vertices}\label{self}
\end{table}

Suppose on $V_i$
 there are $m_1, m_2, m_3$ and $m_4$ many vertices assigned $f$,
 $f^{\frac{\pi}{2}}$,  $f^{\pi}$ and $f^{\frac{3\pi}{2}}$ respectively.
 Then, we have
$$M(h_i)=[\begin{matrix}
a^{m_1}y^{m_2}x^{m_3}b^{m_4} &  a^{m_3}y^{m_4}x^{m_1}b^{m_2}
\end{matrix}].$$
It follows that $h_i \in \mathcal{H}_f$.

For any vertex $v\in V$, it is either in some $V_{i, j}$ or some $V_i$.
Thus, 
\begin{equation*}
\begin{aligned}
\operatorname{Pl-Holant}_{\Omega}  &=  \sum_{\sigma': \{e_1, \cdots, e_k\}\rightarrow \{0, 1\}}\Bigg(\prod_{\substack{ v\in V_{i, j} \\1\leqslant i < j \leqslant k}}f_v(\sigma'|E_{(v)})\Bigg) \Bigg(\prod_{\substack{v\in V_i \\1\leqslant i \leqslant k}}f_v(\sigma'|E_{(v)})\Bigg) \\
&= \sum_{\sigma': \{e_1, \cdots, e_k\}\rightarrow \{0, 1\}}\Bigg(\prod_{1\leqslant i < j \leqslant k}g_{i, j}(\sigma'(e_i), \sigma'(e_j))\Bigg) \Bigg(\prod_{1\leqslant i \leqslant k}h_i(\sigma'(e_i))\Bigg), 
\end{aligned}
\end{equation*}
where $g_{i,j} \in \mathcal{G}_f \text{ and } h_{i}\in \mathcal{H}_f$.
Therefore,  Pl-Holant$(\neq_{2}\mid f)\leqslant_{T}$\#CSP$(\mathcal{G}_f\cup \mathcal{H}_f).$

\vspace{.1in}

Here, we give some examples for the reduction (\ref{<}).

\vspace{1ex}

\begin{figure}[!htbp]
\centering
		\includegraphics[height=1.35in]{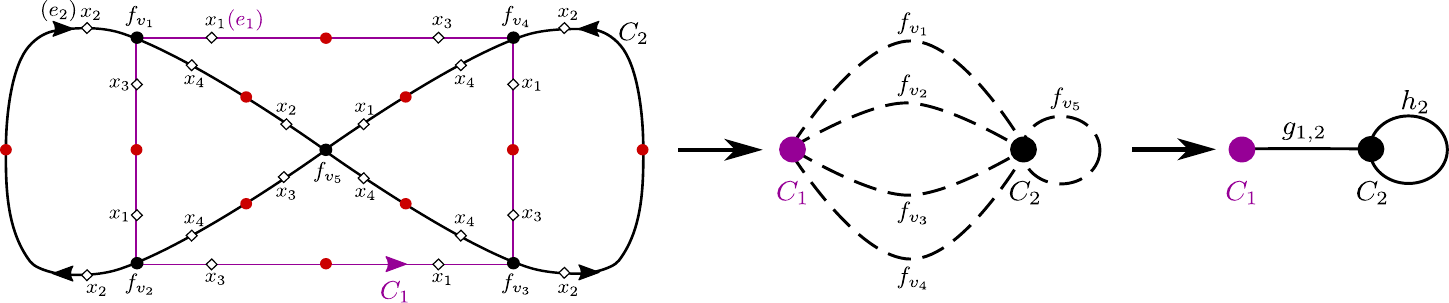}
	\caption{A simple example for the reduction (\ref{<})}
	\label{example}
	\end{figure}

\noindent{\bf Example 1.}
	The signature grid $\Omega=(G, \pi)$ 
for \plholant{\neq_2}{f}  in Figure~\ref{example}
has two circuits $C_1$ (the {\sc Square}) and
 $C_2$ (the {\sc Horizontal Eight}) in $G$.  
	We have chosen  (arbitrarily)  a leader edge $e_i$ for
each circuit $C_i$.
In Figure~\ref{example} they are near the top left corner.
 Given the leader, the direction from its endpoint of degree $2$ to the endpoint of degree $4$ gives a default orientation of the circuit.
Given a nonzero term in the sum Pl-Holant$_{\Omega}$, as a consequence of the support of $f$, the assignment of edges in each circuit is uniquely determined by the assignment of its leader. That is, any assignment of the leaders  $\sigma': \{e_1, e_2\} \rightarrow \{0, 1\}$ can be uniquely extended to an assignment of all edges $\sigma: E \rightarrow \{0, 1\} $ such that on each circuit 
the values of $0, 1$  alternate.

Consider the signatures  $f_{v_1}, f_{v_2}, f_{v_3} \text{ and } f_{v_4}$
on the intersection vertices 
between $C_1$ and $C_2$. 
Assume $C_1$ does not have self-intersection (as is {\sc The Square}); 
otherwise, we will decompose $C_1$ further and reason inductively.  
 Without self-intersection, $C_1$ has an interior and exterior region 
by the Jordan Curve Theorem.
For the chosen orientation of $C_1$, its left side happens to
be the interior region.
With respect to $C_1$, the circuit $C_2$ enters and exits the interior of $C_1$ alternately. Thus, we can divide the intersection vertices into an equal number of  ``entry-vertices'' and ``exit-vertices''.
In this example, $f_{v_1}$ and $f_{v_4}$ are on ``entry-vertices'', while $f_{v_2}$ and $f_{v_3}$  are on ``exit-vertices''.
By analyzing the values of each $f$ when $e_1$ and $e_2$ take assignment $0$ or $1$, we can view each $f$ as a binary constraint on $(C_1, C_2)$. 
Depending on the $4$ different rotation forms of $f$ and whether $f$ is on ``entry-vertices'' or ``exit-vertices'', the resulting binary constraint has $8$ different forms (See Table \ref{gvalue1}).  By multiplying these constraints, we get the binary constraint $g_{1, 2}$. This can be viewed as a binary edge function on the circuits $C_1$ and $C_2$. The property of $g_{1, 2}$ 
crucially depends on there are an equal number of ``entry-vertices''
and ``exit-vertices''.
For any $b, b' \in \{0, 1\}$,
$$g_{1,2}(b, b') 
=\prod_{1 \leqslant i \leqslant 4}f_{v_{i}}(\widehat{\sigma'_{(e_1, e_2)}}
 \mid_{E_{(v_i)}}),$$
where $\widehat{\sigma'_{(e_1, e_2)}}$ uniquely extends to $C_1$ and $C_2$
the assignment 
$\sigma'_{(e_1, e_2)}(e_1) = b$ and $\sigma'_{(e_1, e_2)}(e_2) = b'$.

If the placement of $f_{v_1}$ were to be rotated clockwise $\frac{\pi}{2}$,
then $f_{v_1}$ will be changed to $f^{\frac{\pi}{2}}_{v_1}$ in the above formula, where $M_{x_1x_2, x_4x_3}(f^{\frac{\pi}{2}}_{v_1})=M_{x_2x_3, x_1x_4}(f_{v_1})$.

For the self-intersection vertex $f_{v_5}$, the notions of ``entry-vertex'' and ``exit-vertex'' do not apply.  $f_{v_5}$ gives rise to a unary constraint
 $h_2$ on $e_2$.  Depending on the $4$ different rotation forms of $f$, $h_2$ has $4$ different forms  (see Table 3). 
For any $b  \in \{0, 1\}$,
$$h_{2}(b)=f_{v_5}(\widehat{\sigma'_{(e_2)}} \mid_{E_{(v_5)}}),$$
where $\widehat{\sigma'_{(e_2)}}$ uniquely extends to $C_2$
the assignment 
$\sigma'_{(e_2)}(e_2) = b$.

 Therefore, we have
\begin{equation*}
\begin{aligned}
\operatorname{Pl-Holant}_{\Omega}  
&=\sum_{\sigma:E\rightarrow \{0, 1\}}\prod_{v\in V(G)}f_v(\sigma \mid_{E_{(v)}})\\
&=  \sum_{\sigma': \{e_1, e_2\}\rightarrow \{0, 1\}}\Bigg(\prod_{1 \leqslant i \leqslant 4 }f_{v_i}(\sigma'|E_{(v_i)})\Bigg)f_{v_5}(\sigma'|E_{(v_5)}) \\
&= \sum_{\sigma': \{e_1, e_2\}\rightarrow \{0, 1\}}g_{1, 2}(\sigma'(e_1),
\sigma'(e_2))h_2(\sigma'(e_2)).
\end{aligned}
\end{equation*}

\begin{figure}[h]
\centering
		\includegraphics[height=1.4in]{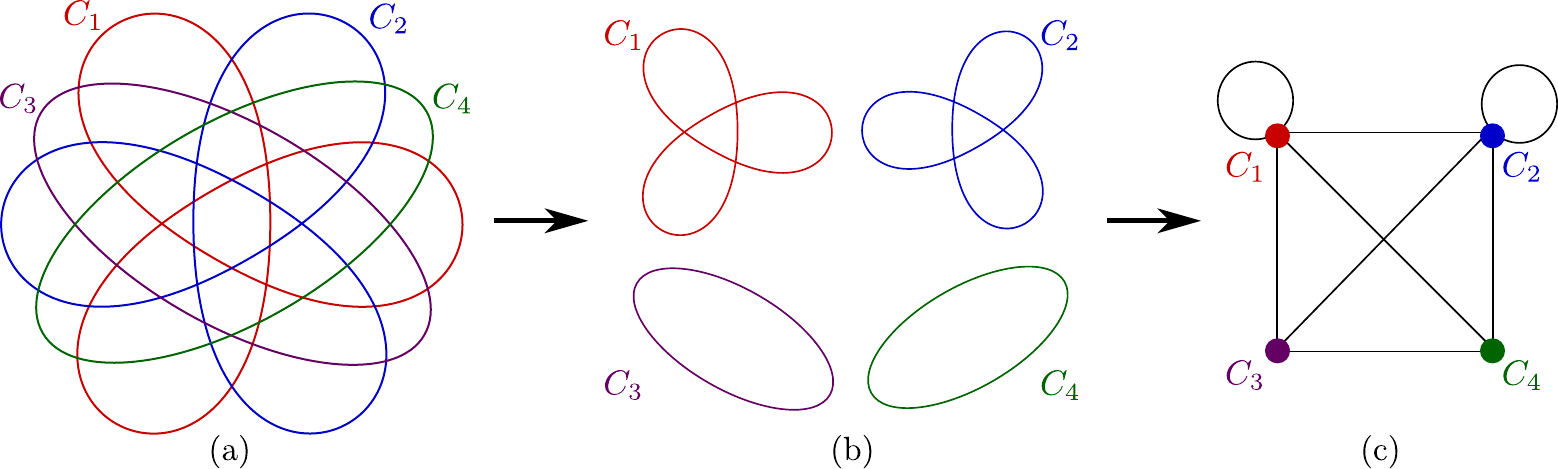}
	\caption{A complicated example for the reduction (\ref{<}).}
	\label{shortexample2}
	\end{figure}

\noindent{\bf Example 2.} 
Figure \ref{shortexample2} is a more complicated example for the reduction (\ref{<}). The graph in Figure \ref{shortexample2} (a) is an instance of $\plholant{\neq_2}{f}$, where all intersection points are degree 4 vertices labeled by $f$ and we omit the degree 2 vertices labeled by $\neq_2$. 
This graph can be divided into 4 circuits colored by red, blue, purple and green (see Figure \ref{shortexample2} (b)). 
Each circuit can be viewed as a Boolean variable for a \#CSP problem (see Figure \ref{shortexample2} (c), edges are constraints).

The $0$-$1$ assignment of edges on a circuit $C_i$ is uniquely determined
 by the assignment of its leader edge $e_i$, corresponding to two orientations of this circuit. 
The binary constraint $g_{i, j}$ on $C_i$ and $C_j$ (for $i<j$) is determined by the placement of signatures $f$ on intersection vertices between circuits $C_i$ and $C_j$.

\vspace{1ex}

{\bf Part \Rmnum{2}:}  
Suppose $I$ is a given instance of $\CSP(g_{1_f}, g_{2_f})$.
Each constraint $g_{1_f}$ and $g_{2_f}$ is applied on certain
pairs of variables. 
It is possible that they are applied to a single variable, resulting in two unary constraints. We will deal with such constraints later. 
 We first consider the case that every constraint is applied on two
distinct variables.

For any pair $i < j$,
consider all binary constraints on variables  $x_i$ and $x_j$ $(i< j)$.
 Note that $g_{1_f}$ is symmetric, that is, $g_{1_f}(x_i, x_j)=g_{1_f}(x_j, x_i)$. We assume all the constraints between $x_i$ and $x_j$
are: $s_{i,j}$ many constraints $g_{1_f}(x_i, x_j)$, $t_{i,j}$ many constraints $g_{2_f}(x_i, x_j)$ and $t'_{i,j}$ many constraints $g_{2_f}(x_j, x_i)$.
Let $g_{i, j}(x_i, x_j)$ be the function product of these constraints. That is, $$g_{i, j}(x_i,x_j)=g_{1_f}^{s_{i, j}}(x_i, x_j)g_{2_f}^{t_{i, j}}(x_i, x_j)g_{2_f}^{t'_{i, j}}(x_j, x_i).$$
Then, we have
$$\CSP(I)=\sum_{\sigma: \{x_1, \ldots, x_k\}\rightarrow \{0, 1\}}\prod_{1\leqslant i < j \leqslant n}g_{i, j}(\sigma(x_i), \sigma(x_j)).$$

We prove the reduction~(\ref{>}) in two steps. 
 We first reduce $\CSP(I)$ to both instances $\Omega_i$ (for $i=1,  2$)
 of $\plholant{\neq_2}{f, \chi_i}$ respectively, where $\chi_1=$
$\left[\begin{smallmatrix}
0 & 0 & 0 & 1\\
0 & 1 & 0 & 0\\
0 & 0 & 1 & 0\\
1 & 0 & 0 & 0\\
\end{smallmatrix}\right]$ and $\chi_2=$
$\left[\begin{smallmatrix}
0 & 0 & 0 & 1\\
0 & 1 & 0 & 0\\
0 & 0 & 1 & 0\\
-1 & 0 & 0 & 0\\
\end{smallmatrix}\right]$. 
 The instance $\Omega_i$ is constructed as follows:

\begin{enumerate}

\item Draw a cycle  $D_1$, i.e.,
a homeomorphic image of $S^1$, on the plane. 
For $2 \leqslant j \leqslant k$ successively draw cycles $D_j$, 
and for all $1 \leqslant i < j$
let $D_j$ intersect
transversally with $D_i$ at least $2(s_{i, j}+t_{i, j}+t'_{i, j})$ many times.
This can be done since we can let $D_j$ enter and exit the interiors of $D_i$
successively. 
A concrete realization is as follows:
Place $k$ vertices $D_i$ on a semi-circle in the order of $i=1, \ldots, k$.
For $1 \leqslant i < j  \leqslant k$, 
connect $D_i$ and $D_j$ by a straight line segment $L_{ij}$.
Now thicken each vertex $D_i$ into a small disk, and deform
the boundary circle of $D_j$ so that, for every $1 \leqslant i < j$,
 it reaches across to $D_i$ along the
line segment $L_{ij}$, and intersects the  boundary circle of $D_i$
exactly
$2(s_{i, j}+t_{i, j}+t'_{i, j})$ many times.  (There are
also other  intersections between these cycles $D_i$'s due to
crossing intersections between those line segments. This is why we
say ``at least'' this many intersections in the overall description.
 We will deal with those extra intersection vertices later.)
We can draw these cycles to satisfy the following conditions:
\begin{enumerate}
\item[a.] There is no self-intersection for each $D_i$.

\item[b.] Every intersection point is between exactly two cycles.
They intersect transversally. Each intersection creates a vertex of degree $4$.


\end{enumerate}

These intersecting cycles define a planar 4-regular graph $G'$,
where intersection  points are the vertices.

\item Replace each edge of $G'$ by a path of length two. 
We get a planar bipartite graph $G=(V, E)$. On one side, all vertices have degree 2, and on the other side, all vertices have degree 4. We can still define edge twins as in  Part \Rmnum{1}. Moreover, we still divide the graph into some circuits $C_1, \ldots, C_k$. In fact, $C_i$ is just the cycle $D_i$ after the replacement of each edge by a path of length two.

Let  $V_{i, j}=C_i\cap C_j$ $(i<j)$ be the intersection vertices between
 $C_i$ and $C_j$. Clearly, $|V_{i, j}|$ is even and 
at least $2(s_{i,j}+t_{i,j}+t^\prime_{i,j})$. Since there is no self-intersection, each circuit is a simple cycle. 
As we did in Part \Rmnum{1}, we pick an edge $e_i$ as the leader edge of $C_i$ and this gives an orientation of $C_i$.
We can define ``entry-vertices'' and ``exit-vertices'' as in Part \Rmnum{1}. Among $V_{i, j}$, half are entry-vertices and the other half are exit-vertices.
(This notion is defined in terms of $C_j$ with respect to $C_i$;
the roles of $i$ and $j$ are not symmetric.)
 List the edges in $C_i$ according to the orientation of $C_i$
starting with the  leader edge $e_i$. 
After we place copies of $f$ on each vertex,
the support of $f$, which is contained in $(x_1 \not = x_3) \wedge
(x_2 \not = x_4)$, ensures that every 4-ary twins can only take  
values $(0, 1)$ or $(1, 0)$, since the 4-ary twin edges are non-adjacent. 
Then   all edges in $C_i$ can only take assignment
 $(0, 1, 0, 1, \cdots, 0, 1)$ when $e_i=0$ 
and $(1, 0, 1, 0, \cdots, 1, 0)$ when $e_i=1$.


\item Label all vertices of degree 2 by 
$(\neq_{2})$. 
For any vertex in $V_{i,j}$ ($i<j$), as we showed in Part \Rmnum{1}, we can label the four edges incident to it by variables $(x_1, x_2, x_3, x_4)$ in a way such that when $\sigma': (e_i, e_j) \mapsto (b, b') \in \{0, 1\}^2$, 
we have $(x_1, x_2, x_3, x_4)=(b, b', 1-b, 1-b')$ at any entry-vertex, and $(x_1, x_2, x_3, x_4)=(b, 1-b', 1-b, b')$ at any exit-vertex (See Table \ref{gvalue1}).
Note that $f$ has four rotation forms under this labeling.
We have (at least) $s_{i,j} + t_{i,j} + t'_{i,j}$ many entry-vertices
and as many exit-vertices.
Let $V'_{i,j}$ be the set of these $2(s_{i,j}+t_{i,j}+t^\prime_{i,j})$ vertices.
 For vertices in $V'_{i,j}$,
we label $s_{i,j}$ many entry-vertices by $f$ and $s_{i,j}$ many exit-vertices by $f^{\frac{\pi}{2}}$, $t_{i,j}$ many entry-vertices by $f$ and $t_{i,j}$ many exit-vertices by $f^{\frac{3\pi}{2}}$, and $t'_{i,j}$ many entry-vertices by $f^{\pi}$ and $t'_{i,j}$ many exit-vertices by $f^{\frac{\pi}{2}}$. 
 Refer to Table \ref{gtable}, this choice amounts to taking
\[k_1 = s_{i,j} + t_{i,j},~~~~ k_3 = t'_{i,j},~~~~ 
\mbox{and,}~~~~ 
\ell_1 = s_{i,j} + t'_{i,j},~~~~\ell_3 = t_{i,j},\]
 and all other $k_i$, $\ell_i$'s equal to 0. 
Recall that $g_{1_f}(x_1, x_2)$ corresponds to choosing
$k_1 = \ell_1 = 1$ and the others all 0,
$g_{2_f}(x_1, x_2)$ corresponds to choosing
$k_1 = \ell_3 =1$ and the others all 0,
and $g_{2_f}(x_2, x_1)$ corresponds to choosing
$k_3 = \ell_1 =1$ and the others all 0,
then we have 
$$\prod_{v\in V'_{ij}}f_v({\sigma'_{(e_i, e_j)}} \mid_{E_{(v)}})=g_{1_f}^{s_{i,j}}(e_i, e_j)g_{2_f}^{t_{i,j}}(e_i, e_j)g_{2_f}^{t'_{i,j}}(e_j, e_i)=g_{i, j}(e_i, e_j).$$
For all vertices  in $V_{i,j}\backslash V'_{i,j}$, if
we label them by an auxiliary signature $ \chi_1=
\left[\begin{smallmatrix}
0 & 0 & 0 & 1\\
0 & 1 & 0 & 0\\
0 & 0 & 1 & 0\\
1 & 0 & 0 & 0\\
\end{smallmatrix}\right],$ then, referring to Table \ref{gtable} (Here $a=x=b=y=1$), we have $$\prod_{v\in V_{i, j}\backslash V'_{ij}}\chi_1(\sigma'_{(e_i, e_j)} \mid_{E_{(v)}}) = 1,$$
for all assignments $\sigma'$ on $\{e_i, e_j\}$.
We can also label the vertices in $V_{i,j}\backslash V'_{i, j}$ by
an auxiliary signature $ \chi_2=
\left[\begin{smallmatrix}
0 & 0 & 0 & 1\\
0 & 1 & 0 & 0\\
0 & 0 & 1 & 0\\
-1 & 0 & 0 & 0\\
\end{smallmatrix}\right].$ 
By our (semi-circle) construction,
   in $V_{i,j}\backslash V'_{i, j}$, the number of entry-vertices is equal to the number of exit-vertices. We label all entry-vertices by $\chi_2$ and label all exit-vertices by its rotated form $ \chi^{\frac{\pi}{2}}_2=
\left[\begin{smallmatrix}
0 & 0 & 0 & 1\\
0 & 1 & 0 & 0\\
0 & 0 & -1 & 0\\
1 & 0 & 0 & 0\\
\end{smallmatrix}\right].$ Refer to Table \ref{gtable} (here $a=b=y=1, x=-1,$ and $ k=k_1=\ell_1=\ell$, and the crucial equation is $g_{i,j}(1,1)
= x^{k_1+\ell_1} = (-1)^2 =1$), we have 
 $$\prod_{v\in V_{i, j}\backslash V'_{ij}}\chi_2(\sigma'_{(e_i, e_j)} \mid_{E_{(v)}}) = 1,$$
for all assignments $\sigma'$ on $\{e_i, e_j\}$.
 \end{enumerate}
 
\begin{figure}[!h]
\centering
		\includegraphics[height=1.8in]{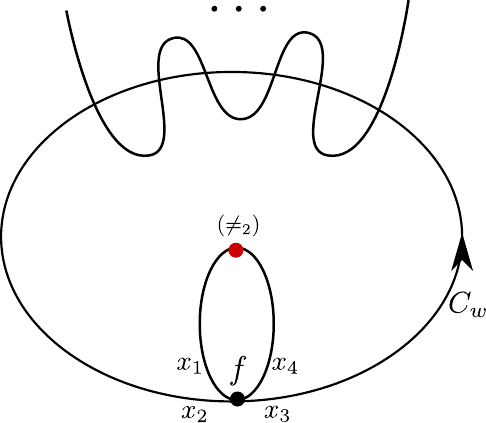}
	\caption{Creating self-loop locally on cycle $C_w$}
	\label{fig:unary-realization-sec4-lm4.5}
	\end{figure}

  Then, consider the case that $g_{1_f}$ and $g_{2_f}$ are applied to
the pair  variables  $(w, w)$,
 in which case $g_{1_f}$ and $g_{2_f}$  effectively become
unary constraints $[a^2, x^2]$ and $[ax, ax]$ on the variable $x_i$. The latter
is a constant multiple of $[1,1]$ and can be ignored.
The unary constraint $[a, x]$, and hence also $[a^2, x^2]$,
 can be easily realized by $f$
in $\operatorname{Pl-Holant}(\neq_2\mid f, \chi_i)$, by creating a self-loop
for the cycle representing the variable $w$, denoted by $C_w$ (See Figure \ref{fig:unary-realization-sec4-lm4.5}). 
Note that the self-loop is created locally on the cycle $C_w$ such that it does not affect other cycles.
As we did in Part I, we label the four edges incident to a self-intersection vertex by $(x_1, x_2, x_3, x_4)$ such that $x_3$ is the successor of $x_1$ and $x_4$ is the successor of $x_2$ depending on the default orientation of $C_w$, and $(x_1, x_2, x_3, x_4)$ are labeled in counterclockwise order. 
Then, we have $(x_1, x_3)=(x_2, x_4)=(0, 1)$ when $w=0$ and $(1, 0)$ when $w=1$.
That is, ${g_1}_f(0, 0)=a^2=f^2_{0011}$ and ${g_1}_f(1, 1)=x^2=f^2_{1100}$.

Now, we get an instance $\Omega_s$ $(s=1, 2)$ for each problem $\plholant{\neq_2}{f, \chi_s}$ respectively. 
Note that $\chi_s$ has the support $(x_1 \neq x_3) \wedge (x_2 \neq x_4)$
 as $f$. As we have showed in Part \Rmnum{1},
 for any nonzero term in the sum $\PlHolant_{\Omega_s}$,  the assignment of all edges $\sigma:E\rightarrow \{0, 1\}$ can be uniquely extended from the assignment of all leader edges $\sigma': \{e_1, e_2, \ldots, e_k\}\rightarrow \{0, 1\}$. Therefore, we have \begin{equation*}
\begin{aligned}
 \CSP(I)&= 
  \sum_{\sigma': \{e_1, \cdots, e_k\}\rightarrow \{0, 1\}}\prod_{1\leqslant i < j \leqslant n}g_{i, j}(\sigma'(e_i), \sigma'(e_j)) \\
&=  \sum_{\sigma': \{e_1, \cdots, e_k\}\rightarrow \{0, 1\}}\Bigg(\prod_{\substack{ v\in V'_{i, j} \\1\leqslant i < j \leqslant n}}f_v(\sigma'|_{E_{(v)}})\Bigg) \Bigg(\prod_{\substack{ v\in V_{i, j}\backslash V'_{i, j} \\1\leqslant i < j \leqslant n}}{\chi_s}_v(\sigma'|_{E_{(v)}})\Bigg) \\
&= \operatorname{Pl-Holant}_{\Omega_s}
\end{aligned}
\end{equation*}
for $s = 1, 2$.
That is, $\CSP(g_{1_f}, g_{2_f})\leqslant_T \plholant{\neq_2}{f, \chi_s}$,
($s = 1, 2$).

From the hypothesis of the reduction (\ref{>}), we have $a = \pm x \not = 0, 
b = \pm y \not = 0$, and $(b/a)^8 \not =1$.
We show by interpolation
 $$\PlHolant(\neq_2 \mid f, \chi_1)\leqslant_{T}\PlHolant(\neq_2 \mid f) $$ when $a=\epsilon x, b=\epsilon y,$ 
and 
$$\PlHolant(\neq_2 \mid f, \chi_2)\leqslant_{T}\PlHolant(\neq_2 \mid f)$$ when $a=\epsilon x, b= -\epsilon y$, where $\epsilon=\pm 1$.

\begin{itemize}

\item If $a=x$ and $b=y$, since they are all nonzero, and $(\frac{b}{a})^8 \neq 1$, by normalization we may assume $M(f)=
\left[\begin{smallmatrix}
0 & 0 & 0 & 1\\
0 & b & 0 & 0\\
0 & 0 & b & 0\\
1 & 0 & 0 & 0\\
\end{smallmatrix}\right]$, where $b\neq 0$ and $b^8\neq 1$.

If $b$ is not a root of unity, by Lemma \ref{1111}, we have $\PlHolant(\neq_2 \mid f,  \chi_1)\leqslant_{T}\PlHolant(\neq_2 \mid f).$
Otherwise, $b$ is a root of unity. Construct
a  gadget $f_{\boxtimes}$ as shown in Figure~\ref{square}. 
Given an assignment $(x_1, x_2, x_3, x_4)$ to $f_{\boxtimes}$,
and suppose $f_{\boxtimes}(x_1, x_2, x_3, x_4)\neq 0$.
Then because of the support of $f_{v_1}, f_{v_5}$ and $f_{v_3}$
we must have $x_1 \neq x_3$. Similarly $x_2  \neq x_4$.
Also $f_{v_5}$ receives the same input as $f_{\boxtimes}$.
 Hence the support of $f_{\boxtimes}$ is
contained in $(x_1 \neq x_3) \wedge (x_2  \neq x_4)$,
i.e., contained in  $\{(0, 0, 1, 1),
(1, 1, 0, 0), (0, 1, 1, 0), (1, 0, 0, 1)\}$.  
In particular,
 the edges on each {\sc Diagonal Line} of this gadget can only take assignments
 $(0, 1, 0, 1, 0, 1)$ or $(1, 0, 1, 0, 1, 0)$, otherwise the we get zero. 
On the other hand,
 the {\sc Square} cycle in this gadget is a circuit itself, so that
the edges in it can only take two assignments $(0, 1, 0, 1, 0, 1, 0, 1)$ or $(1, 0, 1, 0, 1, 0, 1, 0)$. We simplify the notation to $(0, 1)$ and $(1, 0)$
respectively.
On  $(x_1 \neq x_3) \wedge (x_2  \neq x_4)$,
the value of $f_{\boxtimes}$ is the sum over these two terms.

\begin{figure}[!htbp]
\centering
		\includegraphics[height=1.2in]{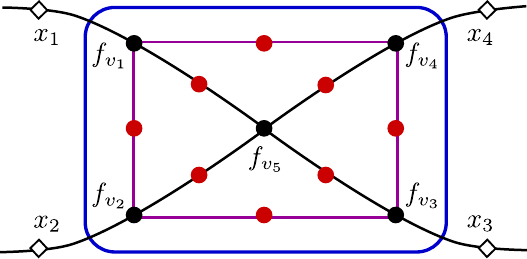}
	\caption{The {\sc Square} gadget}
	\label{square}
	\end{figure}

For the signature $f$, if one pair of its edge twins flips its assignment between $(0, 1)$ and $(1, 0)$, then the value of $f$ changes from $1$ to $b$, or 
from $b$ to $1$. If both pairs of edge twins  flip their assignments,
 then the value of $f$ does not change. 
According to this property, we give the  Table~\ref{square1}. 
Here, we place a suitably rotated copy of $f$
at vertices $v_i$ to get $f_{v_i}$ (for $1 \le i \le 5$)
so that 
 the values of $f_{v_i}$  are all $1$ under the assignment $(x_1, x_2, x_3, x_4)=(0, 0, 1, 1)$ and the {\sc Square} is assigned $=(0, 1)$ (row 2 of Table~\ref{square1}).
When the assignment of {\sc Square} flips from $(0, 1)$ to $(1, 0)$, one pair of edge twins of each vertex except $v_5$ flips its assignment.
So the values of $f$ on these vertices except $v_5$ change from $1$ to $b$ (row 3). 
When $(x_1, x_3)$ flips its assignment, one pair of edge twins of $v_1, v_3$ and $v_5$ flip their assignments.  
When $(x_2, x_4)$ flips its assignment, one pair of edge twins of $v_2, v_4$ and $v_5$ flip their assignments. Using this fact, we get other rows correspondingly.
\begin{table}[!htp]
\renewcommand{\arraystretch}{1.2}
\centering
\begin{tabular}{|c|c|c|c|c|c|c|c|}
\hline
 $(x_1, x_2, x_3, x_4)$ & {\sc Square} & $f_{v_1}$ & $f_{v_2}$ & $f_{v_3}$ & $f_{v_4}$ & $f_{v_5}$ & $f_{\boxtimes}$ \\
 \hline
\multirow{2}{*}{$(0, 0, 1, 1)$} & $(0, 1)$ & $1$ & $1$ & $1$ & $1$ &$1$ & \multirow{2}{*}{$1+b^4$}\\
\cline{2-7}
& $(1, 0)$ & $b$ & $b$ & $b$ & $b$ & $1$ & \\
 \hline
\multirow{2}{*}{$(1, 1, 0, 0)$} & $(0, 1)$ & $b$ & $b$ & $b$ & $b$ &$1$ & \multirow{2}{*}{$1+b^4$}\\
\cline{2-7}
& $(1, 0)$ & $1$ & $1$ & $1$ & $1$ & $1$ & \\
 \hline
\multirow{2}{*}{$(0, 1, 1, 0)$} & $(0, 1)$ & $1$ & $b$ & $1$ & $b$ &$b$ & \multirow{2}{*}{$2b^3$}\\
\cline{2-7}
& $(1, 0)$ & $b$ & $1$ & $b$ & $1$ & $b$ & \\
 \hline
\multirow{2}{*}{$(1, 0, 0, 1)$} & $(0, 1)$ & $b$ & $1$ & $b$ & $1$ &$b$ & \multirow{2}{*}{$2b^3$}\\
\cline{2-7}
& $(1, 0)$ & $1$ & $b$ & $1$ & $b$ & $b$ & \\
\hline
\end{tabular}
\caption{The values of gadget $f_{\boxtimes}$ when $a=x=1$ and $b=y$}\label{square1}
\end{table}

Hence, $f_{\boxtimes}$ has the signature matrix 
$M(f_{\boxtimes})=\left[\begin{smallmatrix}
0 & 0 & 0 & 1+b^4\\
0 & 2b^3 & 0 & 0\\
0 & 0 & 2b^3 & 0\\
1+b^4 & 0 & 0 & 0\\
\end{smallmatrix}\right]$. Since $b^8 \neq 1$, we have
 $1+b^4 \neq 0$, by normalization we  can write
$M(f_{\boxtimes})=\left[\begin{smallmatrix}
0 & 0 & 0 & 1\\
0 & \frac{2b^3}{1+b^4} & 0 & 0\\
0 & 0 & \frac{2b^3}{1+b^4} & 0\\
1 & 0 & 0 & 0\\
\end{smallmatrix}\right]$. 
Since $|b|=1$ and $b^4\neq 1$, we have $|1+b^4|< 2$. Then $ |\frac{2b^3}{1+b^4}|>|b^3|=1$, which means $\frac{2b^3}{1+b^4}$ is not a root of unity. 
By Lemma \ref{1111},  we have $\PlHolant(\neq_2 \mid f, \chi_1) \leqslant_{T} \PlHolant(\neq_2 \mid f, f_{\boxtimes}) $. Since $f_{\boxtimes}$ is constructed by $f$, we have $ \PlHolant(\neq_2 \mid f, \chi_1)\leqslant_{T} \PlHolant(\neq_2 \mid f).$

\item If $a=-x$ and $b=-y$, then $M(f)=\left[\begin{smallmatrix}
0 & 0 & 0 & a\\
0 & b & 0 & 0\\
0 & 0 & -b & 0\\
-a & 0 & 0 & 0\\
\end{smallmatrix}\right]$. Connect the variable $x_4$ with $x_3$ of $f$ using $(\neq_2)$, and we get a binary signature $g'$, where $$g'=M_{x_1x_2, x_4x_3}(0, 1, 1, 0)^T=(0, b, -b, 0)^T.$$  
Since $b \neq 0$, $g'$ can be normalized as $(0, 1, -1, 0)^T$. 
Modifying $x_1=1$ of $f$ by $-1$ scaling, 
we get a signature $f^\prime$ with the signature matrix 
$M(f^\prime)=\left[\begin{smallmatrix}
0 & 0 & 0 & a\\
0 & b & 0 & 0\\
0 & 0 & b & 0\\
a & 0 & 0 & 0\\
\end{smallmatrix}\right].$ 
As we have proved above, $\PlHolant(\neq_2 \mid f, {\chi_1}) \leqslant_{T} \PlHolant(\neq_2 \mid f, f^\prime) $. 
Since $f^\prime$ is constructed by $f$, we have $\PlHolant(\neq_2 \mid f, {\chi_1}) \leqslant_{T} \PlHolant(\neq_2 \mid f).$

\item If $a=-x$, $b=y$ or  $a=x$, $b=-y$, by normalization and rotational symmetry, we may assume $M(f)=
\left[\begin{smallmatrix}
0 & 0 & 0 & 1\\
0 & b & 0 & 0\\
0 & 0 & b & 0\\
-1 & 0 & 0 & 0\\
\end{smallmatrix}\right]$, where $b\neq 0$ and $b^8\neq 1$.


If $b$ is not a root of unity, by Corollary \ref{111-1}, we have  $\PlHolant(\neq_2 \mid f, {\chi_2}) \leqslant_{T} \PlHolant(\neq_2 \mid f)$.
Otherwise, $b$ is a root of unity.
Construct the gadget $f_{\boxtimes}$ in the same way as shown above.
Our discussion on the support of $f_{\boxtimes}$ still holds:
It is contained in $(x_1 \neq x_3) \wedge (x_2  \neq x_4)$;
on  $(x_1, x_2, x_3, x_4)$ with  $(x_1 \neq x_3) \wedge (x_2  \neq x_4)$,
  $f_{v_5}$ receives the same input, and the value of $f_{\boxtimes}$ 
is the sum over two assignments $(0, 1)$ and $(1, 0)$ for the
{\sc Square}.

For the signature $f$, if one pair of its edge twins flips its assignment between $(0, 1)$ and $(1, 0)$, 
then the value of $f$ changes from $\pm 1$ to $b$, or $b$ to $\mp 1$. 
If two pairs of edge twins both flip their assignments, then the value of $f$ 
does not change if the value is $b$, or changes its sign if the value is $\pm 1$. 
According to this property,  we have the following Table~\ref{square2}.
Here, we place a suitably rotated copy of $f$
at vertices $v_i$ to get $f_{v_i}$ (for $1 \le i \le 5$)
so that
the values of $f_{v_i}$ are all $1$ under the assignment 
$(x_1, x_2, x_3, x_4)=(0, 0, 1, 1)$  and the {\sc Square} is assigned $=(0, 1)$ 
(row 2 of Table~\ref{square2}).
When the assignment of {\sc Square} flips from $(0, 1)$ to $(1, 0)$, 
one pair of edge twins at each vertex except $v_5$ flips its assignment. 
So the values of $f$ at these vertices except $v_5$ change from $1$ to $b$ (row 3). 
When $(x_1, x_3)$ flips its assignment, one pair of edge twins at $v_1, v_3$ and $v_5$ flips their assignments. 
When $(x_2, x_4)$ flips its assignment, one pair of edge twins at $v_2, v_4$ and $v_5$ flips their assignments. 
Using this fact, we get other rows correspondingly.

\begin{table}[!htp]
\renewcommand{\arraystretch}{1.2}
\centering
\begin{tabular}{|c|c|c|c|c|c|c|c|}
\hline
 $(x_1, x_2, x_3, x_4)$ & {\sc Square} & $f_{v_1}$ & $f_{v_2}$ & $f_{v_3}$ & $f_{v_4}$ & $f_{v_5}$ & $f_{\boxtimes}$ \\
 \hline
\multirow{2}{*}{$(0, 0, 1, 1)$} & $(0, 1)$ & $1$ & $1$ & $1$ & $1$ &$1$ & \multirow{2}{*}{$1+b^4$}\\
\cline{2-7}
& $(1, 0)$ & $b$ & $b$ & $b$ & $b$ & $1$ & \\
 \hline
\multirow{2}{*}{$(1, 1, 0, 0)$} & $(0, 1)$ & $b$ & $b$ & $b$ & $b$ &$-1$ & \multirow{2}{*}{$-(1+b^4)$}\\
\cline{2-7}
& $(1, 0)$ & $-1$ & $-1$ & $-1$ & $-1$ & $-1$ & \\
 \hline
\multirow{2}{*}{$(0, 1, 1, 0)$} & $(0, 1)$ & $1$ & $b$ & $1$ & $b$ &$b$ & \multirow{2}{*}{$2b^3$}\\
\cline{2-7}
& $(1, 0)$ & $b$ & $-1$ & $b$ & $-1$ & $b$ & \\
 \hline
\multirow{2}{*}{$(1, 0, 0, 1)$} & $(0, 1)$ & $b$ & $1$ & $b$ & $1$ &$b$ & \multirow{2}{*}{$2b^3$}\\
\cline{2-7}
& $(1, 0)$ & $-1$ & $b$ & $-1$ & $b$ & $b$ & \\
\hline
\end{tabular}
\caption{The values of gadget $f_{\boxtimes}$ when $a=-x=1$ and $b=y$}\label{square2}
\end{table}

Hence, $f_{\boxtimes}$ has the signature matrix 
$\left[\begin{smallmatrix}
0 & 0 & 0 & 1+b^4\\
0 & 2b^3 & 0 & 0\\
0 & 0 & 2b^3 & 0\\
-(1+b^4) & 0 & 0 & 0\\
\end{smallmatrix}\right]$. 
Since $|b|=1$ and $b^8\neq 1$, we have  $b^4 \neq \pm 1$,
therefore $0 < | 1+b^4| <2$, and so
$\frac{2b^3}{1+b^4}$ is not a root of unity. 
By Corollary \ref{111-1},  $\PlHolant(\neq_2 \mid f, \chi_2) \leqslant_{T} \PlHolant(\neq_2 \mid f, f_{\boxtimes}) $, 
and hence $ \PlHolant(\neq_2 \mid f, \chi_2)\leqslant_{T} \PlHolant(\neq_2 \mid f).$
\end{itemize}

In summary, we have $$\xymatrix{
  & & \plholant{\neq_2}{f, \chi_1} \ar[dr]|-{\text{$a=\epsilon x, b=\epsilon y$ ($\epsilon=\pm 1$)}}  & \\
 & \CSP(g_{1_f}, g_{2_f}) \ar[ur]^{}\ar[dr]_{} \ar@{}[r]|-{\leqslant_T} & \ar@{}[r]|-{\leqslant_T} &  \plholant{\neq_2}{f}  \\
 & & \plholant{\neq_2}{f, \chi_2} \ar[ur]|-{\text{$a=\epsilon x, b=-\epsilon y$ ($\epsilon =\pm 1)$}}&
}$$
Therefore, we have $\CSP(g_{1_f}, g_{2_f})\leqslant_T \plholant{\neq_2}{f}$ when $a^2= x^2 \neq 0$, $b^2= y^2 \neq 0$ and $(\frac{b}{a})^8\neq 1$.
\qed

\begin{remark}
A crucial point in the  reduction (\ref{<})
is the fact that the given instance graph $G$ of \plholant{\neq_2}{f}
is planar so that $\sum_{i} k_i = \sum_i \ell_i$. Otherwise
this does not hold in general; for example  the 
latitudinal and longitudinal
closed cycles on a torus intersect at  a single  point.
The equation $\sum_{i} k_i = \sum_i \ell_i$ is crucial
to obtain tractability in the following theorem.
\end{remark}

\begin{theorem} \label{inner}
Let $f$ be a 4-ary signature of the form {\rm (\ref{innermatrix})},
where $(a, x) \neq (0, 0)$ and $(b, y) \neq (0, 0)$.
Then $\operatorname{Pl-Holant}(\neq_2\mid f)$
is \#P-hard unless
\begin{enumerate}
\item[$\operatorname{(\rmnum{1})}$] $(ax)^2=(by)^2$, or
\item[$\operatorname{(\rmnum{2})}$] $x=a\frak{i}^\alpha, b=a\sqrt{\frak{i}}^\beta, y=a\sqrt{\frak{i}}^\gamma$, where $\alpha, \beta, \gamma \in \mathbb{N}$, and $\beta\equiv \gamma \pmod 2,$
\end{enumerate}
in which cases, the problem is tractable in polynomial time.
\end{theorem}




{\bf Proof of Tractability:} 
\begin{itemize}

\item In case (\rmnum{1}), if $ax=by=0$, then $f$ has support of size at most $2$. So we have $f \in \mathscr{P}$, and hence $\operatorname{Pl-Holant}(\neq_2\mid f)$ is tractable by Theorem \ref{aptractable}.
Otherwise, $(ax)^2=(by)^2\neq 0$. 
For any signature $g$ in $\mathcal{G}_f$, 
we have $g_{00} \cdot g_{11}=(ax)^{k_1+\ell_1+k_3+\ell_3}(by)^{k_2+\ell_2+k_4+\ell_4}$ and $g_{01} \cdot g_{10}=(ax)^{k_2+\ell_2+k_4+\ell_4}(by)^{k_1+\ell_1+k_3+\ell_3}$. 
Since $(k_1+\ell_1+k_3+\ell_3)-(k_2+\ell_2+k_4+\ell_4)\equiv k+\ell \equiv 0 \pmod 2$, 
we have$$\dfrac{g_{00} \cdot g_{11}}{g_{01} \cdot g_{10}}=\left(\dfrac{ax}{by}\right)^{(k_1+\ell_1+k_3+\ell_3)-(k_2+\ell_2+k_4+\ell_4)}=\left(\dfrac{(ax)^2}{(by)^2}\right)^{\frac{(k_1+\ell_1+k_3+\ell_3)-(k_2+\ell_2+k_4+\ell_4)}{2}}=1.$$
That is, $g \in \mathscr{P}$. Since any signature $h$ in $\mathcal{H}_f$ is unary, $h \in \mathscr{P}$. Hence, we have $\mathcal{G}_f\cup \mathcal{H}_f \subseteq \mathscr{P}$. By Theorem \ref{cspdic}, \#CSP$(\mathcal{G}_f\cup \mathcal{H}_f)$ is tractable.
 By reduction~(\ref{<}) of Lemma \ref{csp}, we have $\operatorname{Pl-Holant}(\neq_2\mid f)$ is tractable.

\item In case (\rmnum{2}), for any signature $g \in \mathcal{G}_f$
defined in Definition~\ref{defn:Gf-Hf}, $M(g)$ is of the form $$a^{k+\ell}\left[ \begin{matrix}
 \sqrt{\frak{i}}^{\beta(k_4+\ell_4)+\gamma(k_2+\ell_2)+2\alpha(k_3+\ell_3)} & \sqrt{\frak{i}}^{\beta(k_1+\ell_3)+\gamma(k_3+\ell_1)+2\alpha(k_4+\ell_2)}\\
 \sqrt{\frak{i}}^{\beta(k_3+\ell_1)+\gamma(k_1+\ell_3)+2\alpha(k_2+\ell_4)} & \sqrt{\frak{i}}^{\beta(k_2+\ell_2)+\gamma(k_4+\ell_4)+2\alpha(k_1+\ell_1)}\\  \end{matrix} \right]=
 a^{k+\ell}\left[\begin{matrix}
 \sqrt{\frak{i}}^{p_{00}} & \sqrt{\frak{i}}^{p_{01}}\\
 \sqrt{\frak{i}}^{p_{10}} & \sqrt{\frak{i}}^{p_{11}}\\
 \end{matrix}
 \right],$$
 where $p_{00}, p_{01}, p_{10}$ and $p_{11}$ denote the integer
exponents of $\sqrt{\frak{i}}$ in the respective entries of $g$.
 Since $\beta\equiv \gamma \pmod 2$, if they are both even, 
 then $p_{00} \equiv p_{01} \equiv p_{10} \equiv p_{11} \equiv 0 \pmod 2;$ 
 if they are both odd, then $p_{00} \equiv p_{11} \equiv 
 k_2+\ell_2+k_4+\ell_4 \equiv
 k_1+\ell_1+k_3+\ell_3 \equiv
 p_{01} \equiv p_{10} \pmod 2.$ 
 If these exponents are all odd, we can take out a $\sqrt{\frak{i}}$. 
 Hence, $g$ is of the form $a'(\frak{i}^{q_{00}}, \frak{i}^{q_{01}}, \frak{i}^{q_{10}}, \frak{i}^{q_{11}})^T$, where $a'=a^{k+\ell}$ or $a^{k+\ell}\sqrt{\frak{i}}$, 
and either $q_{ij}=\frac{p_{ij}}{2}$  for all $i, j \in \{0, 1\}$ are integers,
or $q_{ij}=\frac{p_{ij}-1}{2}$ for all $i, j \in \{0, 1\}$ are integers. Thus, $$q_{00}+q_{01}+q_{10}+q_{11} \equiv (p_{00} + p_{01} + p_{10} + p_{11})/2  \pmod 2.$$ 
 Moreover, since $p_{00} + p_{01} + p_{10} + p_{11}=(k+\ell)(\beta+\gamma+2\alpha) \equiv 0 \pmod 4,$ using the assumption that
$\beta  \equiv \gamma \pmod 2$ and $k  \equiv \ell \pmod 2$,
 we conclude that $q_{00}+q_{01}+q_{10}+q_{11} \equiv 0 \pmod 2$.
Therefore, $g \in \mathscr{A}$  by Lemma \ref{4affine}.

 In this case, for any signature $h$ in $\mathcal{H}_f$, $M(h)$ is of the 
form 
$$a^{m}\left[ \begin{matrix}
\sqrt{\frak{i}}^{\beta m_4+\gamma m_2+2\alpha m_3} & \sqrt{\frak{i}}^{\beta m_2+\gamma m_4+2\alpha m_1} \end{matrix} \right].$$
Since $\beta \equiv \gamma  \pmod 2$, we 
have $\beta m_4+\gamma m_2  \equiv \beta m_2+\gamma m_4 \pmod 2$. Hence, $h$ is of the 
form $a'^\prime [\frak{i}^{q_0}, \frak{i}^{q_1}]$,
for some integers $q_0, q_1$, where $a''=a^{m}$ or $a^{m}\sqrt{\frak{i}}$. That is, $h \in \mathscr{A}$ by Lemma \ref{2affine}. Hence, $\mathcal{G}_f\cup \mathcal{H}_f \subseteq \mathscr{A}$. 
 By Theorem \ref{cspdic}, \#CSP$(\mathcal{G}_f\cup \mathcal{H}_f)$ is tractable.
 By reduction~(\ref{<}) of Lemma \ref{csp}, we have $\operatorname{Pl-Holant}(\neq_2\mid f)$ is tractable.
\end{itemize}


{\bf Proof of Hardness:} 
We are given that $f$ does not belong to
 case (\rmnum{1}) or  case (\rmnum{2}).
Note that $M_{x_4x_1, x_3x_2}(f)=\left[\begin{smallmatrix}
0 & 0& 0& b\\
0 & x& 0& 0\\
0 & 0& a& 0\\
y & 0& 0& 0\\
\end{smallmatrix}\right]$
and $M_{x_2x_3, x_1x_4}(f)=\left[\begin{smallmatrix}
0 & 0& 0& y\\
0 & a& 0& 0\\
0 & 0& x& 0\\
b & 0& 0& 0\\
\end{smallmatrix}\right]$. 
 Connect variables $x_3, x_2$ of a copy of the signature $f$ with variables
 $x_2, x_3$ of another copy of signature $f$ both using $(\neq_2)$. We get a signature $f_1$ with the signature matrix 
$$M(f_1)=M_{x_4x_1, x_3x_2}(f)NM_{x_2x_3, x_1x_4}(f)=\left[\begin{matrix}
0 & 0& 0& by\\
0 & 0& x^2& 0\\
0 & a^2& 0& 0\\
by & 0& 0& 0\\
\end{matrix}\right].$$
Similarly, connect $x_3, x_2$ of a copy of signature $f$ with  $x_4, x_1$ of another copy of signature $f$ both using $(\neq_2)$. We get a signature $f_2$ with the signature matrix 
$$M(f_2)=M_{x_4x_1, x_3x_2}(f)NM_{x_4x_1, x_3x_2}(f)=\left[\begin{matrix}
0 & 0& 0& b^2\\
0 & 0& ax& 0\\
0 & ax& 0& 0\\
y^2 & 0& 0& 0\\
\end{matrix}\right].$$
Notice that $M(f^{\frac{\pi}{2}}_1)=\left[\begin{smallmatrix}
0 & 0& 0& 0\\
0 & by& a^2& 0\\
0 & x^2& by& 0\\
0 & 0& 0& 0\\
\end{smallmatrix}\right]$,
$M(f^{\frac{\pi}{2}}_2)=\left[\begin{smallmatrix}
0 & 0& 0& 0\\
0 & b^2& ax& 0\\
0 & ax& y^2& 0\\
0 & 0& 0& 0\\
\end{smallmatrix}\right]$, $M( g_{1_f})=
\left[\begin{smallmatrix}
a^2 & by\\
by & x^2\\
\end{smallmatrix}\right]$ and $M( g_{2_f})=
\left[\begin{smallmatrix}
ax & b^2\\
y^2 & ax\\
\end{smallmatrix}\right]$. 
Recall that $M\left(\widetilde{f^{\frac{\pi}{2}}_i}_{\rm{In}}\right)=M_{\rm{In}}(f^{\frac{\pi}{2}}_i)\left[\begin{smallmatrix}
0 & 1\\
1 & 0\\
\end{smallmatrix}\right]$.
We have $g_{i_f}= \widetilde{{f}^{\frac{\pi}{2}}_i}_{\rm{In}}$. 
That is, $f_i(x_1, x_2, x_3, x_4)= g_{i_f} (x_2, x_4) \cdot \chi_{x_1\neq x_4} \cdot \chi_{x_2 \neq x_3}$. Now, we analyze $g_{1_f}$ and $g_{2_f}$.

\begin{itemize}

\item If $\{g_{1_f}, g_{2_f}\} \subseteq \mathscr{P}$, then either $(ax)^2=(by)^2$ if
either signature is degenerate, or $ g_{1_f}$ and $ g_{2_f}$ are
each generalized {\sc Equality} or generalized {\sc Disequality} respectively. 
In the latter case, since $(a, x) \neq (0, 0)$ and $(b, y) \neq (0, 0)$, it forces that $ax=by=0$. So we still have $(ax)^2=(by)^2$. That is, $\{a, b, x ,y\}$ belongs to case (\rmnum{1}). A contradiction.

\item If $\{g_{1_f}, g_{2_f}\} \subseteq \mathscr{A}$, there are two subcases. 
Note that the support  of a function in $\mathscr{A}$ has size a power of 2.
\begin{itemize}

\item If both $ g_{1_f}$ and $ g_{2_f}$ have support of size at most $2$, then we have $ax=by=0$ due to $(a, x)\neq (0, 0)$ and $(b, y)\neq (0, 0)$. This 
belongs to case (\rmnum{1}).  A contradiction.
\item Otherwise, at least one of  $ g_{1_f}$ or  $ g_{2_f}$ 
has support of size $4$. Then $abxy \neq 0$
and therefore both $ g_{1_f}$ and   $ g_{2_f}$
  have support of size $4$. Let $x^\prime=\frac{x}{a}, b^\prime=\frac{b}{a}$ and $y^\prime=\frac{y}{a}$. By normalization, we have $$M( g_{1_f})=
a^2\left[\begin{matrix}
1 & b^\prime y^\prime\\
b^\prime y^\prime & x^{\prime 2}\\
\end{matrix}\right].$$
Since $ g_{1_f}\in \mathscr{A}$, by Lemma \ref{4affine}, $x^{\prime 2}$ and $b^{\prime}y^{\prime}$ are both powers of $\frak i$, and the sum of all exponents is even. It forces that $x^{\prime 2}=\frak i^{2\alpha}$ for some $\alpha \in \mathbb{N}$. Then, we can choose $\alpha$ such  $x^{\prime }=\frak{i}^{\alpha}.$ Also, we have  $$M( g_{2_f})=
a^2\left[\begin{matrix}
x^\prime & b^{\prime 2}\\
y^{\prime 2} & x^{\prime}\\
\end{matrix}\right].$$
Since  $ g_{2_f}\in \mathscr{A}$ and $x'$ is already a power of $\frak i$, $ y^{\prime 2}$ and  $b^{\prime 2}$ are both powers of $\frak i$. That is, $b^\prime=\sqrt{\ii}^\beta$ and $y^\prime=\sqrt{\ii}^\gamma$.
Also, since $g_{1_f}\in \mathscr{A}$, $b^{\prime}y^{\prime}=\sqrt{\ii}^{\beta+\gamma}$ is a power of $\ii$, which means $\beta\equiv \gamma \pmod 2.$ That is, $\{a, b, x, y\}$ 
belongs to case (\rmnum{2}). A contradiction.
\end{itemize}
\item If $\{g_{1_f}, g_{2_f}\} \subseteq \mathscr{\widehat{M}}$,
 then by Lemma \ref{binary-m}, we have
both $a^2= \epsilon x^2, by = \epsilon by$
and $ax = \epsilon' ax, y^2 = \epsilon' b^2$, for some
$\epsilon, \epsilon' \in \{1, -1\}$.
If $\epsilon =-1$ then $by = 0$, and then by the second set
of equations $b=y=0$, contrary to assumption that $(b,y) \neq (0,0)$.
So $\epsilon =1$.
Similarly $\epsilon' =1$. Hence
\begin{equation}\label{matchgate-caseinLm4-6}
a^2= x^2~~~~ b^2= y^2,
\end{equation}
and it also follows that all 4 entries are nonzero.
\end{itemize}




Therefore, if $\{a, b, x, y\}$ does not satisfy (\ref{matchgate-caseinLm4-6})
 then $\{g_{1_f}, g_{2_f}\} \nsubseteq \mathscr{P}, \mathscr{A}$ or $\mathscr{\widehat M}$. By Theorem \ref{cspdic}, Pl-\#CSP$(g_{1_f}, g_{2_f})$ is \#P-hard. 
Then by Lemma \ref{outerlemma},  
 $\operatorname{Pl-Holant}(\neq_2\mid f^{\frac{\pi}{2}}_1, f^{\frac{\pi}{2}}_2)$ is \#P-hard, 
and hence $\operatorname{Pl-Holant}(\neq_2\mid f)$ is \#P-hard. 

Otherwise, the 4 nonzero entries
 $\{a, b, x, y\}$ satisfy (\ref{matchgate-caseinLm4-6}). 
If  $(\frac{b}{a})^8 =1$, i.e.,
$b = a \sqrt{\ii}^{\beta}$
for some $\gamma \in \mathbb{N}$, then
$x = \pm a = a {\ii}^{\alpha}$,
and $y = \pm b = a \sqrt{\ii}^{\beta + 4\delta}$ for some $\alpha, \delta
\in \mathbb{N}$. 
It follows that $\{a, b, x, y\}$ satisfies (\rmnum{2}),
a contradiction.

So $(\frac{b}{a})^8 \neq 1$, and we can apply reduction~(\ref{>})
of Lemma~\ref{csp}.
By the reduction~(\ref{>}), 
we have  $\operatorname{\#CSP}(g_{1_f}, g_{2_f})\leqslant_T \operatorname{Pl-Holant}(\neq_2\mid f).$
Moreover, since  $\{a, b, x, y\}$ does not belong to case (\rmnum{1}) or case (\rmnum{2}), we have $\{g_{1_f}, g_{2_f}\} \nsubseteq \mathscr{P}$ or $\mathscr{A}$. By Theorem \ref{cspdic}, $\operatorname{\#CSP}(g_{1_f}, g_{2_f}) $ is \#P-hard.  
Therefore, we have  $\operatorname{Pl-Holant}(\neq_2\mid f)$
is \#P-hard. \qed

\begin{corollary} \label{inner-corollary}
Let $f$ be a 4-ary signature of the form {\rm (\ref{innermatrix})},
where $(a, x) \neq (0, 0)$ and $(b, y) \neq (0, 0)$.
If $|ax| \not =  |by|$ then
$\operatorname{Pl-Holant}(\neq_2\mid f)$
is \#P-hard.
\end{corollary}

\section{Case \Rmnum{3}: ${\sf N} =2$ with No Zero Pair or
${\sf N} =1$ with Zero in an Outer Pair}\label{seconezero}
If there are exactly two zeros ${\sf N} =2$ with no zero pair,
then the two zeros 
are in different pairs, at least one of them must be in an outer pair.
So in Case \Rmnum{3} there is a zero  in an outer pair
regardless ${\sf N} =1$ or ${\sf N} =2$.
 By rotational symmetry, we may assume $a=0$, 
and we prove this case in Theorem~\ref{onezero}. 
We first give the following lemma.

\begin{lemma}\label{fullrank}
Let $f$ be a 4-ary signature with the signature matrix
$M(f)=\left[\begin{smallmatrix}
0 & 0& 0& 0\\
0 & b& c& 0\\
0 & z& y& 0\\
0 & 0& 0& 0\\
\end{smallmatrix}\right]$, where $\det M_{\rm{In}}(f)=by-cz\neq 0$.
Let $g$ be a 4-ary signature with the signature matrix
$M(g)=\left[\begin{smallmatrix}
0 & 0& 0& 0\\
0 & 0& 1& 0\\
0 & 1& 0& 0\\
0 & 0& 0& 0\\
\end{smallmatrix}\right].$
Then for any signature set $\mathcal{F}$ containing $f$, we have $$\operatorname{Pl-Holant}(\neq_2\mid \mathcal{F}\cup \{g\})\leqslant_{T}\operatorname{Pl-Holant}(\neq_2\mid \mathcal{F}).$$
\end{lemma}


{\bf Proof.}
We construct a series of gadgets $f_{s}$ by a chain of $s$ copies of $f$ linked by double {\sc Disequality} $N$. 
$f_{s}$ has the signature matrix
\[M(f_s)=M(f)(NM(f))^{s-1}=N(NM(f))^{s}=
N\left [ \begin{array}{cccc}
0 & \bf{0} & 0 \\
\mathbf{0} & {\left [ \begin{array}{cc}  z & y \\ b & c \end{array} \right ]^s} & \mathbf{0} \\
0 & \bf{0} & 0
\end{array} \right ].\]
The inner matrix of
$NM(f)$ is $N_{\text{In}}M_{\text{In}}(f) = \left[\begin{smallmatrix}
z & y \\ b & c \end{smallmatrix}\right]$. 
Suppose its spectral decomposition is $Q^{-1}\Lambda Q$, where $\Lambda=
[\begin{smallmatrix}
\lambda_1 & \mu\\
0 & \lambda_2 \\
\end{smallmatrix}]$ is the Jordan Canonical Form.
Note that  $\lambda_1 \lambda_2=\det \Lambda=\det (N_{\text{In}}M_{\text{In}}(f))\neq 0$. 
We have $M(f_s)=NP^{-1} \Lambda_s P$, where
\[P=\left [ \begin{array}{ccc}
1 & \bf{0} & 0 \\
\mathbf{0} & Q & \mathbf{0} \\
0 & \bf{0} & 1
\end{array} \right ] ~~~~\mbox{and}~~~~
\Lambda_s=
\left [ \begin{array}{cccc}
0 & \bf{0} & 0 \\
\mathbf{0} & {\left [ \begin{array}{cc}  \lambda_1 & \mu \\ 0 & \lambda_2 \end{array} \right ]^s} & \mathbf{0} \\
0 & \bf{0} & 0 
\end{array} \right ].\]
\begin{enumerate}
\item Suppose $\mu=0$, and $\frac{\lambda_2}{\lambda_1}$ is a root of unity,
with $(\frac{\lambda_2}{\lambda_1})^n=1$. Then 
$\Lambda_n=
\left [ \begin{smallmatrix}
0 & 0 & 0 & 0 \\
0 & \lambda_1^n & 0 & 0\\
0 & 0 & \lambda_2^n & 0\\
0 & 0 & 0 & 0 
\end{smallmatrix} \right ]=
\left [ \begin{smallmatrix}
0 & 0 & 0 & 0 \\
0 & \lambda_1^n & 0 & 0\\
0 & 0 & \lambda_1^n & 0\\
0 & 0 & 0 & 0 
\end{smallmatrix} \right ]$, and
$M(f_n)=
\left [ \begin{smallmatrix}
0 & 0 & 0 & 0 \\
0 & 0 & \lambda_1^n & 0\\
0 & \lambda_1^n & 0 & 0\\
0 & 0 & 0 & 0 
\end{smallmatrix} \right ]=
\lambda_1^n\left [ \begin{smallmatrix}
0 & 0 & 0 & 0 \\
0 & 0 & 1 & 0\\
0 & 1 & 0 & 0\\
0 & 0 & 0 & 0 
\end{smallmatrix} \right ].$ After normalization, we can realize the signature $g$.

\item Suppose $\mu=0$, and $\frac{\lambda_2}{\lambda_1}$ is not a root of unity. The matrix $\Lambda_s=\left [ \begin{smallmatrix}
0 & 0 & 0 & 0 \\
0 & \lambda_1^s & 0 & 0\\
0 & 0 & \lambda_2^s & 0\\
0 & 0 & 0 & 0 
\end{smallmatrix} \right ]$ has a good form for interpolation. 
Suppose $g$ appears $m$ times in an instance $\Omega$ of $\PlHolant(\neq_2 \mid \mathcal{F}\cup \{g\})$.
Replace each appearance of $g$ by a copy of the gadget $f_{s}$ to get an instance $\Omega_{s}$ of $\operatorname{Pl-Holant}(\neq_2\mid \mathcal{F} \cup \{f_{s}\})$, which is also an instance of $\operatorname{Pl-Holant}(\neq_2\mid \mathcal{F})$.
We can treat each of the $m$ appearances of $f_s$ as a new gadget composed of four functions in sequence
$N$, $P^{-1}$, $\Lambda_s$ and $P$, and denote this new instance by $\Omega^\prime_s$.
We divide $\Omega'_{s}$ into two parts.
One part consists of $m$ signatures $\Lambda_{s}^{\otimes{m}}$. Here $\Lambda_{s}^{\otimes{m}}$ is  expressed as a column vector.
The other part is the rest of $\Omega'_{s}$ and its signature is represented by $A$ which is a tensor expressed as a row vector. 
Then the Holant value of $\Omega'_{s}$ is the dot product $\langle A, \Lambda_{s}^{\otimes{m}}\rangle$, which is a summation over $4m$ bits. That is, the value of the $4m$ edges connecting the two parts. 
We can stratify all $0, 1$ assignments of these $4m$ bits having a nonzero evaluation of a term in Pl-Holant$_{\Omega'_{s}}$ into the following categories:
\begin{itemize}
\item There are $i$ many copies of $\Lambda_{s}$ receiving inputs $0110$;
\item There are $j$ many copies of $\Lambda_{s}$ receiving inputs $1001$;
\end{itemize}
where $i+j=m$.

For any assignment in the category with parameter $(i,j)$, the evaluation of
$\Lambda_{s}^{\otimes m}$ is clearly $\lambda^{si}_1\lambda^{sj}_2=\lambda_1^{sm}\left(\frac{\lambda_2}{\lambda_1}\right)^{sj}$. 
Let $a_{ij}$ be the summation of values of the part $A$ over all assignments in the category $(i, j)$. Note that $a_{ij}$ is independent from the value of $s$ since we view the gadget $\Lambda_{s}$ as a block.
Since $i+j=m$, we can denote $a_{ij}$ by $a_j$. Then we rewrite the dot product summation and get 
$$\PlHolant_{\Omega_{s}}=\PlHolant_{\Omega'_{s}}=\langle A, \Lambda_{s}^{\otimes{m}}\rangle=
\lambda_1^{sm} \sum_{0\leqslant j \leqslant m}a_j\left(\frac{\lambda_2}{\lambda_1}\right)^{sj}.$$
Note that $M(g)=NP^{-1}(NM(g))P$, where $NM(g)=\left[ \begin{smallmatrix}
0 & 0 & 0 & 0 \\
0 & 1 & 0 & 0\\
0 & 0 & 1 & 0\\
0 & 0 & 0 & 0 
\end{smallmatrix} \right ]$.
Similarly, divide $\Omega$ into two parts. Under this stratification, 
we have 
$$\PlHolant_{\Omega}=\langle A, (NM(g))^{\otimes{m}}
\rangle=\sum_{0\leqslant j \leqslant m}a_j.$$
Since $\frac{\lambda_2}{\lambda_1}$ is not a root of unity, 
the Vandermonde coefficient matrix 
\[
\left[\begin{matrix}
\rho^0 & \rho^1 & \cdots & \rho^m\\
\rho^0 & \rho^2 &  \cdots & \rho^{2m}\\
\vdots & \vdots & \vdots & \vdots \\
\rho^0 & \rho^{m+1}  &  \cdots & \rho^{(m+1)m}
\end{matrix}
\right],
\]
has full rank,
where $\rho = \frac{\lambda_2}{\lambda_1}$.
 Hence, by oracle querying the values of $\PlHolant_{\Omega_{s}}$, we can solve for $a_j$, and thus
 obtain the value of $\PlHolant_{\Omega}$ in polynomial time.

\item Suppose $\mu=1$, and $\lambda_1=\lambda_2$ denoted by $\lambda$. Then $\Lambda_s=\left [ \begin{smallmatrix}
0 & 0 & 0 & 0 \\
0 & \lambda^s & s\lambda^{s-1} & 0\\
0 & 0 & \lambda^s & 0\\
0 & 0 & 0 & 0 
\end{smallmatrix} \right ].$ We use this form to give a polynomial interpolation. As in the case above, we can stratify the assignments of $\Lambda_{s}^{\otimes{m}}$ of these $4m$ bits having a nonzero evaluation of a term in Pl-Holant$_{\Omega'_{s}}$ into the following categories:
\begin{itemize}
\item There are $i$ many copies of $\Lambda_s$ receiving inputs $0110$ or $1001$;
\item There are $j$ many copies of $\Lambda_s$ receiving inputs $0101$;
\end{itemize}
where $i+j=m$.

For any assignment in the category with parameter $(i,j)$, the evaluation of
$\Lambda_{s}^{\otimes m}$ is clearly $\lambda^{si}(s\lambda^{s-1})^{j}=\lambda^{sm}(\frac{s}{\lambda})^j$. 
Let $a_{ij}$ be the summation of values of the part $A$ over all assignments in the category $(i, j)$. $a_{ij}$ is independent from $s$. Since $i+j=m$, we can denote $a_{ij}$ by $a_j$. Then, we rewrite the dot product summation and get 
$$\PlHolant_{\Omega_{s}}=\PlHolant_{\Omega'_{s}}=\langle A, \Lambda_{s}^{\otimes{m}}\rangle=\lambda^{sm}\sum_{0\leqslant j \leqslant m}a_j\left(\frac{s}{\lambda}\right)^j,$$
for $s \ge 1$.
We consider this as a linear system for $1 \le s \le m+1$.
Similarly, divide $\Omega$ into two parts. 
Under this stratification, 
we have 
$$\PlHolant_{\Omega}=\langle A, (NM(g))^{\otimes{m}}
\rangle=a_0.$$
The Vandermonde coefficient matrix 
\[
\left[\begin{matrix}
\rho_1^0 & \rho_1^1 & \cdots & \rho_1^m\\
\rho_2^0 & \rho_2^1 &  \cdots & \rho_2^{m}\\
\vdots & \vdots & \vdots & \vdots \\
\rho_{m+1}^0 & \rho_{m+1}^1  &  \cdots & \rho_{m}
\end{matrix}
\right],
\]
has full rank, where 
$\rho_s = s/\lambda$ are all distinct.
Hence, we can solve $a_0$ in polynomial time and it is the value of $\PlHolant_{\Omega}$. 
\end{enumerate}
Therefore, we have $\operatorname{Pl-Holant}(\neq_2\mid \mathcal{F}\cup \{g\})\leqslant_{T}\operatorname{Pl-Holant}(\neq_2\mid \mathcal{F}).$ \qed

Theorem~\ref{onezero} gives a classification for Case III.

\begin{theorem}\label{onezero}
Let $f$ be a 4-ary signature with the signature matrix
\begin{center}
$M(f)=\left[\begin{matrix}
0 & 0& 0& 0\\
0 & b& c& 0\\
0 & z& y& 0\\
x & 0& 0& 0\\
\end{matrix}\right]$, 
\end{center} 
where $x\neq 0$ and there is at most one number in $\{b, c, y, z\}$ that is $ 0 $.
Then $\operatorname{Pl-Holant}(\neq_2\mid f)$
is \#P-hard unless
$f \in \mathscr{M}$,
in which case the problem is tractable.
\end{theorem}

{\bf Proof.} Tractability follows from Theorem \ref{mtractable}. 

Suppose $f  \notin \mathscr{M}$.  By Lemma \ref{matchgate4}, $\det M_{\text{{In}}}(f) \neq \det M_{\text{Out}}(f) = 0$, that is $\det\left[\begin{smallmatrix}
b& c\\
z& y\\
\end{smallmatrix}\right]=by-cz\neq 0$.
Note that $M_{x_1x_2, x_4x_3}(f)=\left[\begin{smallmatrix}
0 & 0& 0& 0\\
0 & b& c& 0\\
0 & z& y& 0\\
x & 0& 0& 0\\
\end{smallmatrix}\right]$,  $M_{x_3x_4, x_2x_1}(f)=\left[\begin{smallmatrix}
0 & 0& 0& x\\
0 & y& c& 0\\
0 & z& b& 0\\
0 & 0& 0& 0\\
\end{smallmatrix}\right]$, and $M_{x_2x_3, x_1x_4}(f)=\left[\begin{smallmatrix}
0 & 0& 0& y\\
0 & 0& z& 0\\
0 & c& x& 0\\
b & 0& 0& 0\\
\end{smallmatrix}\right].$
Connect variables $x_4, x_3$ of a copy of signature $f$ with variables $x_3, x_4$ of another copy of signature $f$ both using $(\neq_2)$. We get a signature $f_1$ with the signature matrix
$$M(f_1)=M_{x_1x_2, x_4x_3}(f)NM_{x_3x_4, x_2x_1}(f)=\left[\begin{matrix}
0 & 0& 0& 0\\
0 & b_1& c_1& 0\\
0 & z_1& y_1& 0\\
0 & 0& 0& 0\\
\end{matrix}\right],$$
where $\left[\begin{smallmatrix}
 b_1& c_1\\
 z_1& y_1\\
\end{smallmatrix}\right]= \left[\begin{smallmatrix}
 b& c\\
 z& y\\
\end{smallmatrix}\right]\cdot \left[\begin{smallmatrix}
 z& b\\
 y& c\\
\end{smallmatrix}\right].$ 
This $f_1$  has the form in Lemma~\ref{fullrank}.
Here, $\det\left[\begin{smallmatrix}
 b_1& c_1\\
 z_1& y_1\\
\end{smallmatrix}\right]=-(by-cz)^2\neq 0$. 
By Lemma~\ref{fullrank}, we have $$\operatorname{Pl-Holant}(\neq_2\mid f, g)\leqslant_{T}\operatorname{Pl-Holant}(\neq_2\mid f, f_1),$$
where $g$ has the signature matrix $M(g)=\left[\begin{smallmatrix}
0 & 0& 0& 0\\
0 & 0& 1& 0\\
0 & 1& 0& 0\\
0 & 0& 0& 0\\
\end{smallmatrix}\right].$ 
\begin{itemize}     
\item If $bcyz\neq 0$, connect variables $x_1, x_4$ of signature $f$ with variables $x_1, x_2$ of signature $g$ both using $(\neq_2)$. 
We get a signature $f_2$ with the signature matrix
$$M(f_2)=M_{x_2x_3, x_1x_4}(f)NM_{x_1x_2, x_4x_3}(g)=\left[\begin{matrix}
0 & 0& 0& 0\\
0 & 0& z& 0\\
0 & c& x& 0\\
0 & 0& 0& 0\\
\end{matrix}\right].$$
\item Otherwise, connect variables $x_4, x_3$ of signature $f$ with variables $x_1, x_2$ of signature $g$ both using $(\neq_2)$. 
We get a signature $f_2$ with the signature matrix
$$M(f_2)=M_{x_1x_2, x_4x_3}(f)NM_{x_1x_2, x_4x_3}(g)=\left[\begin{matrix}
0 & 0& 0& 0\\
0 & b& c& 0\\
0 & z& y& 0\\
0 & 0& 0& 0\\
\end{matrix}\right],$$
and there is exactly one entry in $\{b, c, y, z\}$ that is zero.
\end{itemize}
In both cases, the support of $f_2$ has size $3$, which means $f_2 \notin \mathscr{P}, \mathscr{A} $ or $\mathscr{\widehat{M}}$. 
By Theorem~\ref{outer}, $\operatorname{Pl-Holant}(\neq_2\mid f_2)$ is \#P-hard. 
Since 
$$\operatorname{Pl-Holant}(\neq_2\mid f_2)\leqslant_{T}\operatorname{Pl-Holant}(\neq_2\mid f, g)\leqslant_{T}\operatorname{Pl-Holant}(\neq_2\mid f, f_1)\leqslant_{T}\operatorname{Pl-Holant}(\neq_2\mid f),$$ we have $\operatorname{Pl-Holant}(\neq_2\mid f)$ is \#P-hard. \qed


\section{Case \Rmnum{4}: ${\sf N} =1$ with Zero in the 
Inner Pair or ${\sf N} =0$}
By rotational symmetry, \emph{if} there is one zero in the inner pair, we may  
assume it is $c=0$, and $abxyz \not = 0$. We first consider the case 
that $x = \epsilon a, y=\epsilon b$ and $z=\epsilon c$, where $\epsilon=\pm 1$.
\begin{lemma}\label{twins}
Let $f$ be a 4-ary signature with the signature matrix
\[M(f)=\left[\begin{matrix}
0 & 0& 0& a\\
0 & b& c& 0\\
0 & \epsilon c&  \epsilon b& 0\\
 \epsilon a & 0& 0& 0\\
\end{matrix}\right],~~~~\mbox{ where }\epsilon = \pm 1 \mbox{ and } abc\neq 0.\]
Then $\operatorname{Pl-Holant}(\neq_2\mid f)$ is \#P-hard  if $f\notin \mathscr{M}$.
\end{lemma}

{\bf Proof.} 
If $\epsilon =-1$  we first transform the case to $\epsilon =1$
as follows. Connecting the variable $x_4$ with
  $x_3$ of $f$ using $(\neq_2)$ we get a binary signature $g_1$, where 
$$g_1=M_{x_1x_2, x_4x_3}(f)(0, 1, 1, 0)^T=(0, b+c, -(b+c), 0)^T.$$
Also connecting the variable $x_1$ with $x_2$ of $f$ using $(\neq_2)$ we get a binary signature $g_2$, where 
$$g_2=((0, 1, 1, 0)M_{x_1x_2, x_4x_3}(f))^T=(0, b-c, -(b-c),  0)^T.$$
Since $bc\neq 0$, $b+c$ and $b-c$ cannot be both zero. Without loss of generality, suppose $b+c\neq 0$. By normalization, we have $g_1=(0, 1, -1, 0)^T.$
Then, modifying $x_1=1$ of $f$ with $-1$ scaling we get a signature with the signature matrix $\left[\begin{smallmatrix}
0 & 0& 0& a\\
0 & b& c& 0\\
0 &  c&   b& 0\\
  a & 0& 0& 0\\
\end{smallmatrix}\right].$ Therefore, it suffices to show  \#P-hardness for the case $\epsilon=1$.

Since $f \notin \mathscr{M}$, by Lemma \ref{matchgate4}, $c^2-b^2\neq a^2$. We prove \#P-hardness in the following three Cases depending on the values of $a, b$ and $c$.

{\bf Case 1:} Either
$c^2-b^2\neq 0$ and $|c+b|\neq|c-b|$, or $c^2-a^2\neq 0$ and $|c+a|\neq|c-a|$.
By rotational symmetry, we may assume $c^2-b^2\neq 0$ and $|c+b|\neq|c-b|$. 
We may normalize $a=1$ and assume
 $M(f)=\left[\begin{smallmatrix}
0 & 0& 0& 1\\
0 & b& c& 0\\
0 & c& b& 0\\
1 & 0& 0& 0\\
\end{smallmatrix}\right]$, where $ c^2-b^2 \neq 0$ or $1$.

We construct a series of gadgets $f_{s}$ by a chain of $s$ copies of $f$ linked by double {\sc Disequality} $N$. 
$f_{s}$ has the signature matrix 
\[M(f_s)=M(f)(NM(f))^{s-1}=N(NM(f))^{s}=
N\left [ \begin{array}{cccc}
1 & \bf{0} & 0 \\
\mathbf{0} & {\left [ \begin{array}{cc}  c & b \\ b & c \end{array} \right ]^s} & \mathbf{0} \\
0 & \bf{0} & 1
\end{array} \right ].\]
We diagonalize $\left [ \begin{smallmatrix}  c & b \\ b & c \end{smallmatrix} \right ]^s $ using
$H=\frac{1}{\sqrt{2}}
\left[\begin{smallmatrix}
1 & 1 \\
 1 & -1
\end{smallmatrix}\right]
$ (note that $H^{-1} = H$),
 and get $M(f_s)=NP \Lambda_s P$, where
\[P=\left [ \begin{array}{ccc}
1 & \bf{0} & 0 \\
\mathbf{0} & H & \mathbf{0} \\
0 & \bf{0} & 1
\end{array} \right ], ~~~~\mbox{and}~~~~
\Lambda_s=
\left [ \begin{array}{cccc}
1 & 0 & 0 & 0 \\
0 & (c+b)^s & 0 & 0 \\
0 & 0 & (c-b)^s & 0 \\
0 & 0 & 0 & 1
\end{array} \right ].\]
The signature matrix $\Lambda_s$ has a good form for polynomial interpolation. 
In the following, we will
reduce $\PlHolant(\neq_2 \mid \hat{f})$
to $\PlHolant(\neq_2 \mid f)$,
for suitably chosen $M(\hat{f})=\left[\begin{smallmatrix}
0 & 0& 0& 1\\
0 & \hat b& \hat c& 0\\
0 & \hat c& \hat b& 0\\
1 & 0& 0& 0\\
\end{smallmatrix}\right]$,
and use that to prove that $\PlHolant(\neq_2 \mid f)$
is \#P-hard.

Suppose $\hat{f}$ appears $m$ times in an instance $\hat\Omega$ of $\PlHolant(\neq_2 \mid \hat f)$.
We replace each appearance of $\hat f$ by a copy of
the gadget $f_s$ to get an instance $\Omega_{s}$ 
of  $\operatorname{Pl-Holant}(\neq_2\mid f)$.
We can treat each of the $m$ appearances of $f_s$ as a new gadget composed of four functions in sequence
$N$, $P$, $\Lambda_s$ and $P$, and denote this new instance by $\Omega^\prime_s$.
We divide $\Omega'_{s}$ into two parts.
One part consists of $m$ occurrences of 
 $\Lambda_{s}$,
which is  $\Lambda_{s}^{\otimes{m}}$,
and is written as a column vector of dimension $2^{4m}$. 
The other part is the rest of $\Omega'_{s}$ and its signature is expressed by
a tensor  $A$, written as a row vector of dimension $2^{4m}$. 
Then the Holant value of $\Omega'_{s}$ is the dot product $\langle A, \Lambda_{s}^{\otimes{m}}\rangle$, which is a summation over $4m$ bits, i.e., the values of the $4m$ edges connecting the two parts. 
We can stratify all $0, 1$ assignments of these $4m$ bits having a nonzero evaluation of a term in Pl-Holant$_{\Omega'_{s}}$ into the following categories:


\begin{itemize}
\item
There are $i$ many copies of $\Lambda_{s}$ receiving inputs $0000$ or $1111$;
\item
There are $j$ many copies of $\Lambda_{s}$ receiving inputs $0110$;
\item
There are $k$ many copies of $\Lambda_{s}$ receiving inputs $1001$;
\end{itemize}
where $i+j+k=m$.

For any assignment in the category with parameter $(i, j, k)$, the evaluation of
$\Lambda_{s}^{\otimes m}$ is clearly $(c+b)^{sj}(c-b)^{sk}$. 
Let $a_{ijk}$ be the summation of values of the part $A$ over all assignments in the category $(i, j, k)$. Note that $a_{ijk}$ is independent of the value of $s$. Since $i+j+k=m$, we can denote $a_{ijk}$ by $a_{jk}$. Then we rewrite the dot product summation and get 
$$\PlHolant_{\Omega_{s}}=\PlHolant_{\Omega'_{s}}=\langle A, \Lambda_{s}^{\otimes{m}}\rangle=\sum_{0\leqslant j+k \leqslant m}a_{jk}(c+b)^{sj}(c-b)^{sk}.$$
Under this stratification, correspondingly we can define $\hat\Omega'$ and $\hat \Lambda$ from $\hat\Omega$. Then we have 
$$\PlHolant_{\hat\Omega}=\PlHolant_{\hat\Omega'}=\langle A, \hat\Lambda^{\otimes{m}}\rangle=\sum_{0\leqslant j+k \leqslant m}a_{jk}(\hat c+\hat b)^j(\hat c-\hat b)^k,$$
where the same set of values $a_{jk}$ appear.
Let $\phi =\hat c + \hat b$ and $\psi=\hat c - \hat b$. If we can obtain the value of $p(\phi, \psi)=\sum \limits_{0\leqslant j+k \leqslant m}a_{jk}\phi^j \psi^k$ 
from oracle queries to $\PlHolant_{\Omega_{s}}$ (for $s \ge 1$)
in polynomial time, then we will have proved
 $$\PlHolant(\neq_2 \mid \hat f) \leqslant_T \PlHolant(\neq_2 \mid f).$$ 
Let $\alpha=c+b$ and  $\beta=c-b$. Since $c^2-b^2\neq 0$ or $1$, we have $\alpha \neq 0$, $\beta \neq 0$ and $\alpha \beta \neq 1$. Also, by assumption $|c+b|\neq|c-b|$, we have $|\alpha|\neq|\beta|$.  
Define $L=\{(j,k) \in \mathbb{Z}^2 \mid \alpha^j \beta^k=1 \}$.
This is a sublattice of $\mathbb{Z}^2$. Every lattice has a basis.
There are 3 cases depending on the rank of $L$.
\begin{itemize}

  \item $L=\{(0,0)\}$. All $\alpha^j \beta^k$ are distinct. It is an interpolation reduction in full power. That is, we can interpolate $p(\phi, \psi)$ for any $\phi$ and $\psi$ in polynomial time. Let $\phi=4$ and $\psi=0$, that is $\hat b=2$ and $\hat c=2$, and hence $M(\hat f)=\left [ \begin{smallmatrix}
0 & 0 & 0 & 1 \\
0 & 2 & 2 & 0 \\
0 & 2 & 2 & 0 \\
1 & 0 & 0 & 0
\end{smallmatrix} \right ].$
That is, $\hat{f}$ is non-singular redundant. By Theorem~\ref{redundant}, Pl-Holant$(\neq_2 \mid \hat f)$ is \#P-hard, and hence Pl-Holant$(\neq_2 \mid  f)$ is \#P-hard.
\item $L$ contains two independent vectors $(j_1,k_1)$ and $(j_2,k_2)$
 over $\mathbb{Q}$.
Then the nonzero vectors  $j_2 (j_1,k_1)-j_1 (j_2,k_2)=(0,j_2k_1-j_1k_2)$ and
$k_2 (j_1,k_1)-k_1 (j_2,k_2)=(k_2j_1-k_1 j_2,0)$ are in $L$. Hence, both $\alpha$ and $\beta$ are roots of unity. This implies that $|\alpha|=|\beta|=1$,
a contradiction.
\item 
$L=\{(ns, nt) \mid n \in \mathbb{Z}\}$, where $s,t \in \mathbb{Z}$ and
 $(s,t) \neq (0,0)$. Without loss of generality, we may assume $t\geqslant 0$, and  $s> 0$ when $t=0$.
Also, we have $s+t \neq 0$, otherwise $|{\alpha}|=
|{\beta}|$, a  contradiction. 
By Lemma~\ref{interpolation}, for any numbers $\phi$ and $\psi$ satisfying $\phi^s\psi^t=1$, we can obtain $p(\phi, \psi)$ in polynomial time. 
Since $\phi=\hat c+\hat b$ and $\psi=\hat c-\hat b$, we have $\hat b=\frac{\phi-\psi}{2}$ and $\hat c=\frac{\phi+\psi}{2}$. That is $M(\hat f)=\left [ \begin{smallmatrix}
0 & 0 & 0 & 1 \\
0 & \frac{\phi-\psi}{2} &\frac{\phi+\psi}{2}  & 0 \\
0 & \frac{\phi+\psi}{2} &\frac{\phi-\psi}{2} & 0 \\
1 & 0 & 0 & 0
\end{smallmatrix} \right ].$ There are three cases depending on the values of $s$ and $t$.
\begin{itemize}
  \item
  $s\geqslant 0$ and $s+t\geqslant 2$. 
  Consider the function $q(x)=(2-x)^sx^t-1$. Since $s\geqslant 0$ and $t\geqslant 0$, $q(x)$ is a polynomial. Clearly, $1$ is a root and $0$ is not a root.
%
%
 If $q(x)$ has no other roots, then for some constant $\lambda \neq 0$,
 $$q(x)
=\lambda(x-1)^{s+t}=(-1)^{s+t}\lambda((2-x)-1)^{s+t}.$$ 
(In fact by comparing leading coefficients,  $\lambda  = (-1)^s$.)
  Notice that $x^t|q(x)+1$, while $x^t\nmid\lambda(x-1)^{s+t}+1$ for $t\geqslant 2$. 
  Also, notice that $(2-x)^s|q(x)+1$, while $(2-x)^s\nmid (-1)^{s+t}\lambda((2-x)-1)^{s+t}$ for $s\geqslant 2$. 
  Hence, $t=s=1$, which means $\alpha \beta=1$. Contradiction.
  
  Therefore, $q(x)$ has a root $x_0$, with $x_0 \neq 1$ or $0$. Let $\psi=x_0$ and $\phi=2-x_0$. Then $\phi^s\psi^t=1$ and $M(\hat f)=\left [ \begin{smallmatrix}
0 & 0 & 0 & 1 \\
0 & 1-x_0 &1  & 0 \\
0 & 1 &1-x_0 & 0 \\
1 & 0 & 0 & 0
\end{smallmatrix} \right ].$ 
Note that $M_{x_2x_3,x_1x_4}(\hat f)=\left [ \begin{smallmatrix}
0 & 0 & 0 & 1-x_0 \\
0 & 1 &1  & 0 \\
0 & 1 &1 & 0 \\
1-x_0 & 0 & 0 & 0
\end{smallmatrix} \right ]$. Since $1-x_0\neq 0$, $\hat f$ is non-singular 
redundant. 
By Theorem \ref{redundant}, $\PlHolant(\neq_2 \mid \hat f)$ is \#P-hard and hence $\plholant{\neq_2}{ f}$ is \#P-hard.
\item
$s<0$ and $t>0$. (Note that $s<0$ implies $t>0$.) 
Consider the function $q(x)=x^t-(2-x)^{-s}$. Since $t>0$ and $-s>0$, it is a polynomial. Clearly, $1$ is a root, but  neither $0$ nor $2$ is
a root. 
Since $t+s\neq 0$, the highest order term of $q(x)$ is either $x^t$ or $-(-x)^{-s}$, which means the coefficient of the highest order term is $\pm 1$. 
While the constant term of $q(x)$ is $-2^{-s}\neq \pm1$. 
Hence, $q(x)$ cannot be of the form $\lambda(x-1)^{\max(t, -s)}$ for some constant $\lambda \neq 0$. 
Moreover, since $t+s\neq 0$, $\max(t, -s)\geqslant 2$, which means $q(x)$ has a root $x_0$, where $x_0 \neq 0, 1, 2$. 
Dividing by the nonzero term $(2-x_0)^{-s}$ we have
$(2-x_0)^s x_0^t =1$. Now we
 let $\psi=x_0$ and $\phi=2-x_0$, 
and we have $\plholant{\neq_2}{ f}$ is \#P-hard by the same proof as above.
\item
$s\geqslant 0$ and $s+t=1$. In this case, we have $(s, t) = (0,1)$ or $(1,0)$
since $t\geqslant 0$.
\begin{itemize}
  \item $(s, t) = (1,0)$. Let $\phi=1$ and $\psi=\frac{1}{2}$. Then we have $\phi^1\psi^0=1$ and $M(\hat f)=\left [ \begin{smallmatrix}
0 & 0 & 0 & 1 \\
0 & \frac{1}{4} &\frac{3}{4}  & 0 \\
0 & \frac{3}{4} &\frac{1}{4} & 0 \\
1 & 0 & 0 & 0
\end{smallmatrix} \right ]$. Let $M(f')=4M_{x_2x_3, x_1x_4}(\hat f)=\left [ \begin{smallmatrix}
0 & 0 & 0 & 1 \\
0 & 4 &3  & 0 \\
0 & 3 &4 & 0 \\
1 & 0 & 0 & 0
\end{smallmatrix} \right ].$ Clearly, $\PlHolant(\neq_2 \mid f')\leqslant_T \plholant{\neq_2}{f}$. 
For $M(f')$, correspondingly we define $\alpha'=3+4=7$ and $\beta'=3-4=-1$. Obviously, $\alpha'\neq 0$, $\beta' \neq 0$, $\alpha'\beta'\neq 1$, and $|\alpha'|\neq|\beta'|$. Let $L'=\{(j,k) \in \mathbb{Z}^2 \mid \alpha'^j \beta'^k=1 \}$. 
Then we have  $L'=\{(ns',nt')\mid n\in \mathbb{Z}\}$, where $s'=0$ and $t'=2$. Therefore, $s'\geqslant 0$ and $s'+t'\geqslant 2$. As we have showed 
above, we have $\plholant{\neq_2}{ f'}$ is \#P-hard, and hence $\plholant{\neq_2}{ f}$ is \#P-hard.
  \item $(s, t) = (0,1)$. Let $\phi=3$ and $\psi=1$. Then we have $\phi^0\psi^1=1$ and $M(\hat f)=\left [ \begin{smallmatrix}
0 & 0 & 0 & 1 \\
0 & 1 &2  & 0 \\
0 & 2 &1 & 0 \\
1 & 0 & 0 & 0
\end{smallmatrix} \right ]$. 
By Theorem \ref{tutte}, $\PlHolant(\neq_2 \mid \hat f)$ is \#P-hard, and hence $\plholant{\neq_2}{f}$ is \#P-hard.
\end{itemize}
\end{itemize}
\end{itemize}


{\bf Case 2:} If $c^2-b^2\neq 0$ and $|c+b|=|c-b|$, or  $c^2-a^2\neq 0$ and $|c+a|=|c-a|$.
By rotational symmetry, we may assume $c^2-b^2\neq 0$ and $|c+b|=|c-b|$.
Normalizing $f$ by assuming $c=1$, we have $M(f)=\left[\begin{smallmatrix}
0 & 0& 0& a\\
0 & b& 1& 0\\
0 & 1& b& 0\\
a & 0& 0& 0\\
\end{smallmatrix}\right]$, where $ 1^2-b^2 \neq 0$ and $1^2-b^2\neq a^2$ due to $f \notin \mathscr{M}$. Since $|1+b|=|1-b|$, $b$ is a pure imaginary number
(as $b \not =0$). 

 Connect variables $x_4$, $x_3$ of a copy of signature $f$ with variables $x_1$, $x_2$ of another copy of signature $f$ both using $(\neq_2)$.
We get a signature $f_1$ with the signature matrix $$M(f_1)=M_{x_1x_2, x_4 x_3}(f)NM_{x_1x_2, x_4 x_3}(f)=
\left[\begin{matrix}
0 & 0& 0& a^2\\
0 & 2b& b^2+1& 0\\
0 & b^2+1& 2b& 0\\
a^2 & 0& 0& 0\\
\end{matrix}\right].$$


\begin{enumerate}

  \item[\bf a.] If $c^2-a^2=0$, that is $a^2=1$, and then $M(f_1)=\left[\begin{smallmatrix}
0 & 0& 0& 1\\
0 & 2b& b^2+1& 0\\
0 & b^2+1& 2b& 0\\
1 & 0& 0& 0\\
\end{smallmatrix}\right].$ Since $b^2<0$, we have $(b^2+1)^2-(2b)^2=(b^2-1)^2>1=(a^2)^2$, which means $f_1 \notin \mathscr{M}$.

\begin{itemize}
\item If $b^2=-1$, then $M(f_1)=\left[\begin{smallmatrix}
0 & 0& 0& 1\\
0 & \pm2\ii& 0& 0\\
0 & 0& \pm2\ii& 0\\
1 & 0& 0& 0\\
\end{smallmatrix}\right]$. By Corollary~\ref{inner-corollary}, $\PlHolant(\neq_2 \mid f_1)$ is \#P-hard, and hence $\PlHolant(\neq_2 \mid f)$ is \#P-hard.

\item If $b^2=-2$, then $M(f_1)=\left[\begin{smallmatrix}
0 & 0& 0& 1\\
0 & \pm2\sqrt{2}\ii& -1& 0\\
0 & -1& \pm2\sqrt{2}\ii& 0\\
1 & 0& 0& 0\\
\end{smallmatrix}\right].$ Connect two copies of $f_1$,
and we have a signature $f_2$ with the signature matrix $$M(f_2)=M_{x_1x_2, x_4 x_3}(f_1)NM_{x_1x_2, x_4 x_3}(f_1)=
\left[\begin{matrix}
0 & 0& 0& 1\\
0 & \mp4\sqrt{2}\ii& -7& 0\\
0 & -7& \mp4\sqrt{2}\ii& 0\\
1 & 0& 0& 0\\
\end{matrix}\right].$$
It is easy to check that $f_2 \notin \mathscr{M}$, by Lemma~\ref{matchgate4}.
Then, $f_2$ belongs to Case 1. Therefore, $\PlHolant(\neq_2 \mid f_2)$ is \#P-hard, and hence $\PlHolant(\neq_2 \mid f)$ is \#P-hard.

\item If $b^2 \neq -1$ or $-2$, then $b^2+1 \neq \pm 1$ due to $b\neq 0$,
 hence $1^2-(b^2+1)^2\neq 0$.
Also, since $b^2+1$ is a real number 
and $b^2+1\neq 0$, we have 
$|(b^2+1)+1|\neq|(b^2+1)-1|$. 
Then $f_1$, which is not in $\mathscr{M}$ as shown above,
 has a signature matrix of the form $\left[\begin{smallmatrix}
0 & 0& 0& a_1\\
0 & b_1& c_1& 0\\
0 & c_1& b_1& 0\\
a_1 & 0& 0& 0\\
\end{smallmatrix}\right],$ where $a_1 = a^2 =1$, $b_1 = 2b$,
and $c_1 = b^2 +1$, and
 $a_1b_1c_1\neq0$, $c_1^2-a_1^2\neq 0$ and $|c_1+a_1|\neq |c_1-a_1|$.
That is, $f_1$ belongs to Case 1. Therefore, $\PlHolant(\neq_2 \mid f_1)$ is \#P-hard, and hence $\PlHolant(\neq_2 \mid f)$ is \#P-hard.
\end{itemize}
\item[\bf b.] If $c^2-a^2\neq 0$ and $|c+a|=|c-a|$,
i.e., $|1+a| = |1-a|$, then $a \neq 0$ is also a pure imaginary number.
 Connect variables $x_1$, $x_4$ of a copy of signature $f$ with  variables $x_2$, $x_3$ of another copy of signature $f$.
 We get a signature $f_3$ with the signature matrix $$M(f_3)=M_{x_2x_3, x_1 x_4}(f)NM_{x_2x_3, x_1 x_4}(f)=
\left[\begin{matrix}
0 & 0& 0& b^2\\
0 & 2a& a^2+1& 0\\
0 & a^2+1& 2a& 0\\
b^2 & 0& 0& 0\\
\end{matrix}\right].$$
Note that $f_3 \in \mathscr{M}$ implies $(a^2-1)^2=(b^2)^2$. Since $f \notin \mathscr{M}$, $1-a^2\neq b^2$. Hence, $f_3 \in \mathscr{M}$ implies $a^2-1=b^2$. Similarly, $f_1 \in \mathscr{M}$ implies $b^2-1=a^2$. Clearly, $f_1$ and $f_3$ 
cannot both be in $\mathscr{M}$. Without loss of generality, we may assume $f_3\notin \mathscr{M}$.
\begin{itemize}
\item If $a^2\neq-1$, then there are two subcases.
\begin{itemize}
  \item $(a^2+1)^2-(b^2)^2=0$. Since $a$ is a pure imaginary number, $|a^2+1+2a|=|a+1|^2=|a-1|^2=|a^2+1-2a|$. Then $f_3$ has the signature matrix of the form $\left[\begin{smallmatrix}
0 & 0& 0& a_3\\
0 & b_3& c_3& 0\\
0 & c_3& b_3& 0\\
a_3 & 0& 0& 0\\
\end{smallmatrix}\right],$ 
where $a_3b_3c_3\neq0$, $c_3^2-b_3^2
= (a^2-1)^2  \neq 0$ since $a$ is pure imaginary, $|c_3+b_3|= |c_3-b_3|$ and $c_3^2-a_3^2= 0$. That is, $f_3$ belongs to Case 2.a. Therefore, $\PlHolant(\neq_2 \mid f_3)$ is \#P-hard, and hence  $\PlHolant(\neq_2 \mid f)$ is \#P-hard.
\item $(a^2+1)^2-(b^2)^2\neq 0$. Since $a^2+1$ and $b^2$ are both nonzero real numbers due to $a$ and $b$ are both pure imaginary numbers, we have $|a^2+1+b^2|\neq|a^2+1-b^2|$. Then $f_3$ has the signature matrix of the form $\left[\begin{smallmatrix}
0 & 0& 0& a_3\\
0 & b_3& c_3& 0\\
0 & c_3& b_3& 0\\
a_3 & 0& 0& 0\\
\end{smallmatrix}\right],$ where $a_3b_3c_3\neq0$, $c_3^2-a_3^2 \neq 0$ and $|c_3+a_3|\neq |c_3-a_3|$. That is, $f_3$ belongs to Case 1. 
Therefore, $\PlHolant(\neq_2 \mid f_3)$ is \#P-hard, and hence  $\PlHolant(\neq_2 \mid f)$ is \#P-hard.
\end{itemize}
\item If $a^2=-1$ and $b^2\neq -2$, then $M(f_3)=
\left[\begin{smallmatrix}
0 & 0& 0& b^2\\
0 & 2a& 0& 0\\
0 & 0& 2a& 0\\
b^2 & 0& 0& 0\\
\end{smallmatrix}\right]$, where $|2a|=2\neq|b^2|$. By Corollary~\ref{inner-corollary}, $\PlHolant(\neq_2 \mid f_3)$ is  \#P-hard, and hence $\PlHolant(\neq_2 \mid f)$ is \#P-hard.
\item If $a^2=-1$ and $b^2=-2$, then $M(f_1)=\left[\begin{smallmatrix}
0 & 0& 0& -1\\
0 & \pm2\sqrt{2}\ii& -1& 0\\
0 & -1& \pm2\sqrt{2}\ii& 0\\
-1 & 0& 0& 0\\
\end{smallmatrix}\right].$ Note that $M_{x_2 x_3, x_1 x_4}(f_1)=\left[\begin{smallmatrix}
0 & 0& 0& \pm2\sqrt{2}\ii\\
0 & -1& -1& 0\\
0 & -1& -1& 0\\
\pm2\sqrt{2}\ii & 0& 0& 0\\
\end{smallmatrix}\right],$ which means  $f_1$ is non-singular redundant.
Therefore, we have $\PlHolant(\neq_2 \mid f_1)$ is \#P-hard, and hence  $\PlHolant(\neq_2 \mid f)$ is \#P-hard.
  
\end{itemize}
\item[\bf c.] If $c^2-a^2\neq 0$ and $|c+a|\neq |c-a|$. This is Case 1. Done.
\end{enumerate}

{\bf Case 3:} $c^2-b^2=0$ and $c^2-a^2=0$.
If $c=b$ or $c=a$, then $f$ is non-singular redundant, and hence $\PlHolant(\neq_2 \mid f)$ is \#P-hard. Otherwise, $a=b=-c$. By normalization, we have $M(f)=\left[\begin{smallmatrix}
0 & 0& 0& -1\\
0 & -1& 1& 0\\
0 & 1& -1& 0\\
-1 & 0& 0& 0\\
\end{smallmatrix}\right]$, and then $M(f_1)=\left[\begin{smallmatrix}
0 & 0& 0& 1\\
0 & -2& 2& 0\\
0 & 2& -2& 0\\
1 & 0& 0& 0\\
\end{smallmatrix}\right]$. Notice that $2^2-1^2\neq 0$ and $|2+1|\neq|2-1|$. That is, $f_1$ belongs to Case 1.
Therefore, $\PlHolant(\neq_2 \mid f_1)$ is \#P-hard, and hence $\PlHolant(\neq_2 \mid f)$ is \#P-hard. 

Case 1 to Case 3 cover all cases for $(a,b,c)$:
Suppose  $(a,b,c)$ does not satisfy Case 3. Then either 
$c^2-b^2\not =0$ or $c^2-a^2\not =0$.
If $c^2-b^2\not =0$, then
 either $|c+b| \not = |c-b|$ (Case 1)
or 
$|c+b| = |c-b|$ (Case 2).
Similarly if  $c^2-a^2\not =0$ it is either Case 1 or Case 2.
This completes the proof of the lemma.
\qed

\begin{lemma} \label{degenerate}
Let $f$ be a 4-ary signature with the signature matrix
\begin{center}
$M(f)=\left[\begin{matrix}
0 & 0& 0& a\\
0 & b& c& 0\\
0 & z& y& 0\\
x & 0& 0& 0\\
\end{matrix}\right]$,~~~~where~~ $abcxyz\neq 0$.
\end{center}
If $by-cz=0$ or $ax-cz=0$, then $\operatorname{Pl-Holant}(\neq_2\mid f)$ is \#P-hard.
\end{lemma}



{\bf Proof.}
By rotational symmetry, we assume $by-cz=0$. By normalization, we assume $b=1$, and then $y=cz$. That is,
$M_{x_1x_2, x_4 x_3}(f)=\left[\begin{smallmatrix}
0 & 0& 0& a\\
0 & 1& c& 0\\
0 & z& cz& 0\\
x & 0& 0& 0\\
\end{smallmatrix}\right]$. 

\begin{itemize}
\item If $1+c\neq 0$. Connecting the variables $x_4$ with $x_3$ of $f$ using $(\neq_2)$ we get a binary signature $g_1$, where $$g_1=M_{x_1x_2, x_4x_3}(f)(0, 1, 1, 0)^T=(0, 1+c, (1+c)z, 0)^T.$$  
Note that $g_1(x_1,x_2)$ can be normalized as $(0, z^{-1}, 1, 0)^T$. That is $g(x_2, x_1)=(0, 1, z^{-1}, 0)^T$.
Modifying $x_1=1$ of $f$ with $z^{-1}$ scaling
 we get a signature $f_1$ with  the signature matrix $\left[\begin{smallmatrix}
0 & 0& 0& a\\
0 & 1& c& 0\\
0 & 1& c& 0\\
x/z & 0& 0& 0\\
\end{smallmatrix}\right].$ 
Connecting the variable $x_1$ with $x_2$ of $f_1$ using $(\neq_2)$ we get a binary signature $g_2$, where $$g_2=((0, 1, 1, 0)M_{x_1x_2, x_4x_3}(f))^T=(0, 2, 2c, 0)^T,$$
and  $g_2(x_1,x_2)$ can be normalized to $g_2(x_2, x_1) = (0, 1, c^{-1}, 0)^T$.
Modifying $x_4=1$ of $f_1$ with $c^{-1}$ scaling we get a signature $f_2$ with  the signature matrix $\left[\begin{smallmatrix}
0 & 0& 0& a/c\\
0 & 1& 1& 0\\
0 & 1& 1& 0\\
x/z & 0& 0& 0\\
\end{smallmatrix}\right].$ It is non-singular redundant. By Lemma~\ref{redundant},  we have $\operatorname{Pl-Holant}(\neq_2\mid f_2)$ is \#P-hard, and hence $\operatorname{Pl-Holant}(\neq_2\mid f)$ is \#P-hard. 

\item If $1+z\neq 0$, then
connecting the variable $x_1$ with $x_2$ of $f$ using $(\neq_2)$ we get a binary signature $g'_1$, where $$g'_1=((0, 1, 1, 0)M_{x_1x_2, x_4x_3})^T=(0, 1+z, (1+z)c, 0)^T.$$
$g'_1(x_1, x_2)$ can be normalized to $(0, c^{-1}, 1, 0)^T$. 
By the same analysis as in the case $1+c\neq 0$, we  have $\operatorname{Pl-Holant}(\neq_2\mid f)$ is \#P-hard.

\item Otherwise, $1+c=0$ and $1+z=0$, that is $c=z=-1$. Then 
$M_{x_1x_2,x_4x_3}(f)=\left[\begin{smallmatrix}
0 & 0 & 0 & a\\
0 & 1 & -1 & 0\\
0 & -1 & 1 & 0\\
x & 0 & 0 & 0\\
\end{smallmatrix}\right]$, and $M_{x_3x_4, x_2x_1}(f)=\left[\begin{smallmatrix}
0 & 0 & 0 & x\\
0 & 1 & -1 & 0\\
0 & -1 & 1 & 0\\
a & 0 & 0 & 0\\
\end{smallmatrix}\right]$. Connecting variables $x_4, x_3$ of a copy of signature ${f}$ with variables $x_3, x_4$ of another copy of signature $f$,  we get a signature ${f_3}$ with  the signature matrix$$M(f_3)=M_{x_1x_2, x_4x_3}(f)NM_{x_3x_4, x_2x_1}(f)=\left[\begin{matrix}
0 & 0& 0& ax\\
0 & -2& 2& 0\\
0 & 2& -2& 0\\
ax & 0& 0& 0\\
\end{matrix}\right],$$  
Clearly, $ax\neq 0$ and so $f_3 \notin \mathscr{M}$
by Lemma~\ref{matchgate4}. By Lemma \ref{twins}, $\plholant{\neq_2}{ f_3}$ is \#P-hard and hence  $\plholant{\neq_2}{ f}$ is \#P-hard. \qed
\end{itemize}

\vspace{.1in}

In the following Lemmas \ref{t<>1}, \ref{conformal5}, \ref{pin} and Corollaries \ref{conformal4}, \ref{conformal3}, let $f$ be a 4-ary signature with the signature matrix
\begin{equation}\label{sig-f-forafter6-3}
M(f)=\left[\begin{matrix}
0 & 0& 0& a\\
0 & b& c& 0\\
0 & z& y& 0\\
x & 0& 0& 0\\
\end{matrix}\right], 
\end{equation}
where $abxyz\neq 0$, $\det \left[\begin{smallmatrix}
b & c\\ z & y\\
\end{smallmatrix}\right]=
by-cz\neq 0$ and $\det \left[\begin{smallmatrix}
a & z\\ c & x\\
\end{smallmatrix}\right]=ax-cz \neq 0$. Moreover $f\notin \mathscr{M}$, that is $cz-by\neq ax$.
These lemmas handle ``generic'' cases
of this section and will culminate in Theorem~\ref{nonzero},
which is a classification for Case  \Rmnum{4}.
It is here we will use M\"{o}bius transformations
to handle interpolations where it is
 particularly difficult to get desired signatures of ``infinite order''.

\begin{lemma}\label{t<>1}
Let $g=(0, 1, t, 0)^T$ be a binary signature, where $t\neq 0$ is not a root of unity.
Then $\operatorname{Pl-Holant}(\neq_2\mid f, g)$ is \#P-hard.
\end{lemma}

{\bf Proof.}
Let $\mathcal{B}=\{g_1, g_2, g_3\}$ be a set of three binary signatures $g_i=(0, 1, t_i, 0)^T$, for some $t_i \in \mathbb{C}$. By Lemma \ref{binary-interpolation}, we have $\plholant{\neq_2}{\{f\}\cup \mathcal{B}}\leqslant \plholant{\neq_2}{f, g}.$  We will show that $\plholant{\neq_2}{\{f\}\cup \mathcal{B}}$ is \#P-hard and it follows that $\plholant{\neq_2}{f, g}$ is \#P-hard.

Modifying $x_1=1$ of $f$ with $t_i$ $(i=1, 2)$ scaling separately,
we get two signatures $f_{t_i}$ with  the signature matrices
$M(f_{t_i})=\left[ \begin{smallmatrix}
0 & 0 & 0 & a \\
0 & b & c & 0\\
0 & t_iz & t_iy & 0\\
t_ix & 0 & 0 & 0\\
\end{smallmatrix} \right]$. Note that 
\[\det M_{\text{In}}(f_{t_i})=t_i\det M_{\text{In}}(f)~~~\mbox{ and }~~~ 
\det M_{\text{Out}}(f_{t_i})=t_i\det M_{\text{Out}}(f).\]
 Connecting variables $x_4, x_3$ of $f$ with variables $x_1$, $x_2$ of $f_{t_1}$ both using $(\neq_2)$, we get a signature $f_1$ with  the signature matrix
$$M(f_1)=\left[ \begin{matrix}
0 & 0 & 0 & a_1 \\
0 & b_1 & c_1 & 0\\
0 & z_1 & y_1 & 0\\
x_1 & 0 & 0 & 0\\
\end{matrix} \right]=M(f)NM(f_{t_1})=
\left[ \begin{matrix}
0 & 0 & 0 & a^2 \\
0 & t_1bz+bc & t_1by+c^2 & 0\\
0 & t_1z^2+yb & t_1yz+yc & 0\\
t_1x^2 & 0 & 0 & 0\\
\end{matrix} \right].$$

We first show that there is a $t_1\neq0$ such that $b_1y_1c_1z_1\neq 0$ and $(b_1z)(y_1c)- (c_1b)(z_1y)\neq 0$. 
Consider the quadratic polynomial 
\[p(t)=(tbz+bc)(tyz+yc)cz-(tby+c^2)(tz^2+yb)by.\]
 We have $p(t_1) = 
(b_1z)(y_1c)- (c_1b)(z_1y)$. 
Notice that the coefficient of the quadratic term in $p(t)$ is $byz^2(cz-by)$. 
It is not equal to zero since $byz^2\neq 0$ and $cz-by\neq 0$.  That is, $p(t)$ has degree $2$, and hence it has at most two roots. Also we have the following
three implications by the definitions of $b_1, y_1, c_1, z_1$: $b_1y_1=0 \Longrightarrow t_1=-\frac{c}{z}$, 
$c_1=0 \Longrightarrow t_1=-\frac{c^2}{by}$, and $z_1=0
\Longrightarrow t_1=-\frac{yb}{z^2}$. 
Therefore we can choose such a $t_1$ that does not take these values $0, -\frac{c}{z}, -\frac{c^2}{by}$ and $-\frac{yb}{z^2}$, and $t_1$ is not a root of $p(t)$. 
Then, we have $t_1 \neq 0$, $b_1y_1c_1z_1\neq 0$ and $(b_1z)(y_1c)- (c_1b)(z_1y)\neq 0$. 

Connecting variables $x_4, x_3$ of $f_1$ with variables $x_1$, $x_2$ of $f_{t_2}$ both using $(\neq_2)$, we get a signature $f_2$ with  the signature matrix
$$M(f_2)=\left[ \begin{matrix}
0 & 0 & 0 & a_2 \\
0 & b_2 & c_2 & 0\\
0 & z_2 & y_2 & 0\\
x_2 & 0 & 0 & 0\\
\end{matrix} \right]=M(f_1)NM(f_{t_2})=
\left[ \begin{matrix}
0 & 0 & 0 & a_1a \\
0 & t_2b_1z+c_1b & t_2b_1y+c_1c & 0\\
0 & t_2z_1z+y_1b & t_2z_1y+y_1c & 0\\
t_2x_1x & 0 & 0 & 0\\
\end{matrix} \right].$$
Since $b_1z \neq 0$ and $c_1b\neq 0$, we can let $t_2=-\frac{c_1b}{b_1z}$ and $t_2\neq 0$. Then $b_2= t_2b_1z+c_1b =0$.  Since $(b_1z)(y_1c)- (c_1b)(z_1y)\neq 0$, we have $y_2=t_2z_1y+y_1c\neq 0$. 
Notice that 
\begin{equation*}
\begin{aligned}
\det M_{\text{In}}(f_2)&=\det M_{\text{In}}(f_1)\cdot (-1)\cdot \det M_{\text{In}}(f_{t_2})\\
&=\det M_{\text{In}}(f)\cdot(-1)\cdot \det M_{\text{In}}(f_{t_1})\cdot(-1)\cdot \det M_{\text{In}}(f_{t_2})\\
&=t_1t_2\det M_{\text{In}}(f)^3\\
&\neq 0.
\end{aligned}
\end{equation*}
We have $\det M_{\text{In}}(f_2)=b_2y_2-c_2z_2=-c_2z_2\neq 0$.
Similarly, we have $\det M_{\text{Out}}(f_2)=-a_2x_2=t_1t_2\det M_{\text{Out}}(f)^3\neq 0$. Therefore, $M(f_2)$  is of the form $\left[ \begin{smallmatrix}
0 & 0 & 0 & a_2 \\
0 & 0 & c_2 & 0\\
0 & z_2 & y_2 & 0\\
x_2 & 0 & 0 & 0\\
\end{smallmatrix} \right]$, where $a_2x_2y_2c_2z_2\neq 0$. That is, $f_2$ is a signature in Case \Rmnum{3}. If $f_2\notin \mathscr{M}$, then $\plholant{\neq_2}{f_2}$ is \#P-hard by Theorem \ref{onezero}, and hence $\plholant{\neq_2}{\{f\}\cup\mathcal{B}}$ is \#P-hard.

Otherwise, $f_2\in \mathscr{M}$, which means $\dfrac{\det M_{\text{In}}(f_2)}{\det M_{\text{Out}}(f_2)}=1.$ Thus $\dfrac{\det M_{\text{In}}(f)^3}{\det M_{\text{Out}}(f)^3}=1.$ Since $f \notin \mathscr{M}$, $\dfrac{\det M_{\text{In}}(f)}{\det M_{\text{Out}}(f)}\neq 1$, and hence $\dfrac{\det M_{\text{In}}(f)^7}{\det M_{\text{Out}}(f)^7} \neq 1$.  
Similar to the construction of $f_1$, we construct $f_3$. 
First, modify $x_1=1$ of $f_1$ with $t_3$ scaling.
We get a signature $f_{1t_3}$ with  the signature matrix
$M(f_{1t_3})=\left[ \begin{smallmatrix}
0 & 0 & 0 & a_1 \\
0 & b_1 & c_1 & 0\\
0 & t_3z_1 & t_3y_1 & 0\\
t_3x_1 & 0 & 0 & 0\\
\end{smallmatrix} \right]$. Note that $\det M_{\text{In}}(f_{1t_3})=t_3\det M_{\text{In}}(f_1)$ and $\det M_{\text{Out}}(f_{1t_3})=t_3\det M_{\text{Out}}(f_1)$. Then connect variables $x_4, x_3$ of $f_1$ with variables $x_1$, $x_2$ of $f_{1t_3}$ both using $(\neq_2)$. We get a signature $f_3$ with  the signature matrix
$$M(f_3)=\left[ \begin{matrix}
0 & 0 & 0 & a_3 \\
0 & b_3 & c_3 & 0\\
0 & z_3 & y_3 & 0\\
x_3 & 0 & 0 & 0\\
\end{matrix} \right]=M(f_1)NM(f_{1t_3})=
\left[ \begin{matrix}
0 & 0 & 0 & a^2 \\
0 & t_3b_1z_1+b_1c_1 & t_3b_1y_1+c_1^2 & 0\\
0 & t_3z_1^2+y_1b_1 & t_3y_1z_1+y_1c_1 & 0\\
t_3x_1^2 & 0 & 0 & 0\\
\end{matrix} \right].$$
Since $c_1\neq 0$ and $z_1 \neq 0$, we can define $t_3=-\frac{c_1}{z_1}$ and $t_3\neq 0$. 
Then $b_3=b_1(t_3z_1+c_1)=0$ and $y_3=y_1(t_3z_1+c_1)=0$.
Notice that 
\begin{equation*}
\begin{aligned}
\det M_{\text{In}}(f_3)&=\det M_{\text{In}}(f_1)\cdot(-1)\cdot \det M_{\text{In}}(f_{1t_3})\\
&=-\det M_{\text{In}}(f_1)\cdot t_3\det M_{\text{In}}(f_{1})\\
&=-t_3 
\left[ 
\det M_{\text{In}}(f)\cdot(-1)\cdot\det M_{\text{In}}(f_{t_1})
\right]^2\\
&=-t_3t_1^2\det M_{\text{In}}(f)^4  \\ 
&\neq 0\\
\end{aligned}
\end{equation*}
We have $\det M_{\text{In}}(f_3)=-c_3z_3\neq 0$ and similarly, $\det M_{\text{Out}}(f_3)=-a_3x_3 =-t_3t_1^2\det M_{\text{Out}}(f)^4\neq 0$.
That is, $M(f_3)$ is of the form $\left[ \begin{smallmatrix}
0 & 0 & 0 & a_3 \\
0 & 0 & c_3 & 0\\
0 & z_3 & 0 & 0\\
x_3 & 0 & 0 & 0\\
\end{smallmatrix} \right]$ where $a_3x_3c_3z_3\neq 0$.

Connect variables $x_4, x_3$ of $f_2$ with variables $x_1$, $x_2$ of $f_{3}$ both using $(\neq_2)$. We get a signature $f_4$ with  the signature matrix
$$M(f_4)=\left[ \begin{matrix}
0 & 0 & 0 & a_4 \\
0 & b_4 & c_4 & 0\\
0 & z_4 & y_4 & 0\\
x_4 & 0 & 0 & 0\\
\end{matrix} \right]=M(f_2)NM(f_3)=
\left[ \begin{matrix}
0 & 0 & 0 & a_2a_3 \\
0 & 0 & c_2c_3 & 0\\
0 & z_2z_3 & y_2c_3 & 0\\
x_2x_3 & 0 & 0 & 0\\
\end{matrix} \right].$$
Clearly, $f_4$ is a signature in Case \Rmnum{3}. Also, notice that 
\begin{equation*}
\begin{aligned}
\det M_{\text{In}}(f_4)&=\det M_{\text{In}}(f_2)\cdot(-1)\cdot \det M_{\text{In}}(f_{3})\\
&=t_1t_2\det M_{\text{In}}(f)^3\cdot t_3t_1^2\det M_{\text{In}}(f)^4\\
&=t_3t_2t_1^3\det M_{\text{In}}(f)^7.  \\ 
\end{aligned}
\end{equation*}
and $$\det M_{\text{Out}}(f_4)=t_3t_2t_1^3\det M_{\text{Out}}(f)^7.$$
We have $$\dfrac{\det M_{\text{In}}(f_4)}{\det M_{\text{Out}}(f_4)}=\dfrac{\det M_{\text{In}}(f)^7}{\det M_{\text{Out}}(f)^7}\neq 1,$$
which means $f_4 \notin \mathscr{M}$. By Theorem \ref{onezero}, $\plholant{\neq_2}{f_4}$ is \#P-hard, and hence $\plholant{\neq_2}{\{f\}\cup\mathcal{B}}$ is \#P-hard. \qed


\begin{lemma}\label{conformal5}

Let $g=(0, 1, t, 0)^T$ be a binary signature where $t$ is an $n$-th primitive root of unity, and $n\geq 5$.
Then $\operatorname{Pl-Holant}(\neq_2\mid f, g)$ is \#P-hard.
\end{lemma}

{\bf Proof.} 
Note that $M_{x_1, x_2}(g)=\left[\begin{smallmatrix}
 0& 1\\
 t& 0\\
\end{smallmatrix}\right]$. 
 Connect the variable $x_2$ of a copy of signature $g$ with the variable $x_1$ of another copy of signature $g$ using $(\neq_{2})$. We get a signature $g_2$ with the signature matrix
$$
M_{x_1, x_2}(g_2)=
\left[\begin{matrix}
 0& 1\\
 t& 0\\
\end{matrix}\right]
\left[\begin{matrix}
 0& 1\\
 1& 0\\
\end{matrix}\right]
\left[\begin{matrix}
 0& 1\\
 t& 0\\
\end{matrix}\right]=
\left[\begin{matrix}
 0& 1\\
 t^2& 0\\
\end{matrix}\right]
.$$
That is, $g_2=(0, 1, t^2, 0)^T.$ 
Similarly, we can construct $g_i=(0, 1, t^i, 0)^T$ for $1 \leqslant i \leqslant 5$. Here, $g_1$ denotes $g$. Since the order $n \geqslant 5$, $g_i$ are 
 distinct ($1 \le i \le 5$). 
Connect variables $x_4$, $x_3$ of signature $f$ with variables $x_1$, $x_2$ of $g_i$ for $1 \leqslant i \leqslant 5$ respectively. We get binary signatures ${h}_i$, where 
$$h_i=M_{x_1x_2, x_4x_3}(f)g_i=
\left[\begin{matrix}
0 & 0& 0& a\\
0 & b& c& 0\\
0 & z& y& 0\\
x & 0& 0& 0\\
\end{matrix}\right]
\left(\begin{matrix}
0\\
1\\
t^i\\
0\\
\end{matrix}\right)
=\left(\begin{matrix}
0\\
b+ct^i\\
z+yt^i\\
0\\
\end{matrix}\right).
$$
Let $\varphi(\mathfrak z)=\dfrac{z+y\mathfrak z}{b+c\mathfrak z}$. Since $\det \left[\begin{smallmatrix}b &c \\ z & y \end{smallmatrix}\right]=by-cz\neq 0$, $\varphi(\mathfrak z)$ is a M\"{o}bius transformation  of the  extended  complex  plane $\mathbb{\widehat{C}}$. 
We rewrite ${h}_{i}$ in the form of  $(b+ct^i)(0, 1, \varphi(t^i), 0)^T$, with the understanding that if $b+ct^i=0$, then $\varphi(t^i)=\infty$, and we define $(b+ct^i)(0, 1, \varphi(t^i), 0)^T$ to be $(0, 1, z+yt^i, 0)^T$.
If there is a $t^i$ such that $\varphi(t^i)$ is not a root of unity, and $\varphi(t^i)\neq 0$ and  $\varphi(t^i)\neq\infty$, by Lemma~\ref{t<>1}, 
we have $\plholant{\neq_2}{f, {h}_i}$ is \#P-hard, and hence
$\plholant{\neq_2}{f, {g_1}}$ is \#P-hard. 
Otherwise, $\varphi(t^i)$ is $0$, $\infty$ or a root of unity for $1\leqslant i\leqslant 5$. 
Since $\varphi(\mathfrak z)$ is a bijection of $\mathbb{\widehat{C}}$, 
there is at most one $t^i$ such that $\varphi(t^i)=0$ and at most one $t^i$ such that $\varphi(t^i)=\infty$. That means, there are at least three $t^i$ such that $|\varphi(t^i)|=1$. 
Since a M\"{o}bius transformation is determined by any $3$ distinct points, mapping $3$ distinct points from $S^1$ to $S^1$ implies that this $\varphi(\mathfrak{z})$ maps $S^1$ homeomorphically onto $S^1$
(so in fact there is no $t^i$ such that
$\varphi(t^i) = 0$ or $\infty$). 
Such a M\"{o}bius transformation has a special form: $\mathcal{M}(\alpha, e^{\ii\theta})=e^{\ii\theta}\dfrac{(\mathfrak{z}+\alpha)}{1+\bar{\alpha}\mathfrak{z}}$, where $|\alpha| \neq 1$.
(It cannot be of the form $e^{\ii\theta}/\mathfrak{z}$,
since $b \not = 0$.)
 

By normalization in signature $f$,
we may assume $b=1$. Compare the coefficients, we have $c=\bar{\alpha}$, $y=e^{\ii\theta}$ and $z=\alpha e^{\ii\theta}$. 
Here $\alpha \neq 0$ due to $z\neq 0$.
Also, since 
$M_{x_2x_3, x_1x_4}(f)=\left[\begin{smallmatrix}
0 & 0& 0& y\\
0 & a& z& 0\\
0 & c& x& 0\\
b & 0& 0& 0\\
\end{smallmatrix}\right]$
and $\det \left[\begin{smallmatrix}a &z \\ c & x \end{smallmatrix}\right]=ax-cz\neq 0$,
we have another  M\"{o}bius transformation $\psi(\mathfrak z)=\dfrac{c+x\mathfrak{z}}{a+z\mathfrak{z}}$.
Plug in $c=\bar{\alpha}$ and $z=\alpha e^{\ii\theta}$, we have 
$$\psi(\mathfrak z)=\dfrac{\bar{\alpha}+x\mathfrak{z}}{a +\alpha e^{\ii\theta} \mathfrak{z}}= 
\dfrac{\frac{\bar{\alpha}}{a}+\frac{x}{a}\mathfrak{z}}{1+ \frac{\alpha e^{\ii\theta} }{a}\mathfrak{z}}.$$ 
By the same proof for $\varphi({\mathfrak z})$, we get \plholant{\neq_2}{f, g} is \#P-hard, unless $\psi({\mathfrak z})$ also maps $S^1$ to $S^1$. Hence, we can assume $\psi({\mathfrak z})$ has the form $\mathcal{M}(\beta, e^{\ii\theta'})=e^{\ii\theta'}\dfrac{(\mathfrak{z}+\beta)}{1+\bar{\beta}\mathfrak{z}}$, where $|\beta|\neq 1$. (It is clearly not of the form $e^{\ii\theta'}/\mathfrak{z}$.)
Compare the coefficients, we have 
\begin{equation*}
\left\{
\begin{aligned}
{\alpha e^{\ii\theta} }/{a}   &=  \bar{\beta} \\
{ \bar{\alpha}}/{a}   &=  e^{\ii\theta'}\beta  \\
{x}/{a} \  &=  e^{\ii\theta'}
\end{aligned}
\right..
\end{equation*}
Solving these equations, we get $a=e^{\ii\theta}{\alpha}/{\bar{\beta}}$ 
and $x={\bar{\alpha}}/{\beta}$. Let $\gamma={\alpha}/{\bar{\beta}}$, and we have $a=\gamma e^{\ii\theta}$ and $x=\bar{\gamma}$,
where $|\gamma|\neq|\alpha|$ since $|\beta|\neq 1$ and $\gamma \neq 0$ since $x \neq 0$.
Then, we have signature matrices $M_{x_1x_2, x_4x_3}(f)= \left[\begin{smallmatrix}
0 & 0& 0& \gamma e^{\ii\theta}\\
0 & 1& \bar{\alpha}& 0\\
0 & \alpha e^{\ii\theta}& e^{\ii\theta}& 0\\
\bar{\gamma} & 0& 0& 0\\
\end{smallmatrix}\right],$   
$M_{x_2x_3, x_1x_4}(f)=\left[\begin{smallmatrix}
0 & 0& 0& e^{\ii\theta} \\
0 & \gamma e^{\ii\theta}& \alpha e^{\ii\theta}& 0\\
0 & \bar{\alpha} & \bar{\gamma}& 0\\
1 & 0& 0& 0\\
\end{smallmatrix}\right],$
$M_{x_3x_4, x_2x_1}(f)=\left[\begin{smallmatrix}
0 & 0& 0& \bar{\gamma} \\
0 & e^{\ii\theta}& \bar{\alpha}& 0\\
0 & \alpha e^{\ii\theta}& 1& 0\\
\gamma e^{\ii\theta} & 0& 0& 0\\
\end{smallmatrix}\right]$
and
$M_{x_4x_1, x_3x_2}(f)=\left[\begin{smallmatrix}
0 & 0& 0& 1 \\
0 & \bar{\gamma} & \alpha e^{\ii\theta}& 0\\
0 & \bar{\alpha} & \gamma e^{\ii\theta}& 0\\
e^{\ii\theta} & 0& 0& 0\\
\end{smallmatrix}\right]$.
Connect variables $x_4, x_3$ of a copy of signature $f$ with variables $x_3, x_4$ of another copy of signature $f$ using $(\neq_{2})$. We get a signature $f_1$ with the signature matrix 
$$M(f_1)=M_{x_1x_2, x_4x_3}(f)NM_{x_3x_4, x_2x_1}(f)=
\left[\begin{matrix}
0 & 0& 0& \gamma \bar{\gamma} e^{\ii\theta}\\
0 & (\alpha+\bar{\alpha})e^{\ii\theta}& 1+{\bar{\alpha}}^2& 0\\
0 & (1+\alpha^2)e^{\ii2\theta} & (\alpha+\bar{\alpha})e^{\ii\theta}& 0\\
\gamma \bar{\gamma}e^{\ii\theta} & 0& 0& 0\\
\end{matrix}\right].$$

\begin{itemize}
\item If $\alpha+\bar{\alpha}\neq 0$, 
normalizing $M_{x_1x_2, x_4x_3}(f_1)$ by dividing by $(\alpha+\bar{\alpha})e^{\ii\theta}$, 
we have 
$$M(f_1)=
\left[\begin{matrix}
0 & 0& 0& \dfrac{\gamma \bar{\gamma} }{(\alpha+\bar{\alpha})}\\
0 & 1 & \dfrac{(1+{\bar{\alpha}}^2)e^{-\ii\theta}}{(\alpha+\bar{\alpha})}& 0\\
0 & \dfrac{(1+\alpha^2)e^{\ii\theta}}{(\alpha+\bar{\alpha})} &1& 0\\
\dfrac{\gamma \bar{\gamma}}{(\alpha+\bar{\alpha})} & 0& 0& 0\\
\end{matrix}\right].$$
Note that $\dfrac{(1+\alpha^2)e^{\ii\theta}}{(\alpha+\bar{\alpha})}$ and $\dfrac{(1+{\bar{\alpha}}^2)e^{-\ii\theta}}{(\alpha+\bar{\alpha})}$ are conjugates.
Let $\delta=\dfrac{(1+\alpha^2)e^{\ii\theta}}{(\alpha+\bar{\alpha})}$,
and then $\bar{\delta}=\dfrac{(1+{\bar{\alpha}}^2)e^{-\ii\theta}}{(\alpha+\bar{\alpha})}$. 
We have $|\delta|^2=\delta \bar{\delta}=\dfrac{(1+\alpha^2)(1+{\bar{\alpha}^2)}}{(\alpha+\bar{\alpha})^2}\neq 1$ due to $\det M_{\text{In}}(f_1) \neq 0$, and $\delta \neq 0$ due to $|\alpha|\neq1$. 
Consider the inner matrix  of $M(f_1)$, we have $M_{\text{In}}(f_1)=
\left[\begin{smallmatrix}
1 & \bar{\delta}\\
{\delta} & 1\\
\end{smallmatrix}\right].$
Notice that the two eigenvalues of $M_{\text{In}}(f_1)$ are $1+|\delta|$ and $1-|\delta|$, 
and obviously $\left| \frac{1-|\delta|}{1+|\delta|} \right| \neq 1$, 
which means there is no integer $n>0$ and complex number $C$ such that $M^n_{\text{In}}(f_1)=CI$. Note that $\varphi_{1}(\mathfrak{z})=\dfrac{\delta+\mathfrak{z}}{1+\bar{\delta}\mathfrak{z}}$ is a M\"{o}bius transformation of the form $\mathcal{M}(\delta, 1)$ mapping $S^1$ to $S^1$. 

Connect variables $x_4$, $x_3$ of signature $f_1$ with variables $x_1$, $x_2$ of signatures $g_i$. 
We get binary signatures ${g_{(i,{\varphi_1})}}$, where 
$$g_{(i,{\varphi_1})}=M_{x_1x_2, x_4x_3}(f_1)g_i=
\left[\begin{matrix}
0 & 0& 0& *\\
0 & 1& \bar{\delta}& 0\\
0 & \delta& 1& 0\\
* & 0& 0& 0\\
\end{matrix}\right]
\left(\begin{matrix}
0\\
1\\
t^i\\
0\\
\end{matrix}\right)
=\left(\begin{matrix}
0\\
1+\bar{\delta} t^i\\
\delta+t^i\\
0\\
\end{matrix}\right)
=(1+\bar{\delta} t^i)\left(\begin{matrix}
0\\
1\\
\varphi_1(t^i)\\
0\\
\end{matrix}\right).$$
Since $\varphi_1$ is a M\"{o}bius transformation mapping $S^1$ to $S^1$ and $|t^i|=1$, we have $|\varphi_1(t^i)|=1$, which means $1+\bar{\delta} t^i\neq 0$. Hence, ${g_{(i, \varphi_1)}}$ can be normalized as $(0, 1, \varphi_1(t^i), 1)^T$.
Successively construct binary signatures $g_{(i, \varphi_1^n)}$ by connecting $f_1$ with $g_{(i, \varphi_1^{n-1})}$. We have
$$g_{(i, \varphi_1^n)}=M(f_1)g_{(i, \varphi_1^{n-1})}=M^{n}(f_1)g_i=C_{(i,n)}(0, 1, \varphi_1^n(t^i), 0)^T,$$
where $C_{(i,n)}=\prod\limits_{0\leqslant k \leqslant n-1}\left(1+\bar{\delta}\varphi_{1}^{k}(t^i)\right)$. We know $C_{(i,n)}\neq0$, because for any $k$, $1+\bar{\delta}\varphi_{1}^{k-1}(t^i)\neq 0$ due to
 $|\varphi_1^k(t^i)|=\dfrac{|\delta+\varphi_{1}^{k-1}(t^i)|}{|1+\bar{\delta}\varphi_{1}^{k-1}(t^i)|}= 1$. 
Hence, $g_{(i, \varphi_1^{n})}$ can be normalized as $(0, 1, \varphi_1^n(t^i), 0)^T$. 
Notice that the nonzero entries $(1, \varphi_1^n(t^i))^T$ of ${g_{(i, \varphi_1^{n})}}$ are completely  decided by the inner matrix $M_{\text{In}}(f_1)$.
That is $$M^{n}_{\text{In}}(f_1)
\left(\begin{matrix}
1\\
t^i\\
\end{matrix}\right)=
C_{(i,n)}\left(\begin{matrix}
1\\
\varphi_1^n(t^i)\\
\end{matrix}\right).$$
If for each $i\in\{1, 2, 3\}$, there is some $n_i \geqslant 1$ such that $(1, \varphi_1^{n_i}(t^i))^T= (1, t^i)^T$,
then $\varphi_1^{n_0}(t^i)=t^i$, where $n_0=n_1n_2n_3$ for $1\leqslant i\leqslant 3$, i.e.,
 the M\"{o}bius transformation $\varphi_1^{n_0}$ fixes  three distinct complex numbers $t, t^2, t^3$.
 So the M\"{o}bius transformation is the identity map,
i.e., $\varphi_1^{n_0}(\mathfrak z)=\mathfrak z$ for all $\mathfrak z\in\mathbb{C}$.
 This implies that $M^{n_0}_{\text{In}}(f_1)=C\left[\begin{smallmatrix}
 1 & 0\\
0 & 1\\
\end{smallmatrix}\right]$ for some constant $C \not =0$.
This contradicts the fact that the ratio of the eigenvalues of $M_{\text{In}}$ is not a root of unity.
 Therefore, there is an $i$ such that  $(1, \varphi_1^{n}(t^i))^T$ are all distinct for $n\in \mathbb{N}$.
Then, we can realize polynomially many distinct binary signatures of the form $(0, 1, \varphi_1^{n}(t^i), 1)^T$.
By  Lemma~\ref{any-interpolation}, we have $\operatorname{Pl-Holant}(\neq_2\mid f, g)$ is \#P-hard.

\item Otherwise $\alpha + \bar{\alpha} =0$, which means $\alpha$ is a pure imaginary number. We already have $\alpha \neq 0$ due to $z \neq 0$.
Also $|\alpha| \neq 1$ from the form of 
$\mathcal{M}(\alpha, e^{\ii\theta})$.
Let  $\alpha=r \ii$, where $r  \in \mathbb R$ and $|r |\neq 0$ or $1$.  
Connect  variables $x_1$, $x_4$ of a copy of signature $f$ with variables $x_4$,  $x_1$ of another copy of signature $f$, we get a signature $f_2$ with the signature matrix 
\begin{align*}
M(f_2) &= M_{x_2x_3, x_1x_4}(f)NM_{x_4x_1, x_3x_2}(f) \\
& = \left[\begin{matrix}
0 & 0& 0& e^{\ii\theta} \\
0 & \gamma e^{\ii\theta}& r \ii e^{\ii\theta}& 0\\
0 & -r \ii & \bar{\gamma}& 0\\
1 & 0& 0& 0\\
\end{matrix}\right] 
\left[\begin{matrix}
0 & 0 & 0 & 1 \\
0 & 0 & 1 & 0 \\
0 & 1 & 0 & 0 \\
1 & 0 & 0 & 0 \\
\end{matrix}\right]
\left[\begin{matrix}
0 & 0& 0& 1 \\
0 & \bar{\gamma} & \ r \ii e^{\ii\theta}& 0\\
0 & -r \ii & \gamma e^{\ii\theta}& 0\\
e^{\ii\theta} & 0& 0& 0\\
\end{matrix}\right] \\
& =
\left[\begin{matrix}
0 & 0 & 0 & e^{\ii\theta} \\
0 & (-\gamma+ \bar{\gamma})r \ii e^{\ii\theta} & (\gamma^2-r^2)e^{\ii 2\theta} & 0 \\
0 & \bar\gamma^2-r^2 & (-\gamma+ \bar{\gamma})r\ii e^{\ii\theta} & 0\\
e^{\ii\theta} & 0 & 0 & 0\\
\end{matrix}\right].
\end{align*}
\begin{itemize}
\item If $-\gamma+ \bar{\gamma}\neq 0$,
 normalizing $M(f_2)$ by dividing the quantity
 $(-\gamma+ \bar{\gamma})r\ii e^{\ii\theta}$, we have $$M_{\text{In}}(f_2)=
\left[\begin{matrix}
1 & \dfrac{ (\gamma^2-r^2)e^{\ii\theta}}{(-\gamma+ \bar{\gamma})r\ii}\\
\dfrac{ (\bar\gamma^2-r^2)e^{-\ii\theta}}{(-\gamma+ \bar{\gamma})r\ii} & 1\\
\end{matrix}\right].$$
Note that $\dfrac{ (\gamma^2-r^2)e^{\ii\theta}}{(-\gamma+ \bar{\gamma})r\ii}$ and $\dfrac{ (\bar\gamma^2-r^2)e^{-\ii\theta}}{(-\gamma+ \bar{\gamma})r\ii}$ are conjugates. 
Let $\zeta=\dfrac{ (\bar\gamma^2-r^2)e^{-\ii\theta}}{(-\gamma+ \bar{\gamma})r\ii}$, 
and then $|\zeta| \neq 1$ due to $\det M_{\text{In}}(f_2) \neq 0$, and $\zeta
 \neq 0$ due to $|\gamma| \neq |\alpha| = |r|$ (as $|\beta|  \neq 1$).
With the same analysis as for $M_{\text{In}}(f_1)$ in the case $\alpha
+ \bar{\alpha} \neq 0$,
the ratio of the two eigenvalues of $M_{\text{In}}(f_2)
= \left[\begin{smallmatrix}
1 & \bar{\zeta}\\
\zeta& 1 
\end{smallmatrix}
\right]$ is also not equal to 1, which means there is no integer $n$ and complex number $C$ such that $M^n_{\text{In}}(f_2)=CI$. 
Notice that $\varphi_{2}(\mathfrak{z'})=\dfrac{\zeta+\mathfrak{z'}}{1+\bar{\zeta}\mathfrak{z'}}$ is also a M\"{o}bius transformation of the form $\mathcal{M}(\zeta, 1)$ mapping $S^1$ to $S^1$. 
Similarly, we can realize polynomially many distinct binary signatures, and hence $\operatorname{Pl-Holant}(\neq_2\mid f, g)$ is \#P-hard.

\item Otherwise, $-\gamma+ \bar{\gamma}= 0$, which means $\gamma$ is a real number. 
We have  $\gamma \in \mathbb{R}$,
$|\gamma|\neq 0$ or $|r|$. Connect variables $x_4$, $x_3$ of a copy of signature $f$ with variables $x_1$, $x_2$ of another copy of signature $f$, we get a signature $f^\prime$ with the signature matrix 
\begin{align*}
M(f^\prime) &= M_{x_1x_2, x_4x_3}(f)NM_{x_1x_2, x_4x_3}(f) \\
& = \left[\begin{matrix}
0 & 0& 0& \gamma e^{\ii\theta} \\
0 & 1& -r\ii& 0\\
0 & r\ii e^{\ii\theta} & e^{\ii\theta}& 0\\
\gamma & 0& 0& 0\\
\end{matrix}\right] 
\left[\begin{matrix}
0 & 0 & 0 & 1 \\
0 & 0 & 1 & 0 \\
0 & 1 & 0 & 0 \\
1 & 0 & 0 & 0 \\
\end{matrix}\right]
\left[\begin{matrix}
0 & 0& 0& \gamma e^{\ii\theta} \\
0 & 1 &  -r\ii& 0\\
0 & r\ii e^{\ii\theta} & e^{\ii\theta}& 0\\
\gamma  & 0& 0& 0\\
\end{matrix}\right] \\
& =
\left[\begin{matrix}
0 & 0 & 0 & \gamma ^2e^{\ii 2\theta} \\
0 & (e^{\ii\theta}-1)r\ii & e^{\ii\theta}-r^2 & 0 \\
0 & e^{\ii\theta}-e^{\ii2\theta}r^2 & (e^{\ii2\theta}-e^{\ii\theta})r\ii & 0\\
\gamma ^2 & 0 & 0 & 0\\
\end{matrix}\right].
\end{align*}

\begin{itemize}

\item If $e^{\ii\theta}=1$, then $M(f)=\left[\begin{smallmatrix}
0 & 0& 0& \gamma \\
0 & 1& \bar{\alpha}& 0\\
0 & \alpha & 1& 0\\
\bar{\gamma} & 0& 0& 0\\
\end{smallmatrix}\right]$, and $M_{\text{In}}(f)=
\left[\begin{smallmatrix}
1& \bar{\alpha}\\
\alpha & 1\\
\end{smallmatrix}\right].$ Since $|\alpha|\neq 1$, same as the analysis of $M_{\text{In}}(f_1)$, we can realize polynomially many binary signatures, and hence $\operatorname{Pl-Holant}(\neq_2\mid f, g)$ is \#P-hard.

\item Otherwise $e^{\ii\theta}\neq 1$, normalizing $M(f^\prime)$ by dividing by $(e^{\ii\theta}-1)r\ii$, we have $$M(f^\prime)=
\left[\begin{matrix}
0 & 0 & 0 &\dfrac{\gamma^2e^{\ii\theta}}{(e^{\ii\theta}-1)r\ii}\cdot e^{\ii\theta} \\
0 & 1 & \dfrac{e^{\ii\theta}-r^2}{(e^{\ii\theta}-1)r\ii} &0\\
0 & \dfrac{1-e^{\ii\theta}r^2}{ (e^{\ii\theta}-1)r\ii} \cdot e^{\ii\theta} & e^{\ii\theta} & 0\\
\dfrac{\gamma^2}{(e^{\ii\theta}-1)r\ii} &0 & 0 & 0\\
\end{matrix}\right].$$
Note that $\dfrac{1-e^{\ii\theta}r^2}{ (e^{\ii\theta}-1)r\ii}$ and $\dfrac{e^{\ii\theta}-r^2}{(e^{\ii\theta}-1)r\ii}$ are conjugates, and  $\dfrac{\gamma^2e^{\ii\theta}}{(e^{\ii\theta}-1)r\ii}$ and $\dfrac{\gamma^2}{(e^{\ii\theta}-1)r\ii}$ are conjugates. 
Let $\alpha^\prime=\dfrac{1-e^{\ii\theta}r^2}{(e^{\ii\theta}-1)r\ii}$ and $\gamma^\prime=\dfrac{\gamma^2e^{\ii\theta}}{(e^{\ii\theta}-1)r\ii}$. 
Then 
\[
M(f^\prime)=\left[\begin{matrix}
0 & 0& 0& {\gamma^\prime}e^{\ii\theta} \\
0 & 1& \bar{\alpha^\prime}& 0\\
0 & \alpha^\prime e^{\ii\theta}& e^{\ii\theta}& 0\\
\bar{\gamma^\prime}  & 0& 0& 0\\
\end{matrix}\right].\]
 Notice that $M(f^\prime)$ and $M(f)$ have the same form. Similar to the construction of $f_2$, we can construct a signature $f^\prime_2$ using $f^\prime$ instead of $f$. 
 Since $-\gamma^\prime+ \bar{\gamma^\prime}= -\dfrac{\gamma^2e^{\ii\theta}}{(e^{\ii\theta}-1)r\ii}+\dfrac{\gamma^2}{(e^{\ii\theta}-1)r\ii}=-\dfrac{\gamma^2}{r\ii}\neq 0$,
  by the analysis of $f_2$, we can still realize polynomially many binary signatures and hence $\operatorname{Pl-Holant}(\neq_2\mid f, g)$ is \#P-hard. \qed
\end{itemize}
\end{itemize}
\end{itemize}
\begin{remark}
The order $n\geqslant 5$ promises that there are at least three points mapped to points on $S^1$, since at most one point can be mapped to $0$ and at most one can be mapped to $\infty$. When the order $n$ is $3$ or $4$, if no point is mapped to $0$ or $\infty$, then there are  still at least three points mapped to points on $S^1$. So, we have the following corollary.
\end{remark}
\begin{corollary}\label{conformal4}
Let $g=(0, 1, t, 0)^T$ be a  binary signature where $t$ is an $n$-th primitive root of unity, and $n=3$ or $4$.
 Let $g_m$ denote $(0, 1, t^m, 0)^T$.
For any cyclic permutation  $(i, j, k ,\ell)$ of $(1, 2, 3 ,4)$,
if there is no $g_m$ such that $M_{x_ix_j,x_\ell x_k}(f)g_m=d_1(0, 1, 0, 0)^T$ or $d_2(0, 0 , 1, 0)^T$, where $d_1, d_2 \in \mathbb{C}$ , 
then $\operatorname{Pl-Holant}(\neq_2\mid f, g)$ is \#P-hard.
\end{corollary}

We normalize $f$ by setting $b=1$ in  Lemma \ref{sig-f-forafter6-3}.

\begin{lemma}\label{pin}
Let $g=(0, 1, 0, 0)^T$ be a binary signature.
Then $\operatorname{Pl-Holant}(\neq_2\mid f, g)$ is \#P-hard.
\end{lemma}

{\bf Proof.}
Connecting variables $x_4$, $x_3$ of the signature $f$ with variables $x_2$ and $x_1$ of $g$ both using $(\neq_2)$ we get a binary signature ${g_1}$, where
$$g_1=M_{x_1x_2, x_4x_3}(f)(0, 1, 0, 0)^T=(0, 1, z, 0)^T.$$
$g_1(x_1,x_2)$ can be normalized to $(0, z^{-1}, 1, 0)^T$ since $z\neq 0$.
So we have $g_1(x_2,x_1)=(0, 1, z^{-1}, 0)$.
Then, modifying $x_1=1$ of $f$ with $z^{-1}$ scaling,
we get a signature $f_1$ with  the signature matrix $M(f_1)=\left[\begin{smallmatrix}
0 & 0 & 0 & a\\
0 & 1 & c & 0\\
0 & 1 & y/z & 0\\
x/z & 0 & 0 & 0\\
\end{smallmatrix}\right]$. We denote it  by $\left[\begin{smallmatrix}
0 & 0 & 0 & a\\
0 & 1 & c & 0\\
0 & 1 & y_1 & 0\\
x_1 & 0 & 0 & 0\\
\end{smallmatrix}\right]$, where $x_1y_1\neq 0$. 

\begin{itemize}

\item If $c=0$, connecting variables $x_4$, $x_3$ of $f_1$ with variables $x_1$, $x_2$ of $g$ both using $(\neq_2)$ we get a binary signature ${h_1}$, where
$$h_1=M_{x_1x_2, x_4x_3}(f_1)(0, 0, 1, 0)^T=(0, 1, y_1, 0)^T.$$
Also, connecting the variable $x_4$ with $x_3$ of $f_1$ using $(\neq_2)$ we get a binary signature $h_2$, where
$$h_2=M_{x_1x_2, x_4x_3}(f_1)(0, 1, 1, 0)^T=(0, 2, y_1, 0)^T.$$
$h_2$ can be normalized to $(0, 1, \frac{y_1}{2}, 0)^T.$ Clearly, $|y_1|\neq |\frac{y_1}{2}|$, so they cannot both be roots of unity. By Lemma \ref{t<>1}, $\plholant{\neq_2}{f, h_1, h_2}$ is \#P-hard, and we conclude that
 $\plholant{\neq_2}{f, g}$ is \#P-hard.

\item Otherwise $c\neq 0$. Connecting variables $x_2$, $x_1$ of $g$ with variables $x_1$, $x_2$ of $f$ both using $(\neq_2)$ we get a binary signature ${g_2}$, where
$$g_2=((0, 1, 0, 0)M_{x_1x_2, x_4x_3}(f_1))^T=(0, 1, c, 0)^T.$$
which can be normalized to 
$g_2(x_2, x_1)=(0, 1, c^{-1}, 0)^T$.
Then, modifying $x_4=1$ of $f_1$ with $c^{-1}$ scaling,
we get a signature $f_2$ with  the signature matrix 
$M(f_2)=\left[\begin{smallmatrix}
0 & 0 & 0 & \frac{a}{c}\\
0 & 1 & 1 & 0\\
0 & 1 & \frac{y}{zc} & 0\\
\frac{x}{z} & 0 & 0 & 0\\
\end{smallmatrix}\right]$ which we denote by $\left[\begin{smallmatrix}
0 & 0 & 0 & a_2\\
0 & 1 & 1 & 0\\
0 & 1 & y_2 & 0\\
x_2 & 0 & 0 & 0\\
\end{smallmatrix}\right]$, where $a_2x_2y_2\neq 0$. Notice that $M_{x_2x_3, x_1x_4}(f_2)=\left[\begin{smallmatrix}
0 & 0 & 0 & y_2\\
0 & a_2 & 1 & 0\\
0 & 1 & x_2 & 0\\
1 & 0 & 0 & 0\\
\end{smallmatrix}\right]$. Connect variables $x_1$, $x_4$ of signature $f_2$ with variables $x_2$, $x_1$ of $g$ both using $(\neq_2)$. We get a binary signature ${h_3}$, where
$$h_3=M_{x_2x_3, x_1x_4}(f_2)(0, 1, 0, 0)^T=(0, a_2, 1, 0)^T.$$ 
$h_3$ can be normalized as $(0, 1, \frac{1}{a_2}, 0)^T$. Also connect variables $x_1$, $x_4$ of signature $f_2$ with variables $x_1$, $x_2$ of $g$ both using $(\neq_2)$. We get a binary signature ${h_4}$, where
$$h_4=M_{x_2x_3, x_1x_4}(f_2)(0, 0, 1, 0)^T=(0, 1, x_2, 0)^T.$$
If $|a_2|\neq 1$ or $|x_2|\neq 1$, then ${a_2}$ or $x_2$ is not a root of unity. By Lemma \ref{t<>1},  $\plholant{\neq_2}{f, h_3,  h_4}$ is \#P-hard, and hence $\plholant{\neq_2}{f, g}$ is \#P-hard. 
Otherwise, $|a_2|=|x_2|=1$.
Same as the construction of $h_1$ and $h_2$, construct binary signatures $h'_1$ and $h'_2$ using $f_2$ instead of $f_1$. We get 
$$h'_1=M_{x_1x_2, x_4x_3}(f_2)(0, 0, 1, 0)^T=(0, 1, y_2, 0)^T,$$ and
$$h'_2=M_{x_1x_2, x_4x_3}(f_2)(0, 1, 1, 0)^T=(0, 2, 1+y_2, 0)^T.$$
Note that $h'_2$ can be normalized as $(0, 1, \frac{1+y_2}{2}, 0)^T.$
\begin{itemize}
  \item If $y_2$ is not a root of unity, 
then by Lemma \ref{t<>1}, $\plholant{\neq_2}{f, h'_1}$ is \#P-hard, and hence $\plholant{\neq_2}{f, g}$ is \#P-hard.
\item If $y_2$ is an $n$-th primitive root of unity and $n\geqslant 5$, then by Lemma \ref{conformal5}, $\plholant{\neq_2}{f, h'_1}$ is \#P-hard, and hence $\plholant{\neq_2}{f, g}$ is \#P-hard.
\item If $y_2= \frac{-1\pm\sqrt{3}\ii}{2}$ or $\pm \ii$,  then $0<|\frac{1+y_2}{2}|<1$, which means it is not zero neither a root of unity. By Lemma \ref{t<>1}, $\plholant{\neq_2}{f, h'_2}$ is \#P-hard, and hence $\plholant{\neq_2}{f, g}$ is \#P-hard.
\item If $y_2=1$, then $f_2$ is non-singular redundant and hence $\plholant{\neq_2}{f, g}$ is \#P-hard.
\item If $y_2=-1$. Connect two copies of $f_2$, we get a signature $f_3$ with  the signature matrix$$M(f_3)=M_{x_1x_2,x_4x_3}(f_2)NM_{x_1x_2,x_4x_3}(f_2)=\left[\begin{matrix}
0 & 0 & 0 & a^2_2\\
0 & 2 & 0 & 0\\
0 & 0 & -2 & 0\\
x^2_2 & 0 & 0 & 0\\
\end{matrix}\right].$$ 
Since $|a_2|=|x_2|=1$, $|a_2^2x_2^2|=1\neq 4$. 
Therefore, applying Corollary~\ref{inner-corollary} to
 \{$a_2^2, 2, x_2^2, -2\}$, 
we get $\plholant{\neq_2}{f_3}$ is \#P-hard, and hence $\plholant{\neq_2}{f, g}$ is \#P-hard. \qed
\end{itemize}
\end{itemize}

Combining Lemma \ref{conformal5}, Corollary \ref{conformal4} and Lemma \ref{pin}, we have the following corollary.
\begin{corollary}\label{conformal3}

Let $g=(0, 1, t, 0)^T$ be a binary signature where $t$ is an $n$-th primitive root of unity, and $n\geqslant 3$.
Then $\operatorname{Pl-Holant}(\neq_2\mid f, g)$ is \#P-hard.
\end{corollary}

Now, we are able to prove the following theorem for Case \Rmnum{4}.
\begin{theorem}\label{nonzero}
Let $f$ be a 4-ary signature with the signature matrix
\begin{center}
$M(f)=\left[\begin{matrix}
0 & 0& 0& a\\
0 & b& c& 0\\
0 & z& y& 0\\
x & 0& 0& 0\\
\end{matrix}\right]$, 
\end{center}
where $abxyz\neq 0$.
 $\operatorname{Pl-Holant}(\neq_2\mid f)$ is \#P-hard unless $f \in \mathscr{M} $, in which case, $\operatorname{Pl-Holant}(\neq_2\mid f)$ is tractable.
\end{theorem}

{\bf Proof.} 
Tractability follows by \ref{mtractable}.

Now suppose $f \notin \mathscr{M}$.
Connect the variable $x_4$ with $x_3$ of $f$ using $(\neq_2)$, and we get a binary signature $g_1$, where $$g_1=M_{x_1x_2, x_4x_3}(0, 1, 1, 0)^T=(0, b+c, z+y, 0)^T.$$
Connect the variable $x_1$ with $x_2$ of $f$ using $(\neq_2)$, and we get a binary signature $g_2$, where $$g_2=((0, 1, 1, 0)M_{x_1x_2, x_4x_3})^T=(0, b+z, c+y, 0)^T.$$
\begin{itemize}
    \item If one of $g_1$ and $g_2$ is of the form $(0, 0, 0, 0)^T$, then $by=(-c)(-z)=cz$. 
    That is $by-cz=0$. Here $c\neq 0$ due to $by\neq 0$.
    By Lemma~\ref{degenerate},  $\operatorname{Pl-Holant}(\neq_2\mid f)$ is \#P-hard.
    \item If one of $g_1$ and $g_2$ can be normalized as $(0, 1, 0, 0)$ or $(0, 0, 1, 0)$. By Lemma~\ref{pin}, $\operatorname{Pl-Holant}(\neq_2\mid f)$ is \#P-hard.
    \item If one of $g_1$ and $g_2$ can be normalized as $(0, 1, t, 0)^T$, where $t\neq 0$ is not a root of unity, then by Lemma~\ref{t<>1}, $\operatorname{Pl-Holant}(\neq_2\mid f)$ is \#P-hard.
    \item If one of $g_1$ and $g_2$ can be normalized as $(0, 1, t, 0)^T$, where $t$ is an $n$-th primitive root of unity and $n \geqslant 3$, then by Corollary~\ref{conformal3}, $\operatorname{Pl-Holant}(\neq_2\mid f)$ is \#P-hard.

\item Otherwise, $g_1$ and $g_2$ do not belong to those cases above, which means both $g_1$ and $g_2$ both can be normalized as $(0, 1, \epsilon_1, 0)$ and $(0, 1, \epsilon_2, 0)$, where $\epsilon_1=\pm 1$ and $\epsilon_2 =\pm 1$. That is, $b+c=\epsilon_1 (z+y) \neq 0$ and $b+z=\epsilon_2 (c+y) \neq 0$. 
\begin{itemize}
    \item If $b+c=z+y$ and $b+z=c+y$, then $b=y$ and $c=z$. 
This case will be proved below.
    \item If $b+c=-(z+y)$ and $b+z=c+y$, then $b+z=c+y=0$,
so $g_2 = (0, 0, 0, 0)^T$, a contradiction.
    \item If $b+c=z+y$ and $b+z=-(c+y)$, then 
    $b+c=z+y=0$, so $g_1 = (0, 0, 0, 0)^T$, a  contradiction.
    \item If $b+c=-(z+y)$ and $b+z=-(c+y)$,
we get $b+c + y + z =0$.
But $b+c \neq 0$, otherwise $g_1 = (0, 0, 0, 0)^T$, a  contradiction.
So we can normalize $g_1$ to  $(0, 1, -1, 0)^T$. 
    Modify $x_1=1$ of $f$ with $-1$ scaling, and we get a signature $f'$ with the signature matrix $M(f')=\left[\begin{smallmatrix}
0 & 0& 0& a\\
0 & b& c& 0\\
0 & -z& -y& 0\\
-x & 0& 0& 0\\
\end{smallmatrix}\right].$
Connect the variable $x_1$ with $x_2$ of $f'$ using $(\neq_2)$, and we get a binary signature $g^\prime=(0, b-z, c-y, 0)^T$. 
Same as the analysis of $g_1$ and $g_2$ above, we have $\operatorname{Pl-Holant}(\neq_2\mid f')$ is \#P-hard unless $g'$ can be normalized as $(0, 1, \epsilon_3, 0)$, where $\epsilon_3=\pm 1$. That is, $b-z=\epsilon_3 (c-y)\neq 0$, $\epsilon_3=\pm 1$.

\begin{itemize}
\item If $b-z=c-y$, combined with $b+c=-(z+y)$, we have $b=-y$ and $c=-z$. 
This case will be proved below.
\item If $b-z=-(c-y)$, combined with $b+c=-(z+y)$, we have $b+c=z+y=0$,
and so $g_1 = (0, 0, 0, 0)^T$, a   contradiction.
\end{itemize}
Therefore, $\operatorname{Pl-Holant}(\neq_2\mid f')$ is \#P-hard and hence  $\operatorname{Pl-Holant}(\neq_2\mid f)$ is \#P-hard.
\end{itemize}
\end{itemize}

To summarize, except for the cases $b=\epsilon y$ and $c= \epsilon z$, where $\epsilon = \pm 1$,  we have proved that $\operatorname{Pl-Holant}(\neq_2\mid f)$ is \#P-hard.
 We can connect the variable $x_2$ with $x_3$ of $f$ using $(\neq_2)$, and get a binary signature $g_3=(0, a+c, z+x, 0)^T.$  Connect the variable $x_1$ with $x_4$ of $f$ using $(\neq_2)$, and we get a binary signature $g_4=(0, a+z, c+x, 0)^T.$ Same as the analysis of $g_1$ and $g_2$, we have $\operatorname{Pl-Holant}(\neq_2\mid f)$ is \#P-hard unless $a= \epsilon' x$ and $c= \epsilon' z$, where $\epsilon' = \pm 1$. By both $c =  \epsilon z$ and $c =  \epsilon' z$
and $z \neq 0$ we get $\epsilon =  \epsilon'$.
Therefore, $\operatorname{Pl-Holant}(\neq_2\mid f)$ is \#P-hard unless $a=\epsilon x$, $b=\epsilon y$ and $c=\epsilon z$, where $\epsilon = \pm 1$. In this case, since $z\neq 0$, we have $abc\neq 0$.  By Lemma~\ref{twins}, $\operatorname{Pl-Holant}(\neq_2\mid f)$ is \#P-hard, since we have assumed
$f \notin \mathscr{M}$. \qed

\section{Proof of the Main Theorem}
Now we are ready to prove the main theorem, Theorem~\ref{main}.

{\bf Proof of Tractability:}
\begin{itemize}
\item If $f$ satisfies condition \ref{con1} or \ref{con2}, then by Theorem \ref{nonplanardic}, $\operatorname{Holant}(\neq_2\mid f)$ is tractable  without the planarity
restriction. Obviously, $\operatorname{Pl-Holant}(\neq_2\mid f)$ is  tractable.
\item If $f$ satisfies condition \ref{con3}, then by Theorem \ref{mtractable}, $\operatorname{Pl-Holant}(\neq_2\mid f)$ is tractable.
\item If $f$ satisfies condition \ref{con4}, then by Theorem \ref{inner}, $\operatorname{Pl-Holant}(\neq_2\mid f)$ is tractable.
\end{itemize}

{\bf Proof of Hardness:}

Since $f$ does not satisfy condition \ref{con2}, $f$ does not belong to Case \Rmnum{1}.  Therefore it belongs to Cases \Rmnum{2}, \Rmnum{3}, or \Rmnum{4}.
\begin{itemize}
\item Suppose $f$ belongs to Case \Rmnum{2}. 
\begin{itemize}
\item If an outer pair is a zero pair, since $f$ does not satisfy condition \ref{con1} or condition \ref{con3}, then by Theorem \ref{outer}, $\operatorname{Pl-Holant}(\neq_2\mid f)$ is \#P-hard.
\item If the inner pair is a zero pair and no outer pair is zero, since $f$ does not satisfy condition \ref{con4}, then by Theorem \ref{inner}, $\operatorname{Pl-Holant}(\neq_2\mid f)$ is \#P-hard.
\end{itemize}
\item Suppose $f$ belongs to Case \Rmnum{3}. Since $f$ does not satisfy condition \ref{con3}, then by Theorem \ref{onezero}, $\operatorname{Pl-Holant}(\neq_2\mid f)$ is \#P-hard.
\item Suppose $f$ belongs to Case \Rmnum{4}. Since $f$ does not satisfy condition \ref{con3}, then by Theorem \ref{nonzero}, $\operatorname{Pl-Holant}(\neq_2\mid f)$ is \#P-hard. \qed
\end{itemize} 

 \section*{Acknowledgment}
We thank Martin Dyer, Heng Guo, Dana Randall and Mingji Xia for their comments and
interests.

\bibliography{ref}{}


\end{document}